\documentclass[a4paper,11pt]{article}

\usepackage{amsfonts,bm,amssymb,euscript,array,longtable,ltxtable,hhline,cite,lscape,framed}
\usepackage{amsmath}
\usepackage{caption}
\usepackage[dvips]{epsfig}
\usepackage[all]{xy}
\usepackage[symbol]{footmisc}
\usepackage{verbatim}
\usepackage{mcite}
\usepackage{scalefnt}
\usepackage{indentfirst}
\usepackage[toc,page]{appendix}

\newcommand{\be}{\begin{equation}}
\newcommand{\ee}{\end{equation}}
\newcommand{\beq}{\begin{equation}}
\newcommand{\eeq}{\end{equation}}
\newcommand{\beqnn}{\begin{equation*}}
\newcommand{\eeqnn}{\end{equation*}}
\newcommand{\beqa}{\begin{eqnarray}}
\newcommand{\eeqa}{\end{eqnarray}} 
\newcommand{\beqann}{\begin{eqnarray*}}
\newcommand{\eeqann}{\end{eqnarray*}}

\def\nn{{\nonumber}}
\def\bea{\begin{eqnarray}}
\def\eea{\end{eqnarray}}

\def\b{\beta}

 \def\L{\Lambda}

\def\cH{{\cal H}}

\def\cN{{\cal N}}

\def\cR{{\cal R}}

\def\one{\mbox{1 \kern-.59em {\rm l}}}

\def\b#1{\overline {#1}}
\def\mbf#1{\mathbf {#1}}
\def\bb#1{\mathbb {#1}}

\def\lra{{\leftrightarrow}}

\def\AAA{SU(4)}
\def\AA{SU(3)}
\def\A#1{SU^{#1}(2)}
\def\Y{U(1)}

\def\ubr#1{\underbrace{#1}}

\def\bit{\begin{itemize}}
\def\eit{\end{itemize}}

\def\({\left(} \def\){\right)} 

\newcommand{\Dim}{{\rm dim}}
\newcommand{\adj}{{\rm adj}}
\newcommand{\rank}{{\rm rank}}

\begin{document}

\renewcommand{\title}[1]{\vspace{10mm}\noindent{\Large{\bf#1}}\vspace{8mm}} 
\newcommand{\authors}[1]{\noindent{\large#1}\vspace{5mm}}
\newcommand{\address}[1]{{\itshape #1\vspace{2mm}}}

\begin{titlepage}

\begin{center}

\title{{\LARGE {\sc Coset Space Dimensional Reduction}\\
{\sc and} \\
{\sc Wilson Flux Breaking of} \\[1.5ex]
{\sc Ten-Dimensional $\cN=1$, $E_{8}$ Gauge Theory}}}

\vskip 3mm

\authors{George {\sc Douzas}, 
Theodoros {\sc Grammatikopoulos}\footnote{e-mail: \texttt{tgrammat@mail.ntua.gr}}~\footnote{Supported by
the EPEAEK II programme IRAKLEITOS.},\\[1ex]
George {\sc Zoupanos}\footnote{e-mail: \texttt{George.Zoupanos@cern.ch}}~\footnote{Partially supported
by the NTUA programme for basic research "Karatheodoris" and the European Union's RTN programme under
contract MRTN-CT-2006-035505.}}

\vskip 5mm

 
\begin{abstract}
\noindent
We consider a $\cN=1$ supersymmetric $E_{8}$ gauge theory, defined in ten
dimensions and we determine all four-dimensional gauge theories resulting
from the generalized dimensional reduction a la Forgacs-Manton over coset 
spaces, followed by a subsequent application of the Wilson flux 
spontaneous symmetry breaking mechanism. Our investigation is constrained 
only by the requirements that (i) the dimensional reduction leads to 
the potentially phenomenologically interesting, anomaly free, 
four-dimensional $E_{6}$, $SO_{10}$ and $SU_{5}$ GUTs and (ii) the Wilson flux 
mechanism makes use only of the freely acting discrete symmetries of all 
possible six-dimensional coset spaces.
\end{abstract}

\end{center}

\end{titlepage}

\tableofcontents



\section{Introduction}

The LEP era has established the Standard Model (SM) as the present day front of our knowledge concerning the
theory of Elementary Particle Physics. On the other hand, already the plethora of its free parameters
is suggesting the existence of physics beyond the SM that could explain at least some of them and possibly
reduce them. The celebrated proposal of grand unification has inspired the particle physics community since it
was  providing an interesting reduction of couplings in the gauge sector of the SM. However, given that most
of the free parameters of the SM is related to the ad-hoc introduction of the Higgs and Yukawa sectors in the 
theory, it is natural to search for frameworks that could unify them. Then various schemes, with the Coset
Space Dimensional Reduction
(CSDR)~\cite{Forgacs:1979zs,*Witten:1976ck,Kubyshin:1989vd,Kapetanakis:1992hf,Harnad:1979in,
*Harnad:1980ct,Bais:1985yd,Green:1984bx, Chapline:1980mr,
Farakos:1986sm,*Farakos:1986cj,Forgacs:1985vp,Manton:1981es,Chapline:1982wy,Forgacs:1984zx,Olive:1982ai,
*Lust:1985be,
*Kapetanakis:1990rz,Manousselis:2000aj,*Manousselis:2001xb,*Manousselis:2001re,*Manousselis:2004xd}
being pioneer, were suggesting that a unification of gauge and Higgs sectors can be achieved in higher
dimensions. The four-dimensional gauge and Higgs fields are simply the surviving components of the gauge
fields of a pure gauge theory defined in higher dimensions. Moreover the addition of fermions in the
higher-dimensional gauge theory leads naturally after CSDR to Yukawa couplings in four dimensions. A major
achievement in this direction is the possibility to obtain chiral theories in four
dimensions~\cite{Manton:1981es,Chapline:1982wy}. A final step towards unification of the gauge and fermions
introduced in the higher dimensional theory is to demand that they are members of the same vector
supermultiplet of a higher dimensional $\cN=1$ supersymmetric gauge theory. Then another achievement is that
the CSDR over non-symmetric cosets leads to softly broken supersymmetric
theories~\cite{Manousselis:2000aj,*Manousselis:2001xb,*Manousselis:2001re,*Manousselis:2004xd} with all
parameters determined (at the classical level). The latter is in a very interesting contrast to the usual
supersymmetric extensions of the SM, where the soft supersymmetry breaking sector introduces a huge number of
new free parameters.

Concerning supersymmetry, the nature of the four-dimensional theory depends on the corresponding nature of the
compact space used to reduce the higher dimensional theory. Specifically the reduction over CY spaces leads to
supersymmetric theories~\cite{Green:1987sp,*Green:1987mn,*Lust:1989tj} in four dimensions, the reduction over
symmetric coset spaces leads to non-supersymmetric theories, while a reduction over non-symmetric ones leads
to softly broken supersymmetric
theories~\cite{Manousselis:2000aj,*Manousselis:2001xb,*Manousselis:2001re,*Manousselis:2004xd}.

In the spirit described above a very welcome additional input is that string theory suggests furthermore the
dimension and the gauge group of the higher dimensional supersymmetric
theory~\cite{Green:1987sp,*Green:1987mn,*Lust:1989tj}. Further support to this unified description comes from
the fact that the reduction of the theory over coset~\cite{Kapetanakis:1992hf} and CY
spaces~\cite{Green:1987sp,*Green:1987mn,*Lust:1989tj} provides the four-dimensional theory with scalars
belonging in the fundamental representation (rep.) of the gauge group as are introduced in the SM. In addition
the fact that the SM is a chiral theory lead us to consider $D$-dimensional supersymmetric gauge theories with
$D=4n+2$ \cite{Chapline:1982wy,Kapetanakis:1992hf}, which include the ten dimensions suggested by the
heterotic string theory~\cite{Green:1987sp,*Green:1987mn,*Lust:1989tj}.

For many years the studies on the reduction of string theories had as a dominant direction those that consider
the CY spaces as describing the higher compact dimensions. However one should note that there exist some 
problems too, mostly due to the complicated geometry of CY spaces. For instance their metric is not known
explicitly, while their Euler characteristic is usually too large to predict an acceptable number of fermion 
generations. Moreover, in Calabi-Yau compactifications the resulting low-energy field theory in four
dimensions contains a number of massless chiral fields, characteristic of the internal geometry, known as
moduli. These fields correspond to flat directions of the effective potential and therefore their values are
left undetermined. Since these values specify the masses and couplings of the four-dimensional theory, the
theory has limited predictive power.

Fortunately, the moduli problem in the form described above appears only in the simplest choice of string
backgrounds, where out of the plethora of  closed-string fields only the metric is assumed to be non-trivial.
By considering more general backgrounds involving ``fluxes" \cite{Strominger:1986uh, deWit:1986xg} as well as
non-perturbative effects \cite{Dine:1985rz, Derendinger:1985kk}, the four-dimensional theory can be provided
with potentials for some or all moduli. The terminology ``fluxes" refers to the inclusion of non-vanishing
field strengths for the ten-dimensional antisymmetric tensor fields with directions purely inside the internal
manifold.

The presence of fluxes has a dramatic impact on the geometry of the compactification space. Specifically, the
energy carried by the fluxes back-reacts on the geometry of the internal space and the latter is deformed away
from Ricci-flatness. Then, the CY manifolds used so often in string theory compactifications cease to be  true
solutions of the theory. For example, the requirement that some supersymmetry is preserved implies that the
internal manifold is a non-K\"ahler space for heterotic strings with NS-NS
fluxes~\cite{Cardoso:2002hd,Curio:2000dw, Becker:2003yv,*Becker:2003sh}, while it can be
a non-complex manifold for type IIA strings~\cite{Dall'Agata:2003ir,Behrndt:2004mj, Lust:2004ig}.

A considerable amount of literature has been devoted to the problem of including appropriately the
back-reaction of the fluxes on the internal manifold and constructing examples of manifolds which are
true solutions of the theory. In general, these manifolds have non-vanishing torsion. Consequently, demanding
that the low-energy theory is supersymmetric implies that the internal manifold admits a $G$-structure
\cite{Gauntlett:2003cy}. The existence of a $G$-structure is a generalization of the condition of special
holonomy.

The case of SU(3)-structures is of special interest since the structure group SO(6) of the internal space can
be reduced down to SU(3) in a way that a single spinor can be globally defined on it. This spinor is not
necessarily covariantly constant with respect to the Levi-Civita connection, as in the case of Calabi-Yau
manifolds, but it can be constant with respect to a torsionful connection. This condition allows for a wider
class of internal spaces, such as nearly-K\"{a}hler and half-flat manifolds. The Heterotic String theory has
been recently studied in this context in~\cite{Gurrieri:2007jg,*Benmachiche:2008ma}. Simple $G$-structure
manifolds are six dimensional cosets possessing an $SU(3)$-structure. They were identified as supersymmetric
solutions, e.g. in~\cite{Behrndt:2004km,*Koerber:2008rx,House:2005yc} 
for the case of type II theories. In the Heterotic Supergravity cosets were introduced
by~\cite{Lust:1986ix,*Castellani:1986rg,*Govindarajan:1986kb,*Govindarajan:1986iz} and recently studied
in~\cite{Micu:2004tz,*Frey:2005zz,Manousselis:2005xa}. Particularly, in~\cite{Manousselis:2005xa} it was shown
that supersymmetric compactifications of the Heterotic String theory of the form $AdS_4\times S/R$ exist when
background fluxes and general condensates are present. In addition, effective theories where constructed
in~\cite{House:2005yc,KashaniPoor:2007tr,*Caviezel:2008ik} in the case of type II supergravity. For a complete
list of references see~\cite{Grana:2005jc}. 

Due to the above developments and given that the non-symmetric six-dimensional coset spaces are
nearly-K\"{a}hler we plan a detailed investigation of the CSDR of the heterotic string in two directions.
The first concerns the supergravity sector~\cite{Chatzistavrakidis:2008ii}, while the second deals
with the gauge sector.

Here we limit ourselves in the study of the CSDR of the ten-dimensional $E_{8}$ gauge theory under certain
conditions. Specifically in the present work, starting with an $\cN=1$, $E_{8}$ Yang-Mills-Dirac theory
defined in ten dimensions we classify the semi-realistic particle physics models resulting from their CSDR and
a subsequent application of Wilson flux spontaneous symmetry breaking. The space-time on which the theory is
defined can be written in the compactified form $M^{4}\times B$, with $M^{4}$ the ordinary Minkowski spacetime
and $B=S/R$ a six-dimensional homogeneous coset space. We constrain our investigation in those cases that the
dimensional reduction leads in four dimensions to phenomenologically interesting and anomaly-free Grand
Unified Theories (GUTs) such as $E_{6}$, $SO(10)$, $SU(5)$. However since, as already mentioned, the
four-dimensional surviving scalars transform in the fundamental rep. of the resulting gauge group are not
suitable for the GUT breaking towards the SM. As a way out has been
suggested~\cite{Zoupanos:1987wj,Kapetanakis:1992hf} to take advantage of non-trivial topological
properties of the extra compactification coset space, apply the Hosotani-Wilson flux breaking
mechanism~\cite{Hosotani:1983xw,*Hosotani:1983vn,Witten:1985xc} and break the gauge symmetry of the
theory further. The second constraint that we impose in our investigation is that the discrete symmetries,
which we employ when we apply the Wilson flux mechanism, act freely on all possible six-dimensional coset
spaces, i.e. after we mode out the discrete symmetries from a given coset space, there are no points in the
resulting space that remain invariant. This is an obvious requirement to the extent that we deal with field
theory which we assume in the present work. Our main objective is the investigation to which extent applying
both methods namely CSDR and Wilson flux breaking mechanism, one can obtain reasonable low energy models.

In section (sec.)~\ref{sec:CSDR} we present the CSDR scheme in sufficient detail to make the paper
self-contained. We recall some elements of the coset space geometry (sec.~\ref{sec:CosetSpaceGeometry}), the
principle of the CSDR scheme and the constraints that the surviving fields of the four-dimensional theory have
to obey~(secs~\ref{sec:CSDR-rules},~\ref{sec:4D-theory}), and we finally make some remarks on the GUTs that
come from the CSDR scheme~(sec.~\ref{Remarks-on-CSDR}). In sec.~\ref{sec:WilsonFlux-theory} we recall the
Wilson flux breaking mechanism and pave the way for the full investigation which is presented in
sec~\ref{Classification}. More specifically, after recalling the mechanism itself
(sec.~\ref{sec:WilsonFluxBreaking-theory}), we comment on the freely acting discrete symmetries of the coset
spaces we use, which potentially lead to models with phenomenological interest
(sec.~\ref{sec:InterestingDiscreteSymmetries}). In
secs~\ref{TopologicallyInducedGaugeGroupBreaking-E6} and~\ref{TopologicallyInducedGaugeGroupBreaking-SO10}
we determine the topologically induced symmetry breaking patterns of the GUTs of our present interest.
In sec.~\ref{Classification} we present our investigation and a complete list of our results, on which we
comment in sec.~\ref{sec:Conclusions}.

\section{Coset Space Dimensional Reduction\label{sec:CSDR}}

Given a gauge theory defined in higher dimensions the obvious way to dimensionally reduce it is to demand that
the field dependence on the extra coordinates is such that the Lagrangian is independent of them. A crude way
to fulfill this requirement is to discard the field dependence on the extra coordinates, while an elegant one
is to allow for a non-trivial dependence on them, but impose the condition that a symmetry transformation by
an element of the isometry group $S$ of the space formed by the extra dimensions $B$ corresponds to a gauge
transformation. Then the Lagrangian will be independent of the extra coordinates just because it is gauge
invariant. This is the basis of the CSDR scheme~\cite{Forgacs:1979zs,Kubyshin:1989vd,Kapetanakis:1992hf},
which assumes that $B$ is a compact coset space, $S/R$.

In the CSDR scheme one starts with a Yang-Mills-Dirac Lagrangian, with gauge group $G$, defined on a
$D$-dimensional spacetime $M^{D}$, with metric $g^{MN}$, which is compactified to $ M^{4} \times S/R$ with
$S/R$ a coset space. The metric is assumed to have the form
\begin{equation}
g^{MN}=
\left(\begin{array}{cc}\eta^{\mu\nu}&0\\0&-g^{ab}\end{array}
\right),\label{Compactified-metric}
\end{equation}
where $\eta^{\mu\nu}= diag(1,-1,-1,-1)$ and $g^{ab}$ is the coset space metric. The requirement that
transformations of the fields under the action of the symmetry group of $S/R$ are compensated by gauge
transformations lead to certain constraints on the fields. The solution of these constraints provides us with
the four-dimensional unconstrained fields as well as with the gauge invariance that remains in the theory
after dimensional reduction. Therefore a potential unification of all low energy interactions, gauge,
Yukawa and Higgs is achieved, which was the first motivation of this framework.

It is interesting to note that the fields obtained using the CSDR approach are the first terms in the
expansion of the $D$-dimensional fields in harmonics of the internal space $S/R$. The effective field theories
resulting from compactification of higher dimensional theories contain also towers of massive higher harmonics
(Kaluza-Klein) excitations, whose contributions at the quantum level alter the behavior of the running
couplings from logarithmic to power~\cite{Taylor:1988vt}. As a result the traditional picture of unification
of couplings may change drastically~\cite{Dienes:1998vg,*Kim:2007jg}. Higher dimensional theories have also
been studied at the quantum level using the continuous Wilson renormalization
group~\cite{Kobayashi:1998ye,Kubo:1999ua} which can be formulated in any number of space-time dimensions with
results in agreement with the treatment involving massive Kaluza-Klein excitations.

In the following we give a short description of the CSDR scheme, the constraints which have to be satisfied
by the field content of the theory and recall how a four-dimensional gauge theory of unconstrained fields can
be obtained. Complete reviews can be found in~\cite{Kapetanakis:1992hf,Castellani:1999fz}.

\subsection{Coset space geometry}\label{sec:CosetSpaceGeometry}

To recall some aspects of the coset space geometry, we divide the generators of $S$, $ Q_{A}$ in two
sets: the generators of $R$, $Q_{i}$ $(i=1, \ldots,\Dim R)$, and the generators of $S/R$,
$ Q_{a}$($a=\Dim R+1 \ldots,\Dim S)$, and $\Dim(S/R)=\Dim S-\Dim R =d$. Then the commutation relations for the
generators of $S$ are the following
\begin{subequations}
\begin{align}
&\left[ Q_{i},Q_{j} \right] = f_{ij}{}^{k} Q_{k}\,, \\
&\left[ Q_{i},Q_{a} \right]= f_{ia}{}^{b}Q_{b}\,,\\
&\left[ Q_{a},Q_{b} \right]=f_{ab}{}^{i}Q_{i}+f_{ab}{}^{c}Q_{c}\,.\label{S/R-comm-rels}
\end{align}
\end{subequations}
So $S/R$ is assumed to be a reductive but in general non-symmetric coset space. When $S/R$ is symmetric,
the $f_{ab}{}^{c}$ in (\ref{S/R-comm-rels}) vanish. Let us call the coordinates of $M^{4} \times S/R$ space
$x^{M}= (x^{\mu},y^{\alpha})$, where $\alpha$ is a curved index of the coset,  $a$ is a tangent space index
and $y$ defines an element of $S$ which is a coset representative, $L(y)$. The vielbein and the
$R$-connection are defined through the Maurer-Cartan form which takes values in the Lie algebra of $S$
\begin{equation}
L^{-1}(y)dL(y) = e^{A}_{\alpha}Q_{A}dy^{\alpha}\,.\label{R-connection}
\end{equation}
Using (\ref{R-connection}) we can compute that at the origin $y = 0$, $ e^{a}_{\alpha} = \delta^{a}_{\alpha}$
and $e^{i}_{\alpha}= 0$ which is a usefull result in order to determine the constraints that the
four-dimensional matter fields have to obey, as we recall in sec.~\ref{sec:CSDR-rules}.

A connection on $S/R$ which is described by a connection-form $\theta^{a}_{\ b}$, has in general torsion and
curvature. In the general case where torsion may be non-zero, we calculate first the torsionless part
$\omega^{a}_{\ b}$ by setting the torsion form $T^{a}$ equal to zero,
\begin{equation}
T^{a} = de^{a} + \omega^{a}_{\ b} \wedge e^{b} = 0\,,
\end{equation}
while using the Maurer-Cartan equation,
\begin{equation}
de^{a} = \frac{1}{2}f^{a}{}_{bc}e^{b}\wedge e^{c} +f^{a}{}_{bi}e^{b}\wedge e^{i}\,,
\end{equation}
we see that the condition of having vanishing torsion is solved by
\begin{equation}
\omega^{a}{}_{b}= -f^{a}{}_{ib}e^{i}-D^{a}{}_{bc}e^{c}\,,
\end{equation}
where
$$
 D^{a}{}_{bc}=\frac{1}{2}g^{ad}[f_{db}{}^{e}g_{ec}+f_{cb}{}^{e} g_{de}- f_{cd}{}^{e}g_{be}]\,.
$$
Note that the connection-form $\omega^{a}{}_{b}$ is $S$-invariant. This means that parallel transport commutes
with  the $S$ action~\cite{Castellani:1999fz}.

In the case of non-vanishing torsion we have
\begin{equation}
T^{a} = de^{a} + \theta^{a}{}_{b} \wedge e^{b}\,,
\end{equation}
where
$$
\theta^{a}{}_{b}=\omega^{a}{}_{b}+\tau^{a}{}_{b}\,,
$$
with
\begin{equation}
\tau^{a}{}_{b} = - \frac{1}{2} \Sigma^{a}{}_{bc}e^{c}\,,
\end{equation}
while the contorsion $ \Sigma^{a}{}_{bc} $ is given by
\begin{equation}
\Sigma^{a}{}_{bc} = T^{a}{}_{bc}+T_{bc}{}^{a}-T_{cb}{}^{a}
\end{equation}
in terms of the torsion components $ T^{a}{}_{bc} $. Therefore in general and for the case of non-symmetric
cosets the connection-form $ \theta^{a}{}_{b}$ is
\begin{equation}
\theta^{a}{}_{b} = -f^{a}{}_{ib}e^{i} -\left(D^{a}{}_{bc}+\frac{1}{2}\Sigma^{a}{}_{bc}\right)e^{c}
=-f^{a}{}_{ib}e^{i}-G^{a}{}_{bc}e^{c}\,.
\label{Connection}
\end{equation}
The natural choice of torsion which would generalize the case of equal
radii~\cite{Lust:1986ix,*Castellani:1986rg,Gavrilik:1999xr,MuellerHoissen:1987cq,Batakis:1989gb},
$T^{a}{}_{bc}=\eta f^{a}{}_{bc}$ would be $T^{a}{}_{bc}=2\tau D^{a}{}_{bc}$ except that the $D$'s do not have
the required symmetry properties. Therefore we must define $\Sigma$ as a combination of $D$'s which makes
$\Sigma$ completely antisymmetric and $S$-invariant according to the definition given above. Thus we are led
to the definition
\begin{equation}
\Sigma_{abc} \equiv 2\tau(D_{abc}+D_{bca}-D_{cba})\,.\label{Sigma}
 \\
\end{equation}
By choosing vanishing  parameter $\tau$ in the eqs~(\ref{Sigma}) and~(\ref{Connection}) above we obtain the
{ \it Riemannian connection}, $\theta_{R\ \ b}^{\ a}=-f^{a}{}_{ib}e^{i}-D^{a}{}_{bc}e^{c}$. On the other hand,
by adjusting the radii and $\tau$ we can obtain the { \it canonical connection},
$\theta_{C \ \ b}^{\ a} =-f^{a}{}_{bi}e^{i}$ which is an $R$-gauge field
\cite{Lust:1986ix,*Castellani:1986rg}.
In general though the $\theta^{a}{}_{b}$ connection in its general form is an $SO(6)$ field, i.e. lives on the
tangent space of the six-dimensional cosets we consider and describes their general holonomy. In sec.
\ref{sec:4D-theory} we will show how the $G^{ab}{}_{c}$ term of eq.~(\ref{Connection}) it is connected with
the geometrical and torsion contributions that the masses of the surviving four-dimensional gaugini acquire.
Since we are interested here in four-dimensional models without light supersymmetric particles we keep
$\theta^{a}{}_{b}$ general.

\subsection{Reduction of a $D$-dimensional Yang-Mills-Dirac Lagrangian}\label{sec:CSDR-rules}

The group $S$ acts as a symmetry group on the extra coordinates. The CSDR scheme demands that an
$S$-transformation of the extra $d$ coordinates is a gauge transformation of the fields that are defined on
$M^{4}\times S/R$,  thus a gauge invariant Lagrangian written on this space is independent of the extra
coordinates.

To see this in detail we consider a $D$-dimensional Yang-Mills-Dirac theory with gauge group $G$ defined on a
manifold $M^{D}$ which as stated will be compactified to $M^{4}\times S/R$, $D=4+d$, $d=\Dim S-\Dim R$
\begin{equation}
A=\int d^{4}xd^{d}y\sqrt{-g}\Bigl[-\frac{1}{4}
Tr\left(F_{MN}F_{K\Lambda}\right)g^{MK}g^{N\Lambda}
+\frac{i}{2}\overline{\psi}\Gamma^{M}D_{M}\psi\Bigr]\,,
\end{equation}
where
\begin{equation}
D_{M}= \partial_{M}-\theta_{M}-A_{M}\,,
\end{equation}
with
\begin{equation}
\theta_{M}=\frac{1}{2}\theta_{MN\Lambda}\Sigma^{N\Lambda}
\end{equation}
the spin connection of $M^{D}$, and
\begin{equation}
F_{MN}
=\partial_{M}A_{N}-\partial_{N}A_{M}-\left[A_{M},A_{N}\right]\,,
\end{equation}
where $M$, $N$ run over the $D$-dimensional space. The fields $A_{M}$ and $\psi$ are, as explained, symmetric
in the sense that any transformation under symmetries of $S/R$  is compensated by gauge transformations. The
fermion fields can be in any rep. $F$ of $G$ unless a further symmetry is required. Here since we assume
dimensional reductions of $\cN=1$ supersymmetric gauge theory the higher dimensional fermions have to
transform in the adjoint of higher dimensional gauge group. To be more specific let $\xi_{A}^{\alpha}$, $A
=1,\ldots,\Dim S$, be the Killing vectors which generate the symmetries of $S/R$ and $W_{A}$ the compensating
gauge transformation associated with $\xi_{A}$. Define next the infinitesimal coordinate transformation
as $\delta_{A} \equiv L_{\xi_{A}}$, the Lie derivative with respect to $\xi$, then we have for the scalar,
vector and spinor fields,
\begin{align}
&\delta_{A}\phi=\xi_{A}^{\alpha}\partial_{\alpha}\phi=D(W_{A})\phi\,,
\nonumber \\
&\delta_{A}A_{\alpha}=\xi_{A}^{\beta}\partial_{\beta}A_{\alpha}+\partial_{\alpha}
\xi_{A}^{\beta}A_{\beta}=\partial_{\alpha}W_{A}-[W_{A},A_{\alpha}]\,,\label{CSDR-constraints}\\
&\delta_{A}\psi=\xi_{A}^{\alpha}\psi-\frac{1}{2}G_{Abc}\Sigma^{bc}\psi=
D(W_{A})\psi\,. \nonumber
\end{align}
$W_{A}$ depend only on internal coordinates $y$ and $D(W_{A})$ represents a gauge transformation in the
appropriate reps of the fields. $G_{Abc}$ represents a tangent space rotation of the spinor fields. The
variations $\delta_{A}$ satisfy, $[\delta_{A},\delta_{B}]=f_{AB}{}^{C}\delta_{C}$ and lead to the following
consistency relation for $W_{A}$'s,
\begin{equation}
\xi_{A}^{\alpha}\partial_{\alpha}W_{B}-\xi_{B}^{\alpha}\partial_{\alpha}
W_{A}-\left[W_{A},W_{B}\right]=f_{AB}{}^{C}W_{C}\,.
\end{equation}
Furthermore the W's themselves transform under a gauge transformation \cite{Kapetanakis:1992hf} as,
\begin{equation}
{W}^{(g)}_{A} = g\,W_{A}\,g^{-1}+(\delta_{A}g)g^{-1}\,.\label{W-gauge-transf}
\end{equation}
Using (\ref{W-gauge-transf}) and the fact that the Lagrangian is independent of $y$ we can do all calculations
at $y=0$ and choose a gauge where $W_{a}=0$.

The detailed analysis of the constraints (\ref{CSDR-constraints}) given in 
refs~\cite{Forgacs:1979zs,Kapetanakis:1992hf} provides us with the four-dimensional unconstrained fields as
well as with the gauge invariance that remains in the theory after dimensional reduction. Here we give the
results. The components $A_{\mu}(x,y)$ of the initial gauge field $A_{M}(x,y)$ become, after dimensional
reduction, the four-dimensional gauge fields and furthermore they are independent of $y$. In addition one can
find that they have to commute with the elements of the $R_{G}$ subgroup of $G$. Thus the four-dimensional
gauge group $H$ is the centralizer of $R$ in $G$, $H=C_{G}(R_{G})$. Similarly, the $A_{\alpha}(x,y)$
components of $A_{M}(x,y)$ denoted by $\phi_{\alpha}(x,y)$ from now on, become scalars at four dimensions.
These fields transform under $R$ as a vector ${\rm v}$, i.e.
\beq
\begin{array}{r@{}c@{}l@{}}
S~&\supset&~ R  \\
\adj\,S~&=&~ \adj\,R +{\rm v}\label{RinS-embedding}\,.
\end{array}
\eeq
Moreover $\phi_{\alpha}(x,y)$ act as an intertwining operator connecting induced representations (reps) of $R$
acting on $G$ and $S/R$. This implies, exploiting Schur's lemma, that the transformation properties of the
fields $\phi_{\alpha}(x,y)$ under $H$ can be found if we express the adjoint rep. of $G$ in terms of $R_{G}
\times H$
\beq
\begin{array}{r@{}c@{}l@{}}
G~&\supset&~ R_{G} \times H  \\
\adj\,G~&=&~(\adj\,R,1)+(1,\adj\,H)+\sum(r_{i},h_{i})\label{RinG-embedding}\,.
\end{array}
\eeq
Then if ${\rm v}=\sum s_{i}$, where each $s_{i}$ is an irreducible representation (irrep.) of $R$, there
survives an $h_{i}$ multiplet for every pair $(r_{i},s_{i})$, where $r_{i}$ and $s_{i}$ are identical irreps
of $R$.

Turning next to the fermion
fields~\cite{Kapetanakis:1992hf,Manton:1981es,Chapline:1982wy,Wetterich:1982ed,Palla:1983re,Pilch:1984xx,
Forgacs:1985vp,Barnes:1986ea} similarly to scalars, they act as intertwining operators between induced reps
acting on $G$ and the tangent space of $S/R$, $SO(d)$. Proceeding along similar lines as in the case of
scalars to obtain the rep. of $H$ under which the four-dimensional fermions transform, we have to decompose
the rep. $F$ of the initial gauge group in which the fermions are assigned under $R_{G} \times H$, i.e.
\begin{equation}
F= \sum (t_{i},h_{i})\,,\label{Fdecomp}
\end{equation}
and the spinor of $SO(d)$ under $R$
\begin{equation}
\sigma_{d} = \sum \sigma_{j}\,.\label{Frule}
\end{equation}
Then for each pair $t_{i}$ and $\sigma_{i}$, where $t_{i}$ and $\sigma_{i}$ are irreps there is an
$h_{i}$ multiplet of spinor fields in the four-dimensional theory. In order however  to obtain chiral fermions
in the effective theory we have to impose further requirements. We first impose the Weyl condition in $D$
dimensions. In $D = 4n+2$ dimensions which is the case at hand, the decomposition of the left handed, say
spinor under $SU(2) \times SU(2) \times SO(d)$ is
\begin{equation}
\sigma _{D} = (2,1,\sigma_{d}) + (1,2,\overline{\sigma}_{d})\,.
\end{equation}
Furthermore in order to be $\sigma_{d}\neq\b{\sigma}_{d}$ the coset space $S/R$ must be such that
$\rank(R)=\rank(S)$~\cite{Bott:1965,Kapetanakis:1992hf}. The six-dimensional coset spaces which satisfy this
condition are listed in the first column of tables~\ref{SO6VectorSpinorContentSCosets}
and~\ref{SO6VectorSpinorContentNSCosets}. Then under the $SO(d)\supset R$ decomposition we have
\begin{equation}
\sigma_{d} = \sum \sigma_{k}\,,\qquad\overline{\sigma}_{d}= \sum \overline{\sigma}_{k}\,.
\end{equation}

\setlength{\tabcolsep}{2pt} 
\setlength{\arraycolsep}{1.5pt} 
\begin{table}[htbp]
\begin{center}
\begin{scriptsize}
\begin{tabularx}{15.5cm}{|c!{\vrule width 1pt}c!{\vrule width 1pt} c|c!{\vrule width1 pt}
X|l|}\hline
{\bf Case}
&${\mbf 6D}$~{\bf Coset~Spaces} 
& ${\mbf Z(S)}$
& ${\mbf W}$
& ${\mbf V}$
& ${\mbf F}$  \\ \hhline{*{6}{=}}                                             
{\bf a}
& $\frac{SO(7)}{SO(6)}$
& ${\bb Z}_{2}$
& ${\bb Z}_{2}$
&
$\begin{array}{l}
{\mbf 6} \lra {\mbf 6}
\end{array}$
&
$\begin{array}{l}
{\mbf 4} \lra \b{{\mbf 4}}
 \end{array}$
\\\hline
{\bf b}
&$\frac{SU(4)}{SU(3)\times U(1)}$
&${\bb Z}_{4}$
&$\one$
& 
$\begin{array}{l}
{\mbf 6}={\mbf 3}_{(-2)} + \b{{\mbf 3}}_{(2)} \\
                                     -
\end{array}$
& 
$\begin{array}{l}
{\mbf 4}={\mbf 1}_{(3)}+{\mbf 3}_{(-1)}\\
              -
\end{array}$
\\\hline
{\bf c}
& $\frac{Sp(4)}{(SU(2)\times U(1))_{max}}$ 
& ${\bb Z}_{2}$
& ${\bb Z}_{2}$
&
$\begin{array}{l}
{\mbf 6}={\mbf 3}_{(-2)} +{\mbf 3}_{(2)}\\
{\mbf 3}_{(-2)}\lra{\mbf 3}_{(2)}
\end{array}$
&
$\begin{array}{l}
{\mbf 4}={\mbf 1}_{(3)}+{\mbf 3}_{(-1)}\\
{\mbf 1}_{(3)}\lra{\mbf 1}_{(-3)}~~{\mbf 3}_{(-1)}\lra{\mbf 3}_{(1)}
\end{array}$
\\\hline
{\bf d}
& 
$\begin{array}{l}
\left(\frac{SU(3)}{SU(2)\times U(1)}\right)\\
\hfill\times\left(\frac{SU(2)}{U(1)}\right)
\end{array}$
& ${\bb Z}_{2}\times {\bb Z}_{3}$
& ${\bb Z}_{2}$
&
$\begin{array}{l}
\begin{array}{l@{}l@{}l@{}}
{\mbf 6}&=&{\mbf 1}_{(0,2a)}+{\mbf 1}_{(0,-2a)}\\
           &+&{\mbf 2}_{(b,0)}+{\mbf 2}_{(-b,0)}
\end{array}\\
{\mbf 1}_{(0,2a)}\lra{\mbf 1}_{(0,-2a)}
\end{array}$
&
$\begin{array}{l}
{\mbf 4}={\mbf 2}_{(0,a)}+{\mbf 1}_{(b,-a)}+{\mbf 1}_{(-b,-a)}\\
{\mbf 2}_{(0,a)}\lra{\mbf 2}_{(0,-a)}\\
{\mbf 1}_{(b,-a)}\lra{\mbf 1}_{(b,a)}\\
{\mbf 1}_{(-b,-a)}\lra{\mbf 1}_{(-b,a)}
\end{array}$
\\\hline
{\bf e}
&
$\begin{array}{l}
\left(\frac{Sp(4)}{SU(2)\times SU(2)}\right)\\
\hfill\times\left(\frac{SU(2)}{U(1)}\right)
\end{array}$
& $({\bb Z}_{2})^{2}$ 
& $({\bb Z}_{2})^{2}$
&
$\begin{array}{l}
{\mbf 6}=({\mbf 2},{\mbf 2})_{(0)}+({\mbf 1},{\mbf 1})_{(2)}+({\mbf 1},{\mbf 1})_{(-2)}\\
\mbox{(${\bb Z}_{2}$ of $SU(2)/U(1)$)}\\
({\mbf 1},{\mbf 1})_{(2)}\lra({\mbf 1},{\mbf 1})_{(-2)}
\end{array}$
&
$\begin{array}{l}
{\mbf 4}=({\mbf 2},{\mbf 1})_{(1)}+({\mbf 1},{\mbf 2})_{(-1)}\\
({\mbf 2},{\mbf 1})_{(1)}\lra({\mbf 2},{\mbf 1})_{(-1)}\\
({\mbf 1}, {\mbf 2})_{(1)}\lra({\mbf 1},{\mbf 2})_{(-1)}
\end{array}$
\\\hline
{\bf f}
& $\left(\frac{SU(2)}{U(1)}\right)^{3}$
& $({\bb Z}_{2})^{3}$
& $({\bb Z}_{2})^{3}$
&
$\begin{array}{l}
\begin{array}{l@{}l@{}l@{}}
{\mbf 6}&=&(2a,0,0)+(0,2b,0)+(0,0,2c)\\
            &+&(-2a,0,0)+(0,-2b,0)+(0,0,-2c)\\
\end{array}\\
\mbox{each ${\bb Z}_{2}$ changes the sign of $a$,$b$, $c$}
\end{array}$
&
$\begin{array}{l}
\begin{array}{l@{}l@{}l@{}}
{\mbf 4}&=&(a,b,c)+(-a,-b,c)\\
           &+&(-a,b,-c)+(a,-b,-c)
\end{array}\\
\mbox{each ${\bb Z}_{2}$ changes the sign of}\\
\mbox{$a$,$b$, $c$}
\end{array}$
\\\hline
\end{tabularx}
\end{scriptsize}
\end{center}
\caption{\textbf{Six-dimensional symmetric cosets spaces with $\rank(R)=\rank(S)$.}
\emph{The  freely acting discrete symmetries ${\rm Z}(S)$ and ${\rm W}$ for each case are listed.
The transformation properties of the vector and spinor representations under $R$ are also
noted.}\label{SO6VectorSpinorContentSCosets}}
\end{table}

\setlength{\tabcolsep}{2pt}
\setlength{\arraycolsep}{1.5pt}
\begin{table}[htbp]
\begin{center}
\begin{scriptsize}
\begin{tabularx}{15.5cm}{|c!{\vrule width 1pt}c!{\vrule width 1pt}c|c!{\vrule width
1pt}X|l|}\hline
{\bf Case}
& ${\mbf 6D}$~{\bf Coset~Spaces} 
& ${\mbf Z(S)}$
& ${\mbf W}$
& ${\mbf V}$
& ${\mbf F}$  \\ \hhline{*{6}{=}}                                       
{\bf a'}
& $\frac{G_{2}}{SU(3)}$
& ${\mbf 1}$
& ${\bb Z}_{2}$
&
$\begin{array}{l}
{\mbf 6}={\mbf 3}+\b{{\mbf 3}}\\
{\mbf 3}\lra\b{{\mbf 3}}
\end{array}$
&
$\begin{array}{l}
{\mbf 4}={\mbf 1}+{\mbf 3}\\
{\mbf 1}\lra{\mbf 1}\\
{\mbf 3}\lra\b{{\mbf 3}}
\end{array}$
\\\hline
{\bf b'}
& $\frac{Sp(4)}{(SU(2) \times U(1))_{nonmax}}$
& ${\bb Z}_{2}$
& ${\bb Z}_{2}$
&
$\begin{array}{l}
{\mbf 6}={\mbf 1}_{(2)}+{\mbf 1}_{(-2)} + {\mbf 2}_{(1)}+{\mbf 2}_{(-1)}\\
{\mbf 1}_{(2)}\lra{\mbf 1}_{(-2)}\\
{\mbf 2}_{(1)}\lra{\mbf 2}_{(-1)}
\end{array}$
&
$\begin{array}{l}
{\mbf 4}={\mbf 1}_{(0)}+{\mbf 1}_{(2)}+{\mbf 2}_{(-1)}\\
{\mbf 1}_{(2)}\lra{\mbf 1}_{(-2)}~~{\mbf 1}_{(0)}\lra{\mbf 1}_{(0)}\\
{\mbf 2}_{(1)}\lra{\mbf 2}_{(-1)}
\end{array}$
\\\hline
{\bf c'}
&$\frac{SU(3)}{U(1)\times U(1)}$
&${\bb Z}_{3}$
&${\mbf S}_{3}$
&
$\begin{array}{l@{}l@{}l@{}}
{\mbf 6}&=&(a,c)+(b,d)+(a+b,c+d)\\
           &+&(-a,-c)+(-b,-d)\\
           &+&(-a-b,-c-d)
\end{array}$
&
$\begin{array}{l@{}l@{}l@{}}
{\mbf 4}&=&(0,0)\\
           &+&(a,c)+(b,d)+(-a-b,-c-d)
\end{array}$\\[25pt]
\ \ 
& \  \
&\   \ 
& ${\bb Z}_{2}$
&
$\begin{array}{l}
(b,d)\lra(-b,-d)\\
(a+b,c+d)\lra(a,c)\\
(-a,-c)\lra(-a-b,-c-d)
\end{array}$
&
$\begin{array}{l}
(b,d)\lra(-b,-d)\\
(a,c)\lra(a+b,c+d)\\
(-a-b,-c-d)\lra(-a,-c)
\end{array}$\\[25pt]
\  \
& \  \
&\  \
&${\bb Z}_{2}$
&
$\begin{array}{l}
(b,d)\lra(a+b,c+d)\\
(a,c)\lra(-a,-c)\\
(-b,-d)\lra(-a-b,-c-d)
\end{array}$
&
$\begin{array}{l}
(b,d)\lra(a+b,c+d)\\
(a,c)\lra(-a,-c)\\
(-a-c,-b-d)\lra(-b,-d)
\end{array}$\\ [25pt]
\ \
&\  \
& \ \
& ${\bb Z}_{2}$
&
$\begin{array}{l}
(b,d)\lra(-a,-c)\\
(a+b,c+d)\lra(-a-b,-c-d)\\
(a,c)\lra(-b,-d)
\end{array}$
&
$\begin{array}{l}
(b,d)\lra(-a,-c)\\
(a,c)\lra(-b,-d)\\
(-a-b,-c-d)\lra(a+b,c+d)
\end{array}$
\\\hline
\end{tabularx}
\end{scriptsize}
\end{center}
\caption{\textbf{Six-dimensional non-symmetric cosets spaces with $\rank(R)=\rank(S)$.}
\emph{The available freely acting discrete symmetries ${\rm Z}(S)$ and ${\rm W}$ for each case are
listed. The transformation properties of vector and spinor representations under $R$ are also
noted.}\label{SO6VectorSpinorContentNSCosets}}
\end{table}

In the following sections we assume that the higher dimensional theory is $\cN=1$ supersymmetric. Therefore
the higher dimensional fermion fields have to be considered transforming in the adjoint of $E_{8}$ which is
vectorlike. In this case each term $(t_{i},h_{i})$ in eq.~(\ref{Fdecomp}) will be either self-conjugate or it
will have a partner $(\overline{t}_{i},\overline{h}_{i})$. According to the rule described in
eqs~(\ref{Fdecomp}),~(\ref{Frule}) and considering $\sigma_{d}$ we will have in four dimensions left-handed
fermions transforming as $ f_{L} = \sum h^{L}_{k}$. It is important to notice that since $\sigma_{d}$ is non
self-conjugate, $f_{L}$ is non self-conjugate too. Similarly from $\overline{\sigma}_{d}$ we will obtain the
right handed rep. $ f_{R}= \sum \overline{h}^{R}_{k}$ but as we have assumed that $F$ is vector-like,
$\overline{h}^{R}_{k}\sim h^{L}_{k}$. Therefore there will appear two sets of Weyl fermions with the same
quantum numbers under $H$. This is already a chiral theory but still one can go further and try to impose the
Majorana condition in order to eliminate the doubling of the fermion spectrum. However this is not required in
the present case of study, where we apply the Hosotani mechanism for the further breaking of the gauge
symmetry, as we will explain in sec.~\ref{sec:WilsonFlux-theory}.

An important requirement is that the resulting four-dimensional theories should be anoma\-ly free.
Starting with an anomaly free theory in higher dimensions,  Witten~\cite{Witten:1984dg}  has given the
condition to be fulfilled in order to obtain anomaly free four-dimensional theories. The condition restricts
the allowed embeddings of $R$ into $G$ by relating them with the embedding of $R$ into $SO(6)$, the tangent
space of the six-dimensional cosets we consider~\cite{Pilch:1985qf,Kapetanakis:1992hf}. To be more specific if
$\L_{a}$ are the generators of $R$ into $G$ and $T_{a}$ are the generators of $R$ into $SO(6)$ the condition
reads
\beq
Tr(\L_{a}\L_{b})=30\,Tr(T_{a}T_{b})\,.\label{WittenCond}
\eeq
According to ref.~\cite{Pilch:1985qf} the anomaly cancellation condition~(\ref{WittenCond}) is automatically
satisfied for the choice of embedding
\beq
E_{8}\supset SO(6) \supset R\,,\label{RinSO6inGembedding}
\eeq
which we adopt here. Furthermore concerning the abelian group factors of the four-dimensional gauge theory,
we note that the corresponding gauge bosons surviving in four dimensions become massive at the
compactification scale~\cite{Green:1984bx,Witten:1984dg} and therefore, they do not contribute in the
anomalies;  they correspond only to global symmetries.

\subsection{The four-dimensional theory}\label{sec:4D-theory}

Next let us obtain the four-dimensional effective action. Assuming that the metric is block diagonal, taking
into account all the constraints and integrating out the extra coordinates we obtain in four dimensions the
following Lagrangian
\begin{equation}
A=C \int d^{4}x \biggl( -\frac{1}{4} F^{t}_{\mu\nu}{F^{t}}^{\mu\nu}+\frac{1}{2}(D_{\mu}\phi_{a})^{t}
(D^{\mu}\phi^{a})^{t}
+V(\phi)+\frac{i}{2}\overline{\psi}\Gamma^{\mu}D_{\mu}\psi-\frac{i}{2}
\overline{\psi}\Gamma^{a}D_{a}\psi\biggr)\,,\label{YM-Dirac-4Daction}
\end{equation}
where $D_{\mu} = \partial_{\mu} - A_{\mu}$ and $D_{a}= \partial_{a}- \theta_{a}-\phi_{a}$ with 
$\theta_{a}= \frac{1}{2}\theta_{abc}\Sigma^{bc}$ the connection of the coset space and $\Sigma^{bc}$ the 
$SO(6)$ generators. With $C$ we denote the volume of the coset space. The potential $V(\phi)$ is given by
\begin{equation}
V(\phi) =F_{ab}F^{ab}=
 - \frac{1}{4} g^{ac}g^{bd}Tr( f _{ab}{}^{C}\phi_{C} -
[\phi_{a},\phi_{b}] ) (f_{cd}{}^{D}\phi_{D} - [\phi_{c},\phi_{d}] )\,,\label{potential}
\end{equation}
where, $A=1,\ldots,\Dim\,S$ and $f$ ' s are the structure constants appearing in the commutators of the
generators of the Lie algebra of S. The expression (\ref{potential}) for $V(\phi)$ is only formal because the
$\phi_{a}$ must satisfy the constraints coming from eqs~(\ref{CSDR-constraints}),
\begin{equation}
f_{ai}{}^{c}\phi_{c} - [\phi_{a},\phi_{i}] = 0\,,
\label{scalar-consrt}
\end{equation}
where the $\phi_{i}$ generate $R_{G}$. These constraints imply that some components $\phi_{a}$'s are zero, 
some are constants and the rest can be identified with the genuine Higgs fields according with the rules
presented in eqs~(\ref{RinS-embedding}) and (\ref{RinG-embedding}).

When $V(\phi)$ is expressed in terms of the unconstrained independent Higgs fields, it remains a quartic
polynomial which is invariant under gauge transformations of the final gauge group $H$, and its minimum
determines the vacuum expectation values of the Higgs
fields~\cite{Chapline:1980mr,Bais:1985yd,Farakos:1986sm,*Farakos:1986cj}. The minimization of the potential
is in general a difficult problem.  However, when $S$ has an isomorphic image $S_{G}$ in $G$, the
four-dimensional gauge group $H$ will break spontaneously to a subgroup $K$, which is the centralizer of
$S_{G}$ in the group of the higher dimensional theory
$G$, $K=C_{G}(S_{G})$~\cite{Harnad:1979in,*Harnad:1980ct,Kapetanakis:1992hf}. This can be illustrated as
follows:
\beq
\begin{array}{c@{}c@{}c@{}c@{}}
G\supset &S_{G}&~\times~  &K\\
\ \           &\cup  &~             &\cap\\
\ \           &R      &~\times~  &~H\,.
\end{array}
\eeq
Furthermore when $\phi$ acquires v.e.v. the $V(\phi)$ vanishes. It should be stressed that, in this class of
models, the four-dimensional fermions acquire large masses due to geometrical contributions at the
compactification scale~\cite{Kapetanakis:1992hf,Barnes:1986ea}. In general it can be
proven~\cite{Kapetanakis:1992hf} that dimensional reduction over a symmetric coset space always gives a
potential of spontaneous breaking form which is not the case of non-symmetric cosets of more than one radii.

In the fermion part of the Lagrangian the first term is just the kinetic term of fermions, while the second is
the Yukawa term~\cite{Barnes:1986ea,Kapetanakis:1990tk}. The last term in (\ref{YM-Dirac-4Daction}) can be
written as
\beq
L_{D}= -\frac{i}{2}\overline{\psi}\Gamma^{a}\left(\partial_{a}-
\frac{1}{2}f_{ibc}e^{i}_{\gamma}e^{\gamma}_{a}\Sigma^{bc}-
\frac{1}{2}G_{abc}\Sigma^{bc}- \phi_{a}\right) \psi
=\frac{i}{2}\overline{\psi}\Gamma^{a}\nabla_{a}\psi+
\overline{\psi}V\psi\,,\label{Dirac-4Daction}
\eeq
where
\begin{align}
&\nabla_{a} = - \partial_{a} +
\frac{1}{2}f_{ibc}e^{i}_{\gamma}e^{\gamma}_{a}\Sigma^{bc} + \phi_{a}\,,\label{CovDerivative}\\
&V=\frac{i}{4}\Gamma^{a}G_{abc}\Sigma^{bc}\,,\label{GeometricMass}
\end{align}
where $G_{abc}$ is given in eq.~(\ref{Connection}) as $G^{a}{}_{bc}=D^{a}{}_{bc}+\frac{1}{2}\Sigma^{a}{}_{bc}$.
The CSDR constraints tell us that $\partial_{a}\psi= 0$. Furthermore we can
consider the Lagrangian at the point $y=0$, due to its invariance under $S$-transformations, and according to
the discussion in sec.~\ref{sec:CosetSpaceGeometry} $e^{i}_{\gamma}=0$ at that point. Therefore
(\ref{CovDerivative}) becomes just $\nabla_{a}= \phi_{a}$ and the term
$\frac{i}{2}\overline{\psi}\Gamma^{a}\nabla_{a}\psi $ in eq.~(\ref{Dirac-4Daction}) is exactly the Yukawa
term. The last term of eq.~(\ref{Dirac-4Daction}) vanishes in the case of dimensional reduction over
symmetric cosets, whereas in the case of non-symmetric cosets  is responsible for the masses of the 
four-dimensional gaugini~\cite{Manousselis:2000aj,*Manousselis:2001xb,*Manousselis:2001re}. However, as
explained in sec.~\ref{sec:CosetSpaceGeometry}, this mass term can be suitably modified under appropriate
adjustment of the torsion and the radii of the non-symmetric coset in question.

\subsection{Remarks on Grand Unified Theories resulting from CSDR}\label{Remarks-on-CSDR}

Here we make few remarks on models resulting from the coset space dimensional reduction of an $\cN=1$, $E_{8}$
gauge theory which is defined on a ten-dimensional compactified space $M^{D}=M^{4}\times (S/R)$. The coset
spaces $S/R$ we consider are listed in the first column of tables~\ref{SO6VectorSpinorContentSCosets}
and~\ref{SO6VectorSpinorContentNSCosets}. In order to obtain four-dimensional GUTs potentially with
phenomenological interest, namely $\cH=E_{6}$, $SO(10)$ and $SU(5)$, is sufficient to consider only embeddings
of the isotropy group $R$ of the coset space in
\begin{subequations}
\begin{alignat}{2}
\cR&=C_{E_{8}}(\cH)= SU(3)\,,\qquad                       & \mbox{for}\quad &\cH=E_{6}\,,\label{Rcases-1} \\
\cR&=C_{E_{8}}(\cH)=SO(6)\thicksim SU(4)\,,\qquad & \mbox{for}\quad &\cH=SO(10)\,,\label{Rcases-2}\\
\cR&=C_{E_{8}}(\cH)=SU(5)\,,\qquad                       & \mbox{for}\quad  &\cH=SU(5)\,.\label{Rcases-3}
\end{alignat}
\end{subequations}

As it was noted in sec.~\ref{sec:CSDR-rules} the anomaly cancellation condition~(\ref{WittenCond})  is
satisfied automatically for the choice of embedding
\beq
E_{8}\supset SO(6)\supset R\,,\label{RinSO6inE8-embedding}
\eeq
which we adopt here. This requirement is trivially fullfiled for the case of $R\hookrightarrow\cR$ embeddings 
of eq.~(\ref{Rcases-2}) which lead to $SO(10)$ GUTs in four dimensions. It is obviously also satisfied for the
case of $R\hookrightarrow\cR$ embeddings of eq.~(\ref{Rcases-1}) since $SU(3)\subset SO(6)$. The above case
leads to $E_{6}$ GUTs in four dimensions. Finally,  $R\hookrightarrow\cR$ embeddings of eq.~(\ref{Rcases-3})
are excluded since the requirement~(\ref{RinSO6inE8-embedding}) cannot be satisfied.

\section{Wilson flux breaking mechanism in CSDR\label{sec:WilsonFlux-theory}}

The surviving scalars in a four-dimensional GUT, being in the fundamental rep. of the gauge group are not able
to provide the appropriate superstrong symmetry breaking towards the standard model.  As a way out it has been
suggested~\cite{Zoupanos:1987wj} to take advantage of non-trivial topological properties of the
compactification coset space, apply the Hosotani or Wilson flux breaking
mechanism~\cite{Hosotani:1983xw,*Hosotani:1983vn,Witten:1985xc} and break the gauge symmetry of the theory
further. Application of this mechanism imposes further  constraints in the scheme.

In the next subsections we first recall the Wilson flux breaking mechanism, we make some remarks on specific 
cases which potentially lead to interesting models and we finally calculate the actual symmetry breaking
patterns of the GUTs.

\subsection{Wilson flux breaking mechanism}\label{sec:WilsonFluxBreaking-theory}

Let us briefly recall the Wilson flux mechanism for breaking spontaneously a gauge theory. Then instead of 
considering a gauge theory on $M^{4}\times B_{0}$, with $B_{0}$ a simply connected manifold, and in our case a
coset space $B_{0}=S/R$, we consider a gauge theory on $M^{4}\times B$, with $B=B_{0}/F^{S/R}$ and $F^{S/R}$ a
freely acting discrete symmetry\footnote{By freely acting we mean that for every element $g\in F$, except the
identity, there exists no points of $B_0$ that remain invariant.} of $B_{0}$. It turns out that $B$ becomes
multiply  connected, which means that there will be contours not contractible to a point due to holes in the
manifold. For each element $g\in F^{S/R}$, we  pick up an element $U_{g}$ in $H$, i.e. in the four-dimensional
gauge group of the reduced theory, which can be represented as the Wilson loop
\begin{equation}
U_{g}=\mathcal{P}exp\left(-i~\int_{\gamma_g}T^{a}A_{M}^{a}(x)dx^{M}\right)\,,
\end{equation}
where $A_{M}^{a}(x)$ are vacuum $H$ fields with group generators $T^{a}$, $\gamma_g$ is a contour representing
the abstract element $g$ of $F^{S/R}$, and $\mathcal{P}$ denotes the path ordering.

Now if $\gamma_g$ is chosen not to be contractible to a point, then $U_g\neq 1$ although the vacuum field 
strength vanishes everywhere. In this  way an homomorphism of $F^{S/R}$ into $H$ is induced with image
$T^{H}$, which is the subgroup of $H$ generated by $\{U_g\}$. A field $f(x)$ on  $B_0$ is obviously equivalent
to another field on $B_0$ which obeys $f(g(x))=f(x)$ for every $g\in F^{S/R}$. However in the presence of the
gauge group $H$ this statement can be generalized to
\begin{equation}
\label{eq:Wilson-symmetry}
f(g(x))=U_{g}f(x)\,.
\end{equation}
Next, one would like to see which gauge symmetry is preserved by the vacuum. The vacuum has $A_{\mu}^{a}=0$ 
and we represent a gauge transformation by a space-dependent matrix $V(x)$ of $H$. In order to keep
$A_{\mu}^{a}=0$ and leave the vacuum invariant, $V(x)$ must be constant. On the other hand, $f\to Vf$ is
consistent with equation (\ref{eq:Wilson-symmetry}), only if $[V,U_g]=0$ for all $g\in F^{S/R}$. Therefore the
$H$ breaks towards the centralizer of $T^{H}$ in $H$, $K'=C_{H}(T^{H})$. In addition the matter fields have to
be invariant under the diagonal sum
\begin{equation}
F^{S/R}\oplus T^{H}\,,\label{WFluxSurvivingField}
\end{equation}
in order to satisfy eq.~(\ref{eq:Wilson-symmetry}) and therefore survive in the four-dimensional theory.

\subsection{Further remarks concerning the use of the $F^{S/R}$}
\label{sec:InterestingDiscreteSymmetries}

The discrete symmetries $F^{S/R}$, which act freely on coset spaces $B_0=S/R$ are the center of $S$,
$\mathrm{Z}(S)$ and the $\mathrm{W}=\mathrm{W}_{S}/\mathrm{W}_{R}$, with $\mathrm{W}_{S}$ and $\mathrm{W}_{R}$
being the Weyl groups of $S$ and $R$,
respectively~\cite{Kapetanakis:1989gd,*Kozimirov:1989kn,*Kozimirov:1989xp,Kapetanakis:1992hf}. The freely
acting discrete symmetries, $F^{S/R}$, of the specific six-dimensional coset spaces under discussion are
listed in the second and third column of tables~\ref{SO6VectorSpinorContentSCosets}
and~\ref{SO6VectorSpinorContentNSCosets}. The $F^{S/R}$ transformation properties of the vector and spinor
irreps under $R$ are noted in the last two columns of the same tables.

Our approach is to embed the $F^{S/R}$ discrete symmetries into four-dimensional $H=E_{6}$ and $SO(10)$ gauge 
groups. We make this choice only for bookeeping reasons since, according to
sec.~\ref{sec:WilsonFluxBreaking-theory},  the actual topological symmetry breaking takes place in higher
dimensions. Few remarks are in order. In both classes of models, namely $E_{6}$ and $SO(10)$ GUTs, the use of
the discrete symmetry of the center of $S$, ${\rm Z}(S)$, cannot lead to phenomenologically interesting
cases since various components of the irreps of the four-dimensional GUTs containing the SM fermions do not
survive. The reason is that the irreps of $H$ remain invariant under the action of the discrete symmetry,
${\rm Z}(S)$, and as a result the phase factors gained  by the action of $T^{H}$ cannot be compensated.
Therefore the complete SM fermion spectrum cannot be invariant under $F^{S/R}\oplus T^{H}$ and survive. On the
other hand, the use of the Weyl discrete symmetry can lead to better results. Models with potentially
interesting fermion spectrum can be obtained employing at least one $\bb{Z}_{2}\subset {\rm W}$. Then, the
fermion content of the four-dimensional theory is found to transform in linear combinations of the two copies
of the CSDR-surviving left-handed fermions. Details  will be given in sec.~\ref{Classification}.
As we will discuss there employing $\bb{Z}_{2}\times\bb{Z}_{2}\subseteq {\rm W}$ or
$\bb{Z}_{2}\times\bb{Z}_{2}\subseteq {\rm W}\times{\rm Z}(S)$ can also lead to interesting models.

Therefore the interesting cases for further study are
\beq
F^{S/R}=\begin{cases}
             \bb{Z}_{2}\subseteq{\rm W} \\
             \bb{Z}_{2}\times\bb{Z}_{2}\subseteq {\rm W}\\
             \bb{Z}_{2}\times\bb{Z}_{2}\subseteq {\rm W}\times{\rm Z}(S)\,.
             \end{cases}\label{FCases}
\eeq

\subsection{Symmetry breaking patterns of $E_{6}$-like GUTs}\label{TopologicallyInducedGaugeGroupBreaking-E6}

Here we determine the image, $T^{H}$, that each of the discrete symmetries of eq.~(\ref{FCases}) induces in 
the gauge group $H=E_{6}$. We consider embeddings of the $F^{S/R}$ discrete symmetries  into abelian subgroups
of $E_{6}$ and examine their topologically induced symmetry breaking patterns~\cite{Witten:1985xc}. These are
realized by a diagonal matrix $U_{g}$ of unit determinant, which as explained in
sec.~\ref{sec:WilsonFluxBreaking-theory}, has to be homomorphic to the considered discrete symmetry. In
fig.~\ref{DecompsDiscreteSymm} we present those $E_{6}$ decompositions which potentially lead to the SM gauge
group structure~\cite{Slansky:1981yr}.

\begin{figure}[t]
$$
\xymatrix@C=0.3cm{
                                 & F_{4}\ar[r]                    & SU(3)\times SU^{c}(3)~(\mbf{1}) &\\
E_{6}\ar[ur]\ar[r]\ar[dr]& SO(10)\ar[r]                 & SU(5)\ar[r]                & SU^{w}(2)\times SU^{c}(3)~(\mbf{2})\\
                                 & SU^{w}(3)\times SU(3)\times SU^{c}(3)~(\mbf{3}) &           & }
$$
\caption{\emph{$E_{6}$ decompositions leading potentially to SM gauge group structure.}
\label{DecompsDiscreteSymm}}
\end{figure}

\subsubsection{The $\bb{Z}_{2}$ case}
\label{TopologicallyInducedGaugeGroupBreaking-E6-Z2}
\paragraph{Embedding $(\mbf{1})$:~
${\bb{Z}_{2}\hookrightarrow SU(3)}$ of ${E_{6}\supset F_{4}\supset SU(3)\times SU^{c}(3)}$.}

We consider the maximal subgroups of $E_{6}$ and the corresponding decomposition of fundamental and adjoint
irreps
\beq
\begin{array}{l}
E_{6}\supset F_{4}\supset SU(3)\times SU^{c}(3)\\
\begin{array}{r@{}l@{}l@{}}
\mbf{27}&=&(\mbf{1},\mbf{1})+(\mbf{8},\mbf{1})+(\mbf{3},\mbf{3})+(\b{\mbf{3}},\b{\mbf{3}})\,,\\
\mbf{78}&=&(\mbf{8},\mbf{1})+(\mbf{3},\mbf{3})+(\b{\mbf{3}},\b{\mbf{3}})
              +(\mbf{8},\mbf{1})+(\mbf{1},\mbf{8})+(\mbf{6},\b{\mbf{3}})+(\b{\mbf{6}},\mbf{3})
\end{array}\label{E6toF4toSU3SU3}
\end{array}
\eeq
and embed the $F^{S/R}=\bb{Z}_{2}$ discrete symmetry in the $SU(3)$ group factor above. There exist two 
distinct possibilities of embedding, either $\bb{Z}_{2}\hookrightarrow U^{I}(1)$ which appears under the
$SU(3)\supset SU(2)\times U^{II}(1)\supset U^{I}(1)\times U^{II}(1)$ decomposition or
$\bb{Z}_{2}\hookrightarrow U^{II}(1)$. Since the former is trivial, namely cannot break the $SU(3)$ appearing
in eq.~(\ref{E6toF4toSU3SU3}), only the latter is interesting for further investigation. This is realized as
\beq
U_{g}^{(1)}=diag(-1,-1,1)\,.\label{Ug1}
\eeq
Indeed $(U_{g}^{(1)})^{2}=\one_{3}$ as required by the $F^{S/R}\mapsto H$ homomorphism and 
$det(U_{g}^{(1)})=1$ since $U_{g}$ is an $H$ group element.

Then, the various components of the decomposition of $SU(3)$ irreps under $SU(2)\times U(1)$ acquire the
underbraced phase factors in the following list
\beq
\begin{array}{l}
SU(3) \supset  SU(2)\times U(1)\\
\begin{array}{r@{}l@{}l@{}}
\mbf{3}&=&\ubr{\mbf{1}_{(-2)}}_{(+1)}+\ubr{\mbf{2}_{(1)}}_{(-1)}\,,\\
\mbf{6}&=&\ubr{\mbf{1}_{(-4)}}_{(+1)}+\ubr{\mbf{2}_{(-1)}}_{(-1)}+\ubr{\mbf{3}_{(2)}}_{(+1)}\,,\\
\mbf{8}&=&\ubr{\mbf{1}_{(0)}}_{(+1)}+\ubr{\mbf{3}_{(0)}}_{(+1)}+\ubr{\mbf{2}_{(-3)}}_{(-1)}
        +\ubr{\mbf{2}_{(3)}}_{(-1)}\,.
\end{array}\label{SU3toSU2U1HB}
\end{array}
\eeq
Consequently the various components of the decomposition of $E_{6}$ irreps ~(\ref{E6toF4toSU3SU3}) under
$
F_{4}\supset SU(3)\times SU^{c}(3)\supset (SU(2)\times U(1))\times SU^{c}(3)
$
acquire the underbraced phase factors in the following list
\beq
\begin{array}{r@{}l@{}l@{}}
E_{6}&\supset& SU(2)\times SU^{c}(3)\times U(1)\\
\mbf{27}&=&\ubr{(\mbf{1},\mbf{1})_{(0)}}_{(+1)}+\ubr{(\mbf{1},\mbf{1})_{(0)}}_{(+1)}
              +\ubr{(\mbf{3},\mbf{1})_{(0)}}_{(+1)}\\
            &+&\ubr{(\mbf{1},\mbf{3})_{(-2)}}_{(+1)}+\ubr{(\mbf{2},\mbf{3})_{(1)}}_{(-1)}
              +\ubr{(\mbf{1},\b{\mbf{3}})_{(2)}}_{(+1)}+\ubr{(\mbf{2},\b{\mbf{3}})_{(-1)}}_{(-1)}
              + \ubr{(\mbf{2},\mbf{1})_{(-3)}}_{(-1)}+\ubr{(\mbf{2},\mbf{1})_{(3)}}_{(-1)}\,,\\
\mbf{78}&=&\ubr{(\mbf{1},\mbf{1})_{(0)}}_{(+1)}+\ubr{(\mbf{3},\mbf{1})_{(0)}}_{(+1)}
              +\ubr{(\mbf{1},\mbf{1})_{(0)}}_{(+1)}+\ubr{(\mbf{3},\mbf{1})_{(0)}}_{(+1)}
              +\ubr{(\mbf{1},\mbf{8})_{(0)}}_{(+1)}\\
            &+&\ubr{(\mbf{1},\mbf{3})_{(-2)}}_{(+1)}+\ubr{(\mbf{1},\b{\mbf{3}})_{(2)}}_{(+1)}
              +\ubr{(\mbf{1},\b{\mbf{3}})_{(-4)}}_{(+1)}+\ubr{(\mbf{1},\mbf{3})_{(4)}}_{(+1)}\\
           &+&\ubr{(\mbf{2},\mbf{1})_{(-3)}}_{(-1)}+\ubr{(\mbf{2},\mbf{1})_{(3)}}_{(-1)}
             + \ubr{(\mbf{2},\mbf{1})_{(-3)}}_{(-1)}+\ubr{(\mbf{2},\mbf{1})_{(3)}}_{(-1)}\\
           &+&\ubr{(\mbf{2},\mbf{3})_{(1)}}_{(-1)}+\ubr{(\mbf{2},\b{\mbf{3}})_{(-1)}}_{(-1)}
             +\ubr{(\mbf{2},\mbf{3})_{(1)}}_{(-1)}+\ubr{(\mbf{2},\b{\mbf{3}})_{(-1)}}_{(-1)}\\
           &+&\ubr{(\b{\mbf{3}},\mbf{3})_{(-2)}}_{(+1)}+\ubr{(\mbf{3},\b{\mbf{3}})_{(+2)}}_{(+1)}\,.
\end{array}\label{E6toF4toSU3SU3HB}
\eeq
According to the discussion in sec. \ref{sec:WilsonFluxBreaking-theory} the four-dimensional gauge group 
after the topological breaking is given  by $K'=C_{H}(T^{H})$. Counting the number of singlets under the
action of $U_{g}^{(1)}$ in  the $\mbf{78}$ irrep. above suggests that $K'=SO(10)\times U(1)$, a fact which
subsequently is determined according to the following decomposition of the $\mbf{78}$ irrep.
\beq
\begin{array}{l}
E_{6}\supset SO(10)\times U(1)\\
\begin{array}{r@{}l@{}l@{}}
\mbf{27}&=&\ubr{\mbf{1}_{(-4)}}_{(+1)}+\ubr{\mbf{10}_{(-2)}}_{(+1)}+\ubr{\mbf{16}_{(1)}}_{(-1)}\,,\\
\mbf{78}&=&\ubr{\mbf{1}_{(0)}}_{(+1)}+\ubr{\mbf{45}_{(0)}}_{(+1)}+\ubr{\mbf{16}_{(-3)}}_{(-1)}
         +\ubr{\b{\mbf{16}}_{(3)}}_{(-1)}\,.
\end{array}\label{E6toSO10U1}
\end{array}
\eeq

It is interesting to note that although one would naively expect the $E_{6}$ gauge group to break further 
towards the SM one this is not the case. The singlets under the action of $U_{g}^{(1)}$ which occur  in the
adjoint irrep. of $E_{6}$ in eq.~(\ref{E6toF4toSU3SU3HB}) add up to provide a larger final unbroken gauge
symmetry, namely $SO(10)\times U(1)$.

\paragraph{Embedding~$(\mbf{2})$:
$\bb{Z}_{2}\hookrightarrow SU(5)$ of $E_{6}\supset SO(10)\times U(1)\supset SU(5)\times U(1)\times U(1)$.}

Similarly, we consider the maximal subgroups of $E_{6}$ and the corresponding decomposition of the fundamental
and adjoint irreps
\beq
\begin{array}{l}
E_{6}\supset SO(10)\times U(1)\supset SU(5)\times U(1)\times U(1)\\
\begin{array}{r@{}l@{}l@{}}
\mbf{27}&=&\mbf{1}_{(0,-4)}+\mbf{5}_{(2,-2)}+\b{\mbf{5}}_{(-2,-2)}+\mbf{1}_{(-5,1)}+\b{\mbf{5}}_{(3,1)}
         +\mbf{10}_{(-1,1)}\,,\\
\mbf{78}&=&\mbf{1}_{(0,0)}+\mbf{1}_{(0,0)}+\mbf{24}_{(0,0)}+\mbf{1}_{(-5,-3)}+\mbf{1}_{(5,3)}\\
            &+&\mbf{5}_{(-3,3)}+\b{\mbf{5}}_{(3,-3)}
              +\mbf{10}_{(4,0)}+\b{\mbf{10}}_{(-4,0)}+\mbf{10}_{(-1,-3)}+\b{\mbf{10}}_{(1,3)}\,.
\end{array}\label{E6toSO10U1toSU5U1U1}
\end{array}
\eeq
Our choice is to embed the $\bb{Z}_{2}$ discrete symmetry in an abelian $SU(5)$ subgroup in a way that is 
realized by the diagonal matrix
\beq
U_{g}^{(2)}=diag(-1,-1,1,1,1)\,.\label{Ug2}
\eeq
Then the various components of the $SU(5)$ irreps decomposed under the $SU(2)\times SU(3)\times U(1)$ 
decomposition acquire the underbraced phase factors in the following list
\beq
\begin{array}{l}
SU(5)\supset SU(2)\times SU(3)\times U(1)\\
\begin{array}{r@{}l@{}l@{}}
\mbf{5}&=&\ubr{(\mbf{2},\mbf{1})_{(3)}}_{(-1)}+\ubr{(\mbf{1},\mbf{3})_{(-2)}}_{(+1)}\,,\\
\mbf{10}&=&\ubr{(\mbf{1},\mbf{1})_{(6)}}_{(+1)}+\ubr{(\mbf{1},\b{\mbf{3}})_{(-4)}}_{(+1)}
              +\ubr{(\mbf{2},\mbf{3})_{(1)}}_{(-1)}\,,\\
\mbf{24}&=&\ubr{(\mbf{1},\mbf{1})_{(0)}}_{(+1)}+\ubr{(\mbf{3},\mbf{1})_{(0)}}_{(+1)}
         +\ubr{(\mbf{1},\mbf{8})_{(0)}}_{(+1)}+\ubr{(\mbf{2},\mbf{3})_{(-5)}}_{(-1)}
         +\ubr{(\mbf{2},\b{\mbf{3}})_ {(5)}}_{(-1)}\,.
\end{array}\label{SU5toSU2SU3U1HB}
\end{array}
\eeq
It can be proven, along the lines of the previous case $(\mbf{1})$, that $U_{g}^{(2)}$ leads to the breaking 
$E_{6}\to SU(2)\times SU(6)$
\beq
\begin{array}{l}
E_{6}\supset SU(2)\times SU(6)\\
\begin{array}{l@{}l@{}l@{}}
\mbf{27}&=&\ubr{(\mbf{2},\b{\mbf{6}})}_{(-1)}+\ubr{(\mbf{1},\mbf{15})}_{(+1)}\,,\\
\mbf{78}&=&\ubr{(\mbf{3},\mbf{1})}_{(+1)}+\ubr{(\mbf{1},\mbf{35})}_{(+1)}+\ubr{(\mbf{2},\mbf{20})}_{(-1)}\,,
\end{array}\label{E6toSU2SU6HB}
\end{array}
\eeq
i.e. we find again an enhancement of the final gauge group as compared to the naively expected one.

Note that other choices of $\bb{Z}_{2}$ into $SU(5)$ embeddings either lead to trivial or to 
phenomenologically uninteresting results.

\paragraph{Embedding $(\mbf{3})$:~
$\bb{Z}_{2}\hookrightarrow SU(3)$ of $E_{6}\supset SU^{w}(3)\times SU(3)\times SU^{c}(3)$.}

We consider the maximal subgroup of $E_{6}$ and the corresponding decomposition of fundamental and adjoint 
irreps
\beq
\begin{array}{l}
E_{6}\supset SU^{w}(3)\times SU(3)\times SU^{c}(3)\\
\begin{array}{l@{}l@{}l@{}}
\mbf{27}&=&(\b{\mbf{3}},\mbf{3},\mbf{1})+(\mbf{3},\mbf{1},\mbf{3})+(\mbf{1},\b{\mbf{3}},\b{\mbf{3}})\,,\\
\mbf{78}&=&(\mbf{8},\mbf{1},\mbf{1})+(\mbf{1},\mbf{8},\mbf{1})+(\mbf{1},\mbf{1},\mbf{8})+
(\mbf{3},\mbf{3},\b{\mbf{3}})+(\b{\mbf{3}},\b{\mbf{3}},\mbf{3})\,.\label{E6-SU3trits}
\end{array}
\end{array}
\eeq
We furthermore assume an $\bb{Z}_{2}\hookrightarrow SU(3)$ embedding, which is realized by
\beq
U_{g}^{(3)}=(\one_{3})\otimes diag(-1, -1, 1)\otimes(\one_{3})\,.\label{Ug3}
\eeq
Although this choice of embedding is not enough to lead to the SM gauge group structure, our results will be 
usefull for the discussion of the $\bb{Z}_{2}\times\bb{Z}_{2}'$ case which is presented in sec.~\ref{E6-Z2Z2}.
With the choice of embedding realized by the eq.~(\ref{Ug3})  the second $SU(3)$ decomposes under $SU(2)\times
U(1)$ as in eq.~(\ref{SU3toSU2U1HB}) and leads to the breaking~(\ref{E6toSU2SU6HB}), as before.
As was mentioned in case~($\mbf{1}$) the choice of embedding $\bb{Z}_{2}\hookrightarrow U^{I}(1)$, which
appears under the decomposition $SU(3)\supset SU(2)\times U^{II}(1)\supset U^{I}(1)\times U^{II}(1)$ of
eq.~(\ref{E6-SU3trits}), cannot break the $SU(3)$ group factor and it is not an interesting case for further
investigation.

In table~\ref{E6-Breaking-Z2} we summarize the above results, concerning the topologically induced symmetry 
breaking patterns of the $E_{6}$ gauge group.

\begin{table}[t]
\setlength{\tabcolsep}{15pt}
\setlength{\arraycolsep}{15pt}
\begin{center}
\begin{tabular}{|c!{\vrule width 1pt}c|c|}\hline
{\bf Embedd.}
& $U_{g}$
&$\mbf{K'}$
\\ \hhline{*{3}{=}}
$\mbf{1}$
&
$U_{g}^{(1)}$
&
$SO(10)\times U(1)$
\\\hline
$\mbf{2}$
&
$U_{g}^{(2)}$
&
$SU(2)\times SU(6)$
\\\hline
$\mbf{3}$
&
$\one_{3}\otimes U_{g}^{(1)}\otimes\one_{3}$
&
$SU(2)\times SU(6)$
\\\hline
\end{tabular}
\caption{\textbf{Embeddings of $\bb{Z}_{2}$ discrete symmetry in $E_{6}$ GUT and its symmetry breaking
patterns.} \emph{$U_{g}^{(1)}=diag(-1,-1,1)$ and $U_{g}^{(2)}=diag(-1,-1,1,1,1)$ as in text.}
\label{E6-Breaking-Z2}}
\end{center}
\end{table}

\subsubsection{The $\bb{Z}_{2}\times\bb{Z}_{2}'$ case}\label{E6-Z2Z2}
\label{TopologicallyInducedGaugeGroupBreaking-E6-Z2Z2}

\paragraph{Embedding~$(\mbf{2'})$:
$\bb{Z}_{2}\hookrightarrow SO(10)$~and~$\bb{Z}_{2}'\hookrightarrow SU(5)$~of
$E_{6}\supset SO(10)\times U(1)\supset SU(5)\times U(1)\times U(1)$.}

Here we embed the $\bb{Z}_{2}$ of the $\bb{Z}_{2}\times\bb{Z}_{2}'$ discrete symmetry in the $SU(5)$ appearing
under the  decomposition $E_{6}\supset SO(10)\times U(1)\supset SU(5)\times U(1)\times U(1)$ as in case
($\mbf{2}$) above. Furthermore we embed the  $\bb{Z}_{2}'$ discrete symmetry in the $SO(10)$ as
\beq
U_{g}'=-\one_{10}\,. 
\eeq
This leads to the breaking $E_{6}\supset SU(2)\times SU(6)$ as before but with the signs of the phase factors,
which appear  in eq.~(\ref{E6toSU2SU6HB}), being reversed under the action of $U_{g}^{(2)}U_{g}'$.

\paragraph{Embedding~$(\mbf{3'})$:~ $\bb{Z}_{2}\hookrightarrow SU(3)$~and~$\bb{Z}_{2}'\hookrightarrow
SU^{w}(3)$~of~$E_{6}\supset SU^{w}(3)\times SU(3)\times SU^{c}(3)$.}

Here we embed the $\bb{Z}_{2}$ of the $\bb{Z}_{2}\times\bb{Z}_{2}'$ discrete symmetry in the $SU(3)$ group 
factor appearing under the $E_{6}\supset SU(3)^{w}\times SU(3)\times SU(3)^{c}$ as in case ($\mbf{3}$) above.
Furthermore we embed the $\bb{Z}_{2}'$ discrete symmetry in the $SU(3)^{w}$ group factor in a similar way.
Then the embedding ($\mbf{3}'$), which we discuss here, is realized by considering an element of the $E_{6}$
gauge group
\beq
U_{g}'U_{g}^{(3)}=diag(-1,-1,1) \otimes diag(-1,-1,1)\otimes (\one_{3})\label{Ug4}\,,
\eeq
which leads to the breaking $E_{6}\to SU^{(i)}(2)\times SU^{(ii)}(2)\times SU(4)\times U(1)$ as it is clear 
from the following decomposition of $\mbf{78}$ irrep.
\begin{align}
&E_{6}\supset SO(10)\times U(1)\supset  SU^{(i)}(2)\times SU^{(ii)}(2)\times SU(4)\times U(1)\nn\\
&\begin{array}{l@{}l@{}l@{}}
E_{6}&\supset& SU^{(i)}(2)\times SU^{(ii)}(2)\times SU(4)\times U(1)\\
\mbf{27}&=&\ubr{(\mbf{1},\mbf{1},\mbf{1})_{(4)}}_{(+1)}+\ubr{(\mbf{2},\mbf{2},\mbf{1})_{(-2)}}_{(+1)}
              + \ubr{(\mbf{1},\mbf{1},\mbf{6})_{(-2)}}_{(+1)}\\
            &+&\ubr{(\mbf{2},\mbf{1},\mbf{4})_{(1)}}_{(-1)}+\ubr{(\mbf{1},\mbf{2},\b{\mbf{4}})_{(1)}}_{(-1)}\,,\\
\mbf{78}&=&\ubr{(\mbf{1},\mbf{1},\mbf{1})_{(0)}}_{(+1)}
              +\ubr{(\mbf{1},\mbf{3},\mbf{1})_{(0)}}_{(+1)}+\ubr{(\mbf{3},\mbf{1},\mbf{1})_{(0)}}_{(+1)}
              +\ubr{(\mbf{1},\mbf{1},\mbf{15})_{(0)}}_{(+1)}+\ubr{(\mbf{2},\mbf{2},\mbf{6})_{(0)}}_{(-1)}\\
            &+&\ubr{(\mbf{2},\mbf{1},\mbf{4})_{(-3)}}_{(-1)}+\ubr{(\mbf{2},\mbf{1},\b{\mbf{4}})_{(3)}}_{(-1)}
              + \ubr{(\mbf{1},\mbf{2},\mbf{4})_{(3)}}_{(-1) }+\ubr{(\mbf{1},\mbf{2},\b{\mbf{4}})_{(-3)}}_{(-1)}\,.
\end{array}\label{E6toSO10U1toSU4SU2SU2U1HB}
\end{align}

In table~\ref{E6-Breaking-Z2Z2} we summarize the above results, concerning the topologically induced symmetry 
breaking patterns of the $E_{6}$ gauge group.

\setlength{\tabcolsep}{5pt}
\setlength{\arraycolsep}{5pt}
\begin{table}[t]
\begin{center}
\begin{tabular}{|c!{\vrule width 1pt}c|c|c|}\hline
{\bf Embedd.}
& $U_{g}$
& $U_{g}'$
& $\mbf{K'}$
\\ \hhline{*{4}{=}}
  $\mbf{2'}$
&
  $U_{g}^{(2)}$
&
  $-\one_{10}$
&
  $SU(2)\times SU(6)$
\\\hline
  $\mbf{3'}$
&
  $U_{g}^{(1)}\otimes\one_{3}\otimes \one_{3}$
&
  $\one_{3}\otimes U_{g}^{(1)}\otimes \one_{3}$
&
  $SU^{(i)}(2)\times SU^{(ii)}(2)\times SU(4)\times U(1)$
\\\hline
\end{tabular}
\end{center}
\caption{\textbf{Embeddings of $\bb{Z}_{2}\times\bb{Z}_{2}'$ discrete symmetries in $E_{6}$ GUT and its
symmetry breaking patterns.} \emph{$U_{g}^{(1)}=diag(-1,-1,1)$ and $U_{g}^{(2)}=diag(-1,-1,1,1,1)$ as in
text.}\label{E6-Breaking-Z2Z2}}
\end{table}

\subsection{Symmetry breaking pattern of  $SO(10)$-like GUTs}
\label{TopologicallyInducedGaugeGroupBreaking-SO10}
Here we determine the image, $T^{H}$, that each of the discrete symmetries of eq.~(\ref{FCases}) induces in
the gauge group $H=SO(10)$. We consider embeddings of the $F^{S/R}$ discrete symmetries  into abelian
subgroups of $SO(10)$ GUTs and examine their topologically induced symmetry breaking patterns. The interesting
$F^{S/R}\hookrightarrow SO(10)$ embeddings are those which potentially lead to SM gauge group structure, i.e.
$$
SO(10)\supset SU(5)\times U^{II}(1)\supset SU^{w}(2)\times SU^{c}(3)\times U^{I}(1)\times U^{II}(1)\,.
$$

\subsubsection{The $\bb{Z}_{2}$ case}
\label{TopologicallyInducedGaugeGroupBreaking-SO10-Z2}
\paragraph{Embedding~$(\mbf{1})$:~$\bb{Z}_{2}\hookrightarrow SU(5)$ of $SO(10)\supset SU(5)\times U(1)$.}

In the present case we assume the maximal subgroup of $SO(10)$
\beq
\begin{array}{l}
SO(10)\supset SU(5)\times U^{II}(1)\\
\begin{array}{r@{}l@{}l@{}}
\mbf{10}&=&\mbf{5}_{(2)}+\b{\mbf{5}}_{(-2)}\,,\\
\mbf{16}&=&\mbf{1}_{(-5)}+\b{\mbf{5}}_{(3)}+\mbf{10}_{(-1)}\,,\\
\mbf{45}&=&\mbf{1}_{(0)}+\mbf{24}_{(0)}+\mbf{10}_{(4)}+\b{\mbf{10}}_{(-4)}\,,
\end{array}\label{SO10SU5U1}
\end{array}
\eeq
and embed a  $\bb{Z}_{2}\hookrightarrow SU(5)$ which is realized as in eq.~(\ref{Ug2}). Then, the $\mbf{5}$, 
$\mbf{10}$ and $\mbf{24}$ irreps of $SU(5)$ under the $SU(5)\supset SU(2)\times SU(3)\times U(1)$
decomposition read as in eq.~(\ref{SU5toSU2SU3U1HB}) and leads to the breaking $SO(10)\to SU^{a}(2)\times
SU^{b}(2)\times SU(4)$ which is a Pati-Salam type model,
\beq
\begin{array}{l}
SO(10) \supset  SU^{(i)}(2)\times SU^{(ii)}(2)\times SU(4)\\
\begin{array}{r@{}l@{}l@{}}
\mbf{10}&=&\ubr{(\mbf{2},\mbf{2},\mbf{1})}_{(-1)}+\ubr{(\mbf{1},\mbf{1},\mbf{6})}_{(+1)}\,,\\
\mbf{16}&=&\ubr{(\mbf{2},\mbf{1},\mbf{4})}_{(-1)}+\ubr{(\mbf{1},\mbf{2},\b{\mbf{4}})}_{(+1)}\,,\\
\mbf{45}&=&\ubr{(\mbf{3},\mbf{1},\mbf{1})}_{(+1)}+\ubr{(\mbf{1},\mbf{3},\mbf{1})}_{(+1)}
         +\ubr{(\mbf{1},\mbf{1},\mbf{15})}_{(+1)}+\ubr{(\mbf{2},\mbf{2},\mbf {6})}_{(-1)}\,.
\end{array}\label{SO10toSU2SU2SU4HB}
\end{array}
\eeq

Again we notice that although one would naively expect the $SO(10)$ gauge group to break towards SM, this is 
not the case.

For completeness in table~\ref{SO10-Breaking-Z2}  we present the above case.

\setlength{\tabcolsep}{15pt}
\setlength{\arraycolsep}{15pt}
\begin{table}[t]
\begin{center}
\begin{tabular}{|c!{\vrule width 1pt}c|c|}\hline
{\bf Embedd.}
& $U_{g}$
& $\mbf{K'}$
\\ \hhline{*{3}{=}}
  $\mbf{1}$
&
  $U_{g}^{(2)}$
&
  $SU^{(i)}(2)\times SU^{(ii)}(2)\times SU(4)$
\\\hline
\end{tabular}
\end{center}
\caption{\textbf{Embedding of $\bb{Z}_{2}$ discrete symmetry in $SO(10)$ GUT and its symmetry breaking
pattern.}\emph{$U_{g}^{(2)}=diag(-1,-1,1,1,1)$ as in text.}\label{SO10-Breaking-Z2}}
\end{table}

\subsubsection{The $\bb{Z}_{2}\times \bb{Z}_{2}'$ case.}
\label{TopologicallyInducedGaugeGroupBreaking-SO10-Z2Z2}

\paragraph{Embedding~$(\mbf{1'})$:~$\bb{Z}_{2}\hookrightarrow SU(5)$ and $\bb{Z}_{2}'\hookrightarrow SO(10)$ 
of $SO(10)\supset SU(5)\times U(1)$.}

Note that a second $\bb{Z}_{2}$ cannot break the $K'=SU^{a}(2)\times SU^{b}(2)\times SU(4)$ further. However 
by choosing the non-trivial embedding $U_{g}'=-\one_{10}$ of $\bb{Z}_{2}$ in the $SO(10)$ the phase factors
appearing in eq.~(\ref{SO10toSU2SU2SU4HB}) have their signs reversed under the action of $U_{g}^{(2)}U_{g}'$.

Again in table~\ref{SO10-Breaking-Z2Z2} we present the above case.

\setlength{\tabcolsep}{15pt}
\setlength{\arraycolsep}{15pt}
\begin{table}[t]
\begin{center}
\begin{tabular}{|c!{\vrule width 1pt}c|c|c|}\hline
{\bf Embedd.}
& $U_{g}$
& $U_{g}'$
& $\mbf{K'}$
\\ \hhline{*{4}{=}}
  $\mbf{1'}$
&
  $U_{g}^{(2)}$
&
 $-\one_{10}$
&
 $SU^{(i)}(2)\times SU^{(ii)}(2)\times SU(4)$
\\\hline
\end{tabular}
\end{center}
\caption{\textbf{Embedding of $\bb{Z}_{2}\times\bb{Z}_{2}'$ discrete symmetries in 
$SO(10)$ GUT and its symmetry breaking pattern.} \emph{$U_{g}^{(2)}=diag(-1,-1,1,1,1)$ as in text.}
\label{SO10-Breaking-Z2Z2}}
\end{table}

\section{Classification of semi-realistic particle physics models}\label{Classification}

Here starting from an $\cN=1$, $E_{8}$ Yang-Mills-Dirac theory defined in ten dimensions, we provide a
complete classification of the semi-realistic particle physics models resulting from CSDR of the original
theory and a subsequent application of the Wilson flux breaking mechanism. According to our requirements  in
sec.~\ref{Remarks-on-CSDR} the dimensional reduction of this theory over the six-dimensional coset spaces,
leads to anomaly free $E_{6}$ and $SO(10)$ GUTs in four dimensions. {Recall} also that the four-dimensional
surviving scalars transform in the fundamental of the resulting gauge group and are not suitable for the
superstrong symmetry breaking of these GUTs towards the SM. One way out was discussed in
sec.~\ref{sec:WilsonFlux-theory}, namely the Wilson flux breaking mechanism. In the present section we
investigate to which extent applying both methods, CSDR and Wilson flux breaking mechanism one can obtain
reasonable low energy models.

\subsection{Dimensional reduction over symmetric coset spaces}
\label{DimReduction-SCosets}

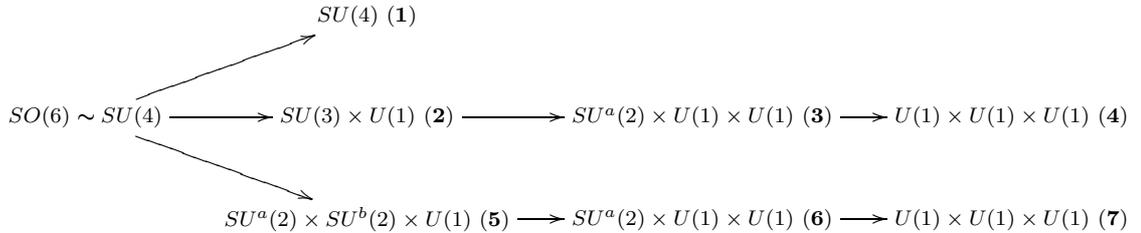
\begin{figure}[htbp]
\begin{scriptsize}
\begin{displaymath}
\xymatrix@C=0.6cm{
&\AAA~(\mbf{1})            &                                                      &\\
SO(6)\thicksim\AAA\ar[ur]\ar[dr]\ar[r]   &\AA\times\Y~(\mbf{2})\ar[r] &\A{a}\times\Y\times\Y~(\mbf{3})\ar[r]
											    & \Y\times\Y\times\Y~(\mbf{4})\\
&\A{a}\times\A{b}\times\Y~(\mbf{5})\ar[r]  &\A{a}\times\Y\times\Y~(\mbf{6})\ar[r] &\Y\times\Y\times\Y~(\mbf{7}) }
\end{displaymath}
\end{scriptsize}
\caption{\emph{Possible $E_{8}\supset SO(6)\supset R$ embeddings for the symmetric coset spaces, $S/R$, of
table~\ref{SO6VectorSpinorContentSCosets}.} 
\label{A3DecompChannels}}
\end{figure}

We consider all the possible embeddings  $E_{8}\supset SO(6)\supset R$ for the six-dimensional
\textit{symmetric} coset spaces, $S/R$, listed in the first column of
table~\ref{SO6VectorSpinorContentSCosets}\footnote{We have excluded the study of dimensional
reduction over the $Sp(4)/(SU(2)\times U(1))_{max}$ coset space which does not admit fermions.}. These
embeddings are presented in fig.~\ref{A3DecompChannels}. It is worth noting that in all cases the dimensional
reduction of the initial gauge theory leads to an $SO(10)$ GUT according to the concluding remarks in
sec.~\ref{sec:CSDR-rules}. The result of our examination in the present section is that the additional use of
the Wilson flux breaking mechanism leads to four-dimensional theories of Pati-Salam type. In the following
sections~\ref{Case-1a}~-~\ref{Case-5e} we present in some detail our examination and the corresponding
results, which we summarize in tables~\ref{SO10ChannelSCosetsTable-CSDR}
and~\ref{SO10ChannelSCosetsTable-CSDR-HOSOTANI} presented in appendix~\ref{CSDR-SCosets-Results}\footnote{For convenience we label the cases examined in the following subsections as `\rm{Case No.x}' with the `\rm{No}' denoting the embedding$R\hookrightarrow E_{8}$ and the `\rm{x}' the coset space we use. The same label is also used in
tables~\ref{SO10ChannelSCosetsTable-CSDR} and~\ref{SO10ChannelSCosetsTable-CSDR-HOSOTANI}.}.

\subsubsection[Reduction of $G=E_{8}$ using $\tfrac{SO(7)}{SO(6)}$]
{Reduction of $G=E_{8}$ over $B=B_{0}/F^{B_{0}}$, $B_{0}=SO(7)/SO(6)$. (Case $\mbf{1a}$)}
\label{Case-1a}

We consider Weyl fermions belonging in the adjoint of $G=E_{8}$  and the embedding of $R=SO(6)$ into $E_{8}$
suggested by the decomposition
\beq
\begin{array}{l@{}l@{}l@{}}                                                     
E_{8}&\supset& SO(16) \supset SO(6)\times SO(10)\\                              
\mbf{248}&=&(\mbf{1}, \mbf{45})+(\mbf{6},\mbf{10})+(\mbf{15},\mbf{1}) 
                +(\mbf{4},\mbf{16})+(\b{\mbf {4}},\b{\mbf {16}})\,.
\end{array}\label{DecompCh1-SO10}
\eeq

If only the CSDR mechanism was applied the resulting four-dimensional gauge group would be
\beqnn
H=C_{E_{8}}(SO(6))=SO(10)\,.
\eeqnn
According to table~\ref{SO6VectorSpinorContentSCosets}, the $R=SO(6)$ content of vector and spinor of $B_{0}=S/R=SO(7)/SO(6)$ is
$\mbf{6}$ and $\mbf{4}$, respectively. Then applying the CSDR rules~(\ref{RinS-embedding}),~(\ref{RinG-embedding})
and~(\ref{Fdecomp}),~(\ref{Frule}) the four-dimensional theory would contain scalars transforming as $\mbf{10}$ under the
$H=SO(10)$ gauge group and two copies of chiral fermion belonging in the $\mbf{16}_{L}$ of $H$.
%

Next we apply in addition the Wilson flux breaking mechanism discussed already in
sec.~\ref{sec:WilsonFlux-theory} and take into account the various observations made there. The freely acting
discrete symmetries, $F^{S/R}$, of the coset space $SO(7)/SO(6)$ (case `$\mbf{a}$' in
table~\ref{SO6VectorSpinorContentSCosets}) are the Weyl, $\rm{W}=\bb{Z}_{2}$ and the center
of $S$, $\rm{Z}(S)=\bb{Z}_{2}$. As it was explained in sec.~\ref{sec:InterestingDiscreteSymmetries} the use of
 ${\rm Z}(S)$ alone is excluded. On the other hand, according to the discussion in
sec.~\ref{TopologicallyInducedGaugeGroupBreaking-SO10-Z2} the ${\rm W}$ discrete symmetry leads to a
four-dimensional theory with gauge group
\beqnn
K'=C_{H}(T^{H})=SU^{(i)}(2)\times SU^{(ii)}(2)\times SU(4)\,.
\eeqnn
Then according to~eq.~(\ref{WFluxSurvivingField}), the surviving field content has to be invariant
under the combined action of the considered discrete symmetry itself, $F^{S/R}$, and its induced image in the
$H$ gauge group, $T^{H}$. Using the ${\rm W}=\bb{Z}_{2}$ discrete symmetry, the decomposition of the irrep.
$\mbf{10}$ of $SO(10)$ under the $K'$ gauge group is given in eq.~(\ref{SO10toSU2SU2SU4HB}),
\beq
\begin{array}{r@{}c@{}l@{}}
SO(10) &~\supset~&  SU^{(i)}(2)\times SU^{(ii)}(2)\times SU(4)\\
\mbf{10} &~=~& \ubr{(\mbf{2},\mbf{2},\mbf{1})}_{(-1)}+\ubr{(\mbf{1},\mbf{1},\mbf{6})}_{(+1)}\,.
\end{array}\label{SO10toSU2SU2SU4HB-10}
\eeq
Then, recalling that the vector $B_{0}=SO(7)/SO(6)$ is invariant under the action of ${\rm W}$
(see table~\ref{SO6VectorSpinorContentSCosets}), we conclude that the four-dimensional theory contains scalars transforming
according to
$$
(\mbf{1},\mbf{1},\mbf{6})
$$
of $K'$.
Similarly, the irrep. $\mbf{16}$ of $SO(10)$ decomposes under the $K'$ as
\beq
\begin{array}{r@{}c@{}l@{}}
SO(10) &~\supset~&  SU^{(i)}(2)\times SU^{(ii)}(2)\times SU(4)\\
\mbf{16}&~=~&\ubr{(\mbf{2},\mbf{1},\mbf{4})}_{(-1)}+\ubr{(\mbf{1},\mbf{2},\b{\mbf{4}})}_{(+1)}\,.
\end{array}\label{SO10toSU2SU2SU4HB-16}
\eeq
In this case the spinor of the tangent space of $SO(7)/SO(6)$ decomposed under $R=SO(6)$ is obviously $\mbf{4}$.
Then, since the ${\rm W}$ transformation property is $\mbf{4}\lra\b{\mbf{4}}$ (see
table~\ref{SO6VectorSpinorContentSCosets}), the fermion content of the four-dimensional theory transforms as
\beq
(\mbf{2},\mbf{1},\mbf{4})_{L}-(\mbf{2},\mbf{1}, \mbf{4})'_{L} \qquad\mbox{and}\qquad
(\mbf{1},\mbf{2},\mbf{4})_{L}+(\mbf{1},\mbf{2}, \mbf{4})'_{L}\label{4d-fermions-H1-1a}
\eeq
under $K'$.

In the present case as far as the spontaneous symmetry breaking of the four-dimensional theory is concerned,
both theorems mentioned in sec~\ref{sec:4D-theory} are applicable. According to the first theorem mentioned
there, dimensional reduction over the $SO(7)/SO(6)$ symmetric coset space leads to a four-dimensional
potential with spontaneously symmetry breaking form. However, since the four-dimensional scalar fields
transform as $(\mbf{1},\mbf{1},\mbf{6})$ under the $K'$ gauge group obtaining a v.e.v. break the $SU(3)$
colour. Therefore, employing the ${\rm W}$ discrete symmetry is not an interesting case for further
investigation.

Next if we use the ${\rm W}\times{\rm Z}(S)=\bb{Z}_{2}\times\bb{Z}_{2}$ discrete symmetry, the Wilson flux
breaking mechanism leads again to the Pati-Salam gauge group, $K'$ (see
sec.~\ref{TopologicallyInducedGaugeGroupBreaking-SO10-Z2Z2}). However in this case, all the underbraced phase
factors of eqs~(\ref{SO10toSU2SU2SU4HB-10}) and~(\ref{SO10toSU2SU2SU4HB-16}) are multiplied
by $-1$. Therefore the four-dimensional theory now contains scalars transforming according to
$$
(\mbf{2},\mbf{2},\mbf{1})
$$
of $K'$, and two copies of chiral fermions transforming as in eq.~(\ref{4d-fermions-H1-1a}) but with the signs
of the linear combinations reversed.

Concerning the spontaneous symmetry breaking of the latter model, we note that the isometry group of the coset,
$SO(7)$, is embeddable in $E_{8}$ as
\beqnn
\begin{array}{c@{}c@{}c@{}}
E_{8}\supset &SO(7)\times &SO(9)\\
\ \                     &\cup              &\cap\\
\ \                     &SO(6) \times  &SO(10)\,,
\end{array}
\eeqnn
and according to the second theorem mentioned in sec.~\ref{sec:4D-theory}, if only the CSDR mechanism was
applied, the final gauge group would be
\beqnn
\cH=C_{E_{8}}(SO(7))=SO(9)\,.
\eeqnn
In other words the $\mbf{10}$ of $SO(10)$ would obtain v.e.v. leading to the spontaneous symmetry breaking
\beq
\begin{array}{r@{}c@{}l@{}}
SO(10)&~\to~& SO(9)\label{SO10SO9-breaking}\\
\mbf{10}&~=~&\langle\mbf{1}\rangle + \mbf{9}\,.
\end{array}
\eeq

However now we employ the Wilson flux breaking mechanism which breaks the gauge symmetry further in higher
dimensions. It is instructive to understand the spontaneous breaking indicated in eq.~(\ref{SO10SO9-breaking})
in this context too. A straightforward examination of the gauge group structure and the reps of the scalars
that are involved, suggests that the breaking indicated in eq.~(\ref{SO10toSU2SU2SU4HB-10}) is realized in the
present context as
\beq
\begin{array}{r@{}l@{}l@{}}
SU^{(i)}(2)\times SU^{(ii)}(2)\times SU(4)&~\to~& SU^{diag}(2)\times SU(4)\\
                        (\mbf{2},\mbf{2},\mbf{1})&~=~&\langle(\mbf{1},\mbf{1})\rangle+(\mbf{3},\mbf{1})\,,
\end{array}\label{SU2SU2SU4toSU2SU4}
\eeq
i.e. the final gauge group of the four-dimensional theory is
$$
K=SU^{diag}(2)\times SU(4)\,.
$$
Accordingly, the fermions transform as
\beqnn
(\mbf{2},\mbf{4})_{L}+(\mbf{2},\mbf{4})'_{L}\qquad\mbox{and}\qquad
(\mbf{2},\mbf{4})_{L}-(\mbf{2},\mbf{4})'_{L}
\eeqnn
under $K$.

\subsubsection[Reduction of $G=E_{8}$ using $\tfrac{SU(4)}{SU(3)\times U(1)}$]
{Reduction of $G=E_{8}$ over $B=B_{0}/F^{B_{0}}$, $B_{0}=SU(4)/(SU(3)\times U(1))$. (Case $\mbf{2b}$)}

We consider again Weyl fermions belonging in the adjoint of  $G=E_{8}$ and the embedding of $R=SU(3)\times U(1)$ into $E_{8}$
suggested by the decomposition\footnote{This decomposition is in accordance with the Slansky tables\cite{Slansky:1981yr} but with
opposite $U(1)$ charge.}
\begin{align}
&E_{8}\supset SO(16)\supset SO(6)\times SO(10)\backsim SU(4)\times SO(10)\supset SU(3)\times U^{I}(1)\times SO(10)\nn\\
&\begin{array}{l@{}l@{}l@{}}
 E_{8}&\supset& (SU(3)\times U^{I}(1))\times SO(10)\\
 \mbf{248}&=&(\mbf{1},\mbf{1})_{(0)}+(\mbf{1}, \mbf{45})_{(0)}+(\mbf{8},\mbf{1})_{(0)}  
                 +(\mbf{3},\mbf{10})_{(-2)}+(\b{\mbf{3}},\mbf{10})_{(2)}\\       
               &+&(\mbf{3},\mbf{1})_{(4)}+(\b{\mbf{3}},\mbf{1})_{(-4)}
                 +(\mbf{1},\mbf{16})_{(-3)}+(\mbf{1},\b{\mbf{16}})_{(3)}\\
               &+&(\mbf{3},\mbf {16})_{(1)}+(\b{\mbf {3}},\b{\mbf {16}})_{(-1)}\,.
 \end{array}\label{DecompCh2-SO10}
\end{align}

If only the CSDR mechanism was applied, the resulting four-dimensional gauge group  would be
\beqnn
H=C_{E_{8}}(SU(3)\times U^{I}(1))=SO(10)\,\Big(\times U^{I}(1)\Big)\,,
\eeqnn
where the additional $U(1)$ factor in the parenthesis corresponds to a global symmetry, according to the
concluding remarks in sec.~\ref{sec:CSDR-rules}. The $R=SU(3)\times U^{I}(1)$ content of the vector and spinor
of $B_{0}=S/R=SU(4)/(SU(3)\times U^{I}(1))$ can be read in the last two columns of
table~\ref{SO6VectorSpinorContentSCosets}. Then according to the CSDR rules, {the theory would contain scalars
belonging in the $\mbf{10}_{(-2)}$, $\mbf{10}_{(2)}$ of $H$ and two copies of chiral fermions transforming as
$\mbf{16}_{L(3)}$ and $\mbf{16}_{L(-1)}$ under the same gauge group.}

The freely acting discrete symmetries of the coset space $SU(4)/(SU(3)\times U(1))$ are not included in the
list~(\ref{FCases}) of those ones that are worth to be examined further.

\subsubsection[Reduction of $G=E_{8}$ using
$\tfrac{SU(3)}{SU(2)\times U(1)}\times\tfrac{SU(2)}{U(1)}$]
{Reduction of $G=E_{8}$ over $B=B_{0}/F^{B_{0}}$, $B_{0}=SU(3)/(SU(2)\times U(1))\times(SU(2)/U(1))$.
(Cases $\mbf{3d}$, $\mbf{6d}$)}
\label{Case-3d}

We consider again Weyl fermions belonging in the adjoint of $G=E_{8}$ and the following decomposition
\beq
\begin{array}{l@{}l@{}l@{}}
 E_{8} \supset SO(16)~&\supset&~ SO(6)\times SO(10)\backsim SU(4)\times SO(10)
                       \supset (SU'(3)\times U^{II}(1))\times SO(10)\\ 
                     ~&\supset&~ (SU^{a}(2)\times U^{I}(1)\times U^{II}(1))\times SO(10)
\end{array}\label{DecompCh3-SO10}
\eeq
or
\beq
\begin{array}{l@{}l@{}l@{}}
E_{8} \supset SO(16)~&\supset&~ SO(6)\times SO(10)\backsim SU(4)\times SO(10)
                      \supset  (SU^{a}(2)\times SU^{b}(2)\times U^{II}(1))\times SO(10)\\
                    ~&\supset&~(SU^{a}(2)\times U^{I}(1) \times U^{II}(1))\times SO(10)
\end{array}\label{DecompCh6-SO10}
\eeq
In both cases we can properly redefine the $U(1)$ charges, and consequently choose an embedding of $R=SU(2)\times
U^{I}(1)\times U^{II}(1)$ into $E_{8}$ as follows
\beq
\begin{array}{l@{}l@{}l@{}}
 E_{8}&\supset& (SU^{a}(2)\times U^{I'}(1)\times U^{II'}(1))\times SO(10)\\
\mbf{248}&=&(\mbf{1}, \mbf{1})_{(0, 0)}+(\mbf{1}, \mbf{1})_{(0, 0)}             
 	        + (\mbf{3}, \mbf{1})_{(0, 0)}+(\mbf{1}, \mbf{45})_{(0, 0)}\\           
  	      &+&(\mbf{1}, \mbf{1})_{(-2 b, 0)}+(\mbf{1}, \mbf{1})_{(2 b, 0)}
                + (\mbf{2}, \mbf{1})_{(-b, 2 a)}+(\mbf{2}, \mbf{1})_{(b, -2 a)}\\
              &+& (\mbf{2}, \mbf{1})_{(-b, -2 a)}+(\mbf{2}, \mbf{1})_{(b, 2 a)}
                + (\mbf{1}, \mbf{10})_{(0, -2 a)}+(\mbf{1}, \mbf{10})_{(0, 2 a)}\\
              &+& (\mbf{2}, \mbf{10})_{(b, 0)}+(\mbf{2}, \mbf{10})_{(-b, 0)}
                + (\mbf{1}, \mbf{16})_{(b, -a)}+(\mbf{1}, \b{\mbf{16}})_{(-b, a)}\\
              &+& (\mbf{1}, \mbf{16})_{(-b, -a)}+(\mbf{1}, \b{\mbf{16}})_{(b, a)}
                + (\mbf{2}, \mbf{16})_{(0, a)}+(\mbf{2}, \b{\mbf{16}})_{(0, -a)}\,.
\end{array}\label{DecompCh3''-SO10}
\eeq

Here, $a$ and $b$ are the $U(1)$ charges of vector and fermion content of the coset space
$B_{0}=S/R=SU(3)/(SU^{a}(2)\times U^{I'}(1))\times (SU(2)/U^{II'}(1))$, shown in the last two columns of
table~\ref{SO6VectorSpinorContentSCosets} (case `$\mbf{d}$'). Then, if only the CSDR mechanism was applied, the resulting
four-dimensional gauge group would be
\beqnn
H=C_{E_{8}}(SU^{a}(2)\times U^{I'}(1) \times U^{II'}(1))=SO(10)\,\Big(\times U^{I'}(1)\times U^{II'}(1)\Big)\,,
\eeqnn
where the additional $U(1)$ factors in the parenthesis correspond to global symmetries. According to the
CSDR rules, the four-dimensional model contains scalars belonging in $\mbf{10}_{(0, -2 a)}$, 
$\mbf{10}_{(0, 2a)}$, $\mbf{10}_{(b, 0)}$ and $\mbf{10}_{(-b, 0)}$ of $H$ and two copies of chiral fermions
transforming as $\mbf{16}_{L(b, -a)}$, $\mbf{16}_{L(-b, -a)}$ and $\mbf{16}_{L(0, a)}$ under the same gauge
group.

The freely acting discrete symmetries of the coset space under discussion are the center of $S$,
$\rm{Z}(S)=\bb{Z}_{3}\times \bb{Z}_{2}$ and the Weyl symmetry, $\rm{W}=\bb{Z}_{2}$. Then according to the
list~(\ref{FCases}) the interesting cases to be examined further are the following two.

In the first case we employ the $\rm{W}=\bb{Z}_{2}$ discrete symmetry which leads to a four-dimensional theory
with gauge symmetry group
\beqnn
K'= C_{H}(T^{H})= SU^{(i)}(2)\times SU^{(ii)}(2)\times SU(4)\,\Big( \times U^{I'}(1)\times U^{II'}(1)\Big)\,.
\eeqnn
Similarly to the case discussed in sec.~\ref{Case-1a}, the surviving scalars transform as
\beq
(\mbf{1},\mbf{1},\mbf{6})_{(b, 0)}\qquad\mbox{and}\qquad (\mbf{1},\mbf{1},\mbf{6})_{(-b, 0)}
\label{Case-3d-W-scalars}
\eeq 
under $K'$ which are the only  ones that are invariant under the action of ${\rm W}$
(table~\ref{SO6VectorSpinorContentSCosets}). Furthermore, taking into account the ${\rm W}$ transformation
properties listed in the last column of table~\ref{SO6VectorSpinorContentSCosets}, as well as the
decomposition of $\mbf{16}$ irrep of $SO(10)$ under $SU^{(i)}(2)\times SU^{(i)}(2)\times SU(4)$ [see
eq.~(\ref{SO10toSU2SU2SU4HB-16})], we conclude that the four-dimensional fermions transform as
\begin{gather}
\begin{aligned}
&(\mbf{2},\mbf{1},\mbf{4})_{L(b,-a)}-(\mbf{2},\mbf{1},\mbf{4})'_{L(b,-a)}\,,\\
&(\mbf{1},\mbf{2},\mbf{4})_{L(b, -a)}+(\mbf{1},\mbf{2},\mbf{4})'_{L(b, -a)}\,,\\
\end{aligned}\nn\\
\begin{aligned}
&(\mbf{2},\mbf{1},\mbf{4})_{L(-b,-a)}-(\mbf{2},\mbf{1},\mbf{4})'_{L(-b, -a)}\,,\\
&(\mbf{1},\mbf{2},\mbf{4})_{L(-b, -a)}+(\mbf{1},\mbf{2},\mbf{4})'_{L(-b, -a)}\,,\\
\end{aligned}\qquad
\begin{aligned}
&(\mbf{2},\mbf{1},\mbf{4})_{L(0, a)}-(\mbf{2},\mbf{1},\mbf{4})'_{L(0, a)}\,,\\
&(\mbf{1},\mbf{2},\mbf{4})_{L(0, a)}+(\mbf{1},\mbf{2},\mbf{4})'_{L(0, a)}\,,
\end{aligned}\label{4d-fermions-H-3d}
\end{gather}
under $K'$.

Once more we have spontaneous symmetry breaking (since the coset space is symmetric) which breaks the
$SU(3)$-colour (since the scalars transform as in~(\ref{Case-3d-W-scalars}) under the $K'$ gauge group).
Therefore, employing the ${\rm W}$ discrete symmetry is not an interesting case for further investigation.

In the second case we use the $\bb{Z}_{2}\times\bb{Z}_{2}$ subgroup of the ${\rm W}\times{\rm Z}(S)$
combination of discrete symmetries. The surviving scalars of the four-dimensional theory belong in the
$(\mbf{2},\mbf{2},\mbf{1})_{(b, 0)}$ and $(\mbf{2},\mbf{2},\mbf{1})_{(-b, 0)}$ 
of the $K'$ gauge group which remains the same as before. The fermions, on the other hand, transform as those in
eq.~(\ref{4d-fermions-H-3d}) but with the signs of the linear combinations reversed. The final gauge group
after the spontaneous symmetry breaking of the theory is found to be
$$
K=SU^{diag}(2)\times SU(4)\,\Big(\times U^{I'}(1)\times U^{II'}(1)\Big),
$$
and its fermions transform as
\begin{gather}
\begin{aligned}
&(\mbf{2},\mbf{4})_{(b,-a)}+(\mbf{2},\mbf{4})'_{(b,-a)}\,,\\
&(\mbf{2},\mbf{4})_{(b, -a)}-(\mbf{2},\mbf{4})'_{(b, -a)}\,,\\
\end{aligned}\nn\\
\begin{aligned}
&(\mbf{2},\mbf{4})_{(-b,-a)}+(\mbf{2},\mbf{4})'_{(-b, -a)}\,,\\
&(\mbf{2},\mbf{4})_{(-b, -a)}-(\mbf{2},\mbf{4})'_{(-b, -a)}\,,\\
\end{aligned}\qquad
\begin{aligned}
&(\mbf{2},\mbf{4})_{(0, a)}+(\mbf{2},\mbf{4})'_{(0, a)}\,,\\
&(\mbf{2},\mbf{4})_{(0, a)}-(\mbf{2},\mbf{4})'_{(0, a)}
\end{aligned}
\end{gather}
under $K$.

\subsubsection[Reduction of $G=E_{8}$ using $\left(\tfrac{SU(2)}{U(1)}\right)^{3}$]
{Reduction of $G=E_{8}$ over $B=B_{0}/F^{B_{0}}$, $B_{0}=(SU(2)/U(1))^{3}$. (Cases $\mbf{4f}$, $\mbf{7f}$)}
\label{Case-4f}

We consider again Weyl fermions belonging in the adjoint of $G=E_{8}$ and the following decomposition
\beq
\begin{array}{l@{}l@{}l@{}}
E_{8}\supset SO(16)~&\supset&~ SO(6)\times SO(10)\backsim SU(4)\times SO(10)
         \supset  SU'(3)\times U^{III}(1)\times SO(10)\\
         ~&\supset&~ (SU^{a}(2)\times U^{II}(1)\times U^{III}(1))\times SO(10)\\
         ~&\supset&~SO(10)\times U^{I}(1)\times U^{II}(1)\times U^{III}(1)
\end{array}\label{DecompCh4-SO10}
\eeq
or
\beq
\begin{array}{l@{}l@{}l@{}}
E_{8}\supset SO(16)~&\supset&~ SO(6)\times SO(10)\backsim SU(4)\times SO(10)\\
                   ~&\supset&~ (SU^{a}(2)\times SU^{b}(2)\times U^{III}(1))\times SO(10)\\
                   ~&\supset&~ SO(10)\times U^{I}(1)\times U^{II}(1)\times U^{III}(1)
\end{array}\label{DecompCh7-SO10}
\eeq
In both cases we can properly redefine the $U(1)$ charges, and consequently choose an embedding of $R=SU(2)\times
U^{I}(1)\times U^{II}(1)$ into $E_{8}$ as follows
\beq
 \begin{array}{l@{}l@{}l@{}}
 E_{8}&\supset& SO(10)\times U^{I'}(1)\times U^{II'}(1)\times U^{III'}(1)\\
\mbf{248}&=&\mbf{1}_{(0, 0, 0)}+\mbf{1}_{(0, 0, 0)}+\mbf{1}_{(0, 0, 0)}   
                + \mbf{45}_{(0, 0, 0)}\\
              &+&\mbf{1}_{(-2 a, 2 b, 0)}+\mbf{1}_{(2 a, -2 b, 0)}
                +  \mbf{1}_{(-2 a, -2 b, 0)}+\mbf{1}_{(2 a, 2 b, 0)}\\
              &+&\mbf{1}_{(-2 a, 0, -2 c)}+\mbf{1}_{(2 a, 0, 2 c)}
                + \mbf{1}_{(0, -2 b, -2 c)}+\mbf{1}_{(0, 2 b, 2 c)}\\
              &+&\mbf{1}_{(-2 a, 0, 2 c)}+\mbf{1}_{(2 a, 0, -2 c)}
                + \mbf{1}_{(0, -2 b, 2 c)}+\mbf{1}_{(0, 2 b, -2 c)}\\
              &+&\mbf{10}_{(0, 0, 2 c)}+\mbf{10}_{(0, 0, -2 c)}
                + \mbf{10}_{(0, 2 b, 0)}+\mbf{10}_{(0, -2 b, 0)}\\
              &+&\mbf{10}_{(2 a, 0, 0)}+\mbf{10}_{(-2 a, 0, 0)}
                + \mbf{16}_{(a, b, c)}+\b{\mbf{16}}_{(-a, -b, -c)}\\
              &+&\mbf{16}_{(-a, -b, c)}+ \b{\mbf{16}}_{(a, b, -c)}
                + \mbf{16}_{(-a, b, -c)}+\b{\mbf{16}}_{(a, -b, c)}\\
              &+&\mbf{16}_{(a, -b, -c)}+\b{\mbf{16}}_{(-a, b, c)}\,.
\end{array}\label{DecompCh4'-SO10}
\eeq

Then, if only the CSDR mechanism was applied, the four-dimensional gauge group would be
\beqnn
H=C_{E_{8}}(U^{I'}(1)\times U^{II'}(1)\times U^{III'}(1))=SO(10)\,\Big(\times U^{I'}(1)\times U^{II'}(1)\times U^{III'}(1)\Big)\,.
\eeqnn
The same comment as in the previous cases holds for the additional $U(1)$ factors in the parenthesis.
The $R=U^{I'}(1)\times U^{II'}(1)\times U^{III'}(1)$ content of vector and spinor of
$B_{0}=S/R=(SU(2)/U^{I'}(1))\times (SU(2)/U^{II'}(1))\times (SU(2)/U^{III'}(1))$ can be read
in the last two columns of table \ref{SO6VectorSpinorContentSCosets}. According to the CSDR rules then, the
resulting four-dimensional theory would contain scalars belonging in $\mbf{10}_{(2 a, 0, 0)}$, 
$\mbf{10}_{(-2 a, 0,0)}$, $\mbf{10}_{(0, 2 b,0)}$, $\mbf{10}_{(0, -2 b, 0)}$, $\mbf{10}_{(0, 0, 2 c)}$ and
$\mbf{10}_{(0, 0, -2 c)}$ of $H$ and two copies of chiral fermions transforming as $\mbf{16}_{L(a, b, c)}$,
$\mbf{16}_{L(-a, -b, c)}$, $\mbf{16}_{L(-a, b, -c)}$ and $\mbf{16}_{L(a, -b, -c)}$ under the same gauge group.

The freely acting discrete symmetries, $F^{S/R}$, of the coset space $(SU(2)/U(1))^{3}\sim (S^{2})^{3}$ are
the center of $S$, $\rm{Z}(S)=(\bb{Z}_{2})^{3}$ and the Weyl discrete symmetry, $\rm{W}=(\bb{Z}_{2})^{3}$.
Then according to the list~(\ref{FCases}) the interesting cases to be examined further are the following.

First, let us mod out the $(S^{2})^{3}$ coset space by the $\bb{Z}_{2}\subset{\rm W}$ and consider the
multiple connected manifold $S^{2}/\bb{Z}_{2}\times S^{2}\times S^{2}$. Then, the resulting four-dimensional
gauge group will be
\beqnn
K'=SU^{(i)}(2)\times SU^{(ii)}(2)\times SU(4)\,\Big(\times U^{I'}(1)\times U^{II'}(1)\times U^{III'}(1)\Big)\,.
\eeqnn
The four-dimensional theory will contain scalar which belong in
\begin{align*}
&(\mbf{1},\mbf{1},\mbf{6})_{(0, 2 b, 0)}\,,\qquad (\mbf{1},\mbf{1},\mbf{6})_{(0, -2 b, 0)}\,,\\
&(\mbf{1},\mbf{1},\mbf{6})_{(0, 0, 2 c)}\,, \qquad (\mbf{1},\mbf{1},\mbf{6})_{(0, 0, -2 c)}
\end{align*}
of $K'$; these are the only ones that are invariant under the action of the considered $\bb{Z}_{2}\subset {\rm W}$.
However, linear combinations between the two copies of the CSDR-surviving left-handed fermions have no definite
properties under the abelian factors of the $K'$ gauge group and they do not survive. As a result, the model is
not an interesting case for further investigation.

Second, if we employ the $\bb{Z}_{2}\times\bb{Z}_{2}\subset{\rm W}$ discrete symmetry and consider the
manifold $S^{2}/\bb{Z}_{2}\times S^{2}/\bb{Z}_{2}\times S^{2}$, the resulting four-dimensional theory has
the same gauge group as before, i.e. $K'$. Similarly as before, scalars transforms as
\beqnn
(\mbf{1},\mbf{1},\mbf{6})_{(0, 0, 2 c)}\,, \qquad (\mbf{1},\mbf{1},\mbf{6})_{(0, 0, -2 c)}
\eeqnn
under $K'$. However, no fermions survive in the four-dimensional theory and the model is again not an interesting case to
examine further.

Finally, if we employ the  $\bb{Z}_{2}\times\bb{Z}_{2}\subset{\rm W}\times{\rm Z}(S)$ discrete symmetry, the
four-dimensional theory contains scalars which belong in
\begin{align*}
&(\mbf{2},\mbf{2},\mbf{1})_{(0, 2 b, 0)}\,, \qquad   (\mbf{2},\mbf{2},\mbf{1})_{(0, -2 b, 0)}\,,\\
&(\mbf{2},\mbf{2},\mbf{1})_{(0, 0, 2 c)}\,,  \qquad  (\mbf{2},\mbf{2},\mbf{1})_{(0, 0, -2 c)}
\end{align*}
of $K'$ but no fermions. The model is again not an interesting case for further study.

Therefore although the above studied cases have been obtained using discrete symmetries which are included in the
list~(\ref{FCases}), no fermion fields survive in the four-dimensional theory. The reason is that we employ
here only a subgroup of the Weyl discrete symmetry ${\rm W}=(\bb{Z}_{2})^{3}$ and we cannot form linear
combinations among the two copies of the CSDR-surviving left-handed fermions which are invariant under
eq.~(\ref{WFluxSurvivingField}). The use of the whole ${\rm W}$ discrete symmetry, on the other hand, would
lead to four-dimensional theories with smaller gauge symmetry than the one of SM.

\subsubsection[Reduction of $G=E_{8}$ using
$\tfrac{Sp(4)}{SU(2)\times SU(2)}\times\tfrac{SU(2)}{U(1)}$]
{Reduction of $G=E_{8}$ over $B=B_{0}/F^{B_{0}}$, $B_{0}=Sp(4)/(SU(2)\times SU(2))\times(SU(2)/U(1))$.
(Case $\mbf{5e}$)}\label{Case-5e}

Finally, we consider Weyl fermions in the adjoint of $G=E_{8}$ and the embedding of $R=SU(2)\times SU(2)\times U(1)$ into $E_{8}$ suggested by the decomposition
\begin{align}
&\begin{array}{l@{}l@{}l@{}}
E_{8} \supset SO(16)~&\supset&~ SO(6)\times SO(10)\backsim SU(4)\times SO(10)\\
                    ~&\supset&~(SU^{a}(2)\times SU^{b}(2)\times U^{I}(1))\times SO(10)\\
\end{array}\nn\\
&\begin{array}{l@{}l@{}l@{}}
E_{8}&\supset& (SU^{a}(2)\times SU^{b}(2)\times U^{I}(1))\times SO(10)\\
 \mbf{248}&=&(\mbf{1},\mbf{1},\mbf{1})_{(0)}+(\mbf{1},\mbf{1},\mbf{45})_{(0)}
                      +(\mbf{3},\mbf{1},\mbf{1})_{(0)}+(\mbf{1},\mbf{3},\mbf{1})_{(0)}\\
                   &+&(\mbf{2},\mbf{2},\mbf{1})_{(2)}+(\mbf{2},\mbf{2},\mbf{1})_{(-2)}
                      +(\mbf{1},\mbf{1},\mbf{10})_{(2)}+(\mbf{1},\mbf{1},\mbf{10})_{(-2)}\\
                   &+&(\mbf{2},\mbf{2},\mbf{10})_{(0)}
                      +(\mbf{2},\mbf{1},\mbf{16})_{(1)}+(\mbf{2},\mbf{1},\b{\mbf{16}})_{(-1)}\\
                    &+&(\mbf{1},\mbf{2},\mbf{16})_{(-1)}+(\mbf{1},\mbf{2},\b{\mbf{16}})_{(1)}\,.
 \end{array}\label{DecompCh5-SO10}
\end{align}

If only the CSDR mechanism was applied, the resulting four-dimensional gauge group would be
\beqnn
H=C_{E_{8}}(SU^{a}(2)\times SU^{b}(2)\times U^{I}(1))=SO(10)\,\Big(\times U^{I}(1)\Big)\,.
\eeqnn
The $R=SU^{a}(2)\times  SU^{b}(2)\times U^{I}(1)$ content of vector and spinor of
$B_{0}=S/R=Sp(4)/(SU^{a}(2)\times SU^{b}(2))\times (SU(2)/U(1))$ can be read in the last two columns of
table \ref{SO6VectorSpinorContentSCosets}. According to the CSDR rules the resulting four-dimensional theory
would contain scalars belonging in $\mbf{10}_{(0)}$, $\mbf{10}_{(2)}$ and $\mbf{10}_{(-2)}$ of $H$ and two
copies of chiral fermions transforming as $\mbf{16}_{L(1)}$ and $\mbf{16}_{L(-1)}$ under the same gauge group.

The freely acting discrete symmetries of the coset space $(Sp(4)/SU(2)\times SU(2))\times(SU(2)/U(1))$ (case
`$\mbf{e}$' in table~\ref{SO6VectorSpinorContentSCosets}), are the the center of $S$,
$\rm{Z}(S)=(\bb{Z}_{2})^{2}$ and the Weyl, $\rm{W}=(\bb{Z}_{2})^{2}$. According to the list~(\ref{FCases}) the
interesting cases to be examined further are the following.

First, if we employ the Weyl discrete symmetry, $\rm{W}=(\bb{Z}_{2})^{2}$ leads to a four-dimensional theory
with a gauge symmetry described by the group
\beqnn
K'=C_{H}(T^{H})=SU^{(i)}(2)\times SU^{(ii)}(2)\times SU(4)\,\Big(\times U^{I}(1)\Big)\,.
\eeqnn
The surviving scalars of the theory belong in
\beqnn
({\mbf 2},{\mbf 2},{\mbf 1})_{(0)}
\eeqnn
of $K'$, whereas the fermion content of the theory transforms as
\beq
\begin{aligned}
&({\mbf 2},{\mbf 1},{\mbf 4})_{L(1)}-({\mbf 2},{\mbf 1},{\mbf 4})'_{L(1)}\,,\\
&({\mbf 1},{\mbf 2},{\mbf 4})_{L(1)}+({\mbf 1},{\mbf 2},{\mbf 4})'_{L(1)}\,,
\end{aligned}\qquad
\begin{aligned}
&({\mbf 2},{\mbf 1},{\mbf 4})_{L(-1)}-({\mbf 2},{\mbf 1},{\mbf 4})'_{L(-1)}\,,\\
&({\mbf 1},{\mbf 2},{\mbf 4})_{L(-1)}+({\mbf 1},{\mbf 2},{\mbf 4})'_{L(-1)}\,
\end{aligned}\label{Case7eFermions}
\eeq
under $K'$.

Second, if we employ a $\bb{Z}_{2}\times\bb{Z}_{2}$ subgroup of the ${\rm W}\times{\rm Z}(S)$
combination of discrete symmetries, leads to a four-dimensional model with scalars belonging in
$({\mbf 2},{\mbf 2},{\mbf 1})_{(0)}$ of $K'$ and fermions transforming as in eq.~(\ref{Case7eFermions}) but
with the signs of the linear combinations reversed.

Finally, in table~\ref{SO10ChannelSCosetsTable-CSDR-HOSOTANI} we also report the less interesting case
$\bb{Z}_{2}\subseteq{\rm W}$.

Concerning the spontaneous symmetry breaking of theory, note that for the interesting cases of the~${\rm
W}=(\bb{Z}_{2})^{2}$~and $\bb{Z}_{2}\times\bb{Z}_{2}\subset {\rm W}\times {\rm Z}(S)$~discrete symmetries, the
final unbroken gauge group in four dimensions is found to be
$$
K=SU^{diag}(2)\times SU(4)\,\Big(\times U(1)\Big).
$$
Then, for the case of ${\rm W}$ discrete symmetry, the fermions of  the model transform as
\beq
\begin{aligned}
&({\mbf 2},{\mbf 4})_{(1)}-({\mbf 2},{\mbf 4})'_{(1)}\,,\\
&({\mbf 2},{\mbf 4})_{(1)}+({\mbf 2},{\mbf 4})'_{(1)}\,,
\end{aligned}\qquad
\begin{aligned}
&({\mbf 2},{\mbf 4})_{(-1)}-({\mbf 2},{\mbf 4})'_{(-1)}\,,\\
&({\mbf 2},{\mbf 4})_{(-1)}+({\mbf 2},{\mbf 4})'_{(-1)}\,,
\end{aligned}
\eeq
under $K$, whereas for the case of $\bb{Z}_{2}\times\bb{Z}_{2}\subset {\rm W}\times {\rm Z}(S)$ the
fermions belong in similar linear combinations as above but with their signs reversed.
 
\subsection{Dimensional reduction over non-symmetric coset spaces}
\label{DimReduction-NSCosets}

According to the discussion in sec.~\ref{Remarks-on-CSDR} we have to consider all the possible embeddings  
$E_{8}\supset SO(6)\supset R$, for the six-dimensional \textit{non-symmetric} cosets, $S/R$, of
table~\ref{SO6VectorSpinorContentNSCosets}. It is worth noting that the embedding of $R$ in all cases of
six-dimensional non-symmetric cosets are obtained by the following chain of maximal subgroups of $SO(6)$
\beq
SO(6)\sim SU(4)\supset SU(3)\times U(1)\supset SU(2)\times U(1)\times U(1)\supset U(1)\times U(1)\times U(1)\,.
\eeq
It is also important to recall from the discussion in secs~\ref{sec:CSDR-rules} and~\ref{Remarks-on-CSDR} that in
all these cases the dimensional reduction of the initial gauge theory leads to an $E_{6}$ GUT. The result of our
examination in the present section is that the additional use of the Wilson flux breaking mechanism leads to
four-dimensional gauge theories based on three different varieties of groups, namely  
$SO(10)\times U(1)$, $SU(2)\times SU(6)$ or $SU^{(i)}(2)\times SU^{(ii)}(2)\times SU(4)\times U(1)$.
In the following sections~\ref{Case-2a'}~-~\ref{Case-4c'} we present  details of our examination.
We summarize our results in tables~\ref{E6ChannelTable-CSDR}
and~\ref{E6ChannelTable-CSDR-HOSOTANI} presented in appendix~\ref{CSDR-NSCosets-Results}\footnote{We follow the same notation as in the examination of the symmetric cosets.}.

\subsubsection[Reduction of $G=E_{8}$ using $\tfrac{G_{2}}{SU(3)}$]
{Reduction of $G=E_{8}$ over $B=B_{0}/F^{B_{0}}$, $B_{0}=G_{2}/SU(3)$. (Case $\mbf{2a'}$)}
\label{Case-2a'}

We consider Weyl fermions belonging in the adjoint of  $G=E_{8}$ and identify the $R$ with the $SU(3)$
appearing in the decomposition~(\ref{DecompCh2-SO10}). Then, if only the CSDR mechanism was applied, the
resulting four-dimensional gauge group would be
$$
H=C_{E_{8}}(SU(3))=E_{6}\,,
$$
i.e. it appears an enhancement of the gauge group, a fact which was noticed earlier in several examples in
secs~\ref{TopologicallyInducedGaugeGroupBreaking-E6} and~\ref{TopologicallyInducedGaugeGroupBreaking-SO10}.
This observation suggests that we could have considered the following more obvious embedding of $R=SU(3)$ into $E_{8}$,
\beq
\begin{array}{l@{}l@{}l@{}}
E_{8}&\supset& SU(3) \times E_{6}\\
\mbf{248}&=&(\mbf{8},\mbf{1})+(\mbf{1},\mbf{78})
                +(\mbf{3},\mbf{27})+(\b{\mbf{3}},\b{\mbf{27}})\,.
\end{array}\label{DecompCh2-SO10-E6-2'}
\eeq
The $R=SU(3)$ content of vector and spinor of $B_{0}=S/R=G_{2}/SU(3)$ is $\mbf{3}+\b{\mbf{3}}$ and
$\mbf{1}+\mbf{3}$, respectively. According to the CSDR rules, the four-dimensional theory would contain
scalars belonging in $\mbf{27}$ and $\b{\mbf{27}}$ of $H=E_{6}$, two copies of chiral fermions transforming as
$\mbf{27}_{L}$ under the same gauge group and a set of fermions in the $\mbf{78}$ irrep., since the
dimensional reduction over non-symmetric coset preserves the supersymmetric
spectrum~\cite{Manousselis:2000aj,*Manousselis:2001xb,*Manousselis:2001re}.

The freely acting discrete symmetry, $F^{S/R}$, of the coset space $G_{2}/SU(3)$ is the Weyl,
$\rm{W}=\bb{Z}_{2}$ (case `$\mbf{a'}$' in table~\ref{SO6VectorSpinorContentNSCosets}). Then, following the
discussion in sec.~\ref{TopologicallyInducedGaugeGroupBreaking-E6-Z2}, the Wilson flux breaking mechanism
leads to a four-dimensional theory either with gauge group
\begin{equation}
(i)\qquad K'^{(\mbf{1})}=C_{H}(T^{H})=SO(10)\times U(1)\,,\label{Case2a'-WFB-1}
\end{equation}
in case we embed the $\bb{Z}_{2}$ into the $E_{6}$ gauge group as in the embedding ($\mbf{1}$) of
sec.~\ref{TopologicallyInducedGaugeGroupBreaking-E6-Z2}, or 
\begin{equation}
(ii)\qquad K'^{(\mbf{2},\mbf{3})}=C_{H}(T^{H})=SU(2)\times SU(6)\,,\label{Case2a'-WFB-2}
\end{equation}
in case we choose to embed the discrete symmetry as in the embeddings ($\mbf{2}$) or ($\mbf{3}$) of the same
subsection [the superscript in the $K'$'s above refer to the embeddings ($\mbf{1}$), ($\mbf{2}$) or
($\mbf{3}$)].

Making an analysis along the lines presented earlier in the case of symmetric cosets, we determine the
particle content of the two models, which is presented in table~\ref{E6ChannelTable-CSDR-HOSOTANI}. In both
cases the gauge symmetry of the four-dimensional theory cannot be broken further due to the absence of
scalars.

\subsubsection[Reduction of $G=E_{8}$ using 
$\left(\tfrac{Sp(4)}{SU(2)\times U(1)}\right)_{nonmax}$]
{Reduction of $G=E_{8}$ over $B=B_{0}/F^{B_{0}}$, $B_{0}=Sp(4)/(SU(2)\times U(1))_{nonmax}$.\\
(Case $\mbf{3b'}$)}

We consider Weyl fermions belonging in the adjoint of $G=E_{8}$ and the decomposition~(\ref{DecompCh3-SO10}).
In order the $R$ to be embedded in $E_{8}$ as  in eq.~(\ref{RinSO6inE8-embedding}), we identify it with the
$SU(2)\times U^{I}(1)$ appearing in the decomposition~(\ref{DecompCh3-SO10}). Then, if only the CSDR mechanism
was applied, the resulting gauge group would be
\beq
H=C_{E_{8}}(SU(2)\times U^{I}(1))=E_{6}\,\Big(\times U^{I}(1)\Big)\,.
\label{Case-3b'-CSDR-4D-gauge-group}
\eeq
Note that again appears an enhancement of the gauge group. Similarly with previously discussed cases, the
additional $U(1)$ factor in the parenthesis corresponds only to a global symmetry. The
observation~(\ref{Case-3b'-CSDR-4D-gauge-group}) suggests that we could have considered the following
embedding of $R=SU(2)\times U(1)$ into $E_{8}$\footnote{This decomposition is in accordance with the Slansky
tables but with opposite $U(1)$ charge.},
\beq
 \begin{array}{l@{}l@{}l@{}}
 E_{8}&\supset& SU(3)\times E_{6}\supset SU(2)\times U^{I}(1) \times E_{6}\\
\mbf{248}&=&(\mbf{1},\mbf{1})_{(0)}+(\mbf{1},\mbf{78})_{(0)}        
                + (\mbf{3},\mbf{1})_{(0)}                                            
                + (\mbf{2},\mbf{1})_{(-3)}+(\mbf{2},\mbf{1})_{(3)}\\
              &+&(\mbf{1},\mbf{27})_{(2)}+(\mbf{1},\b{\mbf{27}})_{(-2)}
                +(\mbf{2},\mbf{27})_{(-1)}+(\mbf{2},\b{\mbf{27}})_{(1)}\,.
\end{array}\label{DecompCh3'-SO10}
\eeq
The $R=SU(2)\times U^{I}(1)$ content of vector and spinor of 
$B_{0}=S/R=Sp(4)/(SU(2)\times U^{I}(1))_{non-max}$ can be read in the last two columns of table
\ref{SO6VectorSpinorContentNSCosets}. According to the CSDR rules then, the surviving scalars in four
dimensions would transform as $\mbf{27}_{(-2)}$, $\mbf{27}_{(1)}$, $\b{\mbf{27}}_{(2)}$ and
$\b{\mbf{27}}_{(-1)}$ under $H=E_{6}(\times U^{I}(1))$. The four-dimensional theory would also contain
fermions belonging  in $\mbf{78}_{(0)}$ of $H$ (gaugini of the model), two copies of left-handed fermions
belonging in $\mbf{27}_{L(2)}$ and $\mbf{27}_{L(-1)}$ and one fermion singlet transforming as $\mbf{1}_{(0)}$
under the same gauge group.

The freely acting discrete symmetries, $F^{S/R}$, of the coset space $Sp(4)/(SU(2)\times U(1))_{non-max}$,
are the center of $S$, $\rm{Z}(S)=\bb{Z}_{2}$ and the Weyl, $\rm{W}=\bb{Z}_{2}$. Then, employing the $\rm{W}$
discrete symmetry, we find that the resulting four-dimensional gauge group is either
\begin{eqnarray}
(i)&\qquad& K'^{(\mbf{1})}=C_{H}(T^{H})=SO(10)\times U(1)\,\Big(\times U^{I}(1)\Big)\,,\qquad\mbox{or}\\[3ex]
(ii)&\qquad& K'^{(\mbf{2},\mbf{3})}=C_{H}(T^{H})=SU(2)\times SU(6)\,\Big(\times U^{I}(1)\Big)\,,
\end{eqnarray}
depending on the embedding of $\bb{Z}_{2}\hookrightarrow E_{6}$ we choose to consider
(see sec.~\ref{TopologicallyInducedGaugeGroupBreaking-E6-Z2}). On the other hand, if we employ  the
$\rm{W}\times\rm{Z}(S)=\bb{Z}_{2}\times\bb{Z}_{2}$ combination of discrete symmetries, the
resulting four-dimensional gauge group is either
\beq
(iii)\qquad K'^{(\mbf{2'})}=SU(2)\times SU(6)\,\Big(\times U^{I}(1)\Big)\,,\label{Case3b'-WFB-3}
\eeq
in case we embed the $(\bb{Z}_{2}\times\bb{Z}_{2})$ into the $E_{6}$ gauge group as in the embedding ($\mbf{2'}$) of
sec.~\ref{TopologicallyInducedGaugeGroupBreaking-E6-Z2Z2}, or 
\beq
(iv)\qquad K'^{(\mbf{3'})}=SU^{(i)}(2)\times SU^{(ii)}(2)\times SU(4)\times U(1)\,\Big(\times U^{I}(1)\Big)\,,\label{Case3b'-WFB-4}
\eeq
in case we choose to embed the discrete symmetry as in the embedding ($\mbf{3'}$) of the same subsection.

Making a similar analysis as before, we determine the particle content of the four different models, which is
presented in table~\ref{E6ChannelTable-CSDR-HOSOTANI}. In all cases the gauge symmetry of the resulting
four-dimensional theory cannot be broken further by a Higgs mechanism due to the absence of scalars.

\subsubsection[Reduction of $G=E_{8}$ using $\tfrac{SU(3)}{U(1)\times U(1)}$]
{Reduction of $G=E_{8}$ over $B=B_{0}/F^{B_{0}}$, $B_{0}=SU(3)/(U(1)\times U(1))$.
(Case $\mbf{4c'}$)}
\label{Case-4c'}

We consider Weyl fermions in the adjoint of $G=E_{8}$ and the decomposition~(\ref{DecompCh4-SO10}). In order the
$R=U(1)\times U(1)$, to be embedded in $E_{8}$ as in eq.~(\ref{RinSO6inE8-embedding}) one has to identify it
with the $U^{I}(1)\times U^{II}(1)$ appearing in the decomposition~(\ref{DecompCh4-SO10}). Then, if only the
CSDR mechanism was applied, the resulting four-dimensional gauge group would be
\beq
H=C_{E_{8}}(U^{I}(1)\times U^{II}(1))=E_{6}\,\Big(\times U^{I}(1)\times U^{II}(1)\Big)\,.
\label{Case-4c'-CSDR-4D-gauge-group}
\eeq
Note again that an enhancement of the gauge group appears, whereas the additional $U(1)$ factors correspond to
global symmetries. The observation~(\ref{Case-4c'-CSDR-4D-gauge-group}) suggests that we could have considered
the following embedding of $R=U(1)\times U(1)$ into $E_{8}$,
\begin{align}
&E_{8}\supset SU(3)\times E_{6}\supset (SU(2)\times U^{II}(1))\times E_{6}\supset E_{6}\times U^{I}(1)\times U^{II}(1)\nn\\
&\begin{array}{l@{}l@{}l@{}}
 E_{8}&\supset&  E_{6}\times U^{I}(1) \times U^{II}(1)\\
\mbf{248}&=&\mbf{1}_{(0,0)}+\mbf{1}_{(0,0)}
                      +\mbf{78}_{(0,0)}
                      +\mbf{1}_{(-2,0)}+\mbf{1}_{(2,0)}
                      +\mbf{1}_{(-1,3)}+\mbf{1}_{(1,-3)}\\
                   &+&\mbf{1}_{(1,3)}+\mbf{1}_{(-1,-3)}
                      +\mbf{27}_{(0,-2)}+\b{\mbf{27}}_{(0,2)}
                      +\mbf{27}_{(-1,1)}+\b{\mbf{27}}_{(1,-1)}\\
                   &+&\mbf{27}_{(1,1)}+\b{\mbf{27}}_{(-1,-1)}\,.
\end{array}\label{DecompCh3-E6}
\end{align}
The $R=U^{I}(1)\times U^{II}(1)$ content of vector and spinor of $B_{0}=S/R=SU(3)/(U^{I}(1)\times U^{II}(1))$
can be read in the last two columns of table~\ref{SO6VectorSpinorContentNSCosets}.
{The embedding $R\hookrightarrow E_{8}$ suggested by the decomposition~(\ref{DecompCh3-E6}) 
corresponds in the following choice of the $U(1)$  charges appearing in the last case of
table~\ref{SO6VectorSpinorContentNSCosets}: $a=0$, $c=-2$, $b=-1$ and $d=1$.} Then, according to the CSDR
rules, the four-dimensional theory would contain scalars which belong in $\mbf{27}_{(0,-2)}$,
$\mbf{27}_{(0,2)}$, $\mbf{27}_{(-1,1)}$, $\mbf{27}_{(1,-1)}$, $\mbf{27}_{(1,1)}$ and $\mbf{27}_{(-1,-1)}$ of
$H=E_{6}(\times U^{I}(1)\times U^{II}(1))$. The resulting four-dimensional theory would also contain gaugini
transforming as $\mbf{78}_{(0,0)}$ under $H$, two copies of left-handed fermions belonging in
$\mbf{27}_{L(0,-2)}$, $\mbf{27}_{L(-1,1)}$, $\mbf{27}_{L(1,1)}$ and two fermion
singlets belonging in $\mbf{1}_{(0,0)}$ and $\mbf{1}_{(0,0)}$ of the same gauge group.

The freely acting discrete symmetries, $F^{S/R}$, of the coset space $SU(3)/(U(1)\times U(1))$ (case 
`$\mbf{c'}$' in table~\ref{SO6VectorSpinorContentNSCosets}), are the center of $S$, $\rm{Z}(S)=\bb{Z}_{3}$ and
the Weyl, $\rm{W}=\mbf{S}_{3}$. Then according to the list~(\ref{FCases}) only the $\bb{Z}_{2}\subset{\rm W}$
discrete symmetry is an interesting case to be examined further.

Then, employing the $\bb{Z}_{2}$ subgroup of the ${\rm W}=\mbf{S}_{3}$ discrete symmetry leads to a
four-dimensional theory either with gauge group
\begin{eqnarray}
(i)&\qquad& K'^{(\mbf{1})}=C_{H}(T^{H})=SO(10)\times U(1)\,\Big(\times U^{I}(1)\times U^{II}(1)\Big)\,,\qquad\mbox{or}\\[3ex]
(ii)&\qquad& K'^{(\mbf{2},\mbf{3})}=C_{H}(T^{H})=SU(2)\times SU(6)\,\Big(\times U^{I}(1)\times U^{II}(1)\Big)
\end{eqnarray}
depending on the embedding of $\bb{Z}_{2}\hookrightarrow E_{6}$ we choose to consider
(see sec.~\ref{TopologicallyInducedGaugeGroupBreaking-E6-Z2}).

Making a similar analysis as before, we determine the particle content of the two models as follows.

\paragraph{Case (i).}
The resulting four-dimensional theory contains gaugini which transform as
\beqnn
\mbf{1}_{(0,0,0)}\,,\qquad\mbf{45}_{(0,0,0)}
\eeqnn
under $K'^{(\mbf{1})}$, a set of fermion singlets which belong in
\beqnn
\mbf{1}_{(0,0,0)}\,,\qquad\mbf{1}_{(0,0,0)}\,,
\eeqnn
of $K'^{(\mbf{1})}$ and a set of chiral fermions which belong in one of the linear combinations
\beqnn
\left\{
\begin{array}{l}
\mbf{1}_{L(-4,0,-2)}+\mbf{1}'_{L(-4,0,-2)}\,,\\
\mbf{10}_{L(-2,0,-2)}+\mbf{10}'_{L(-2,0,-2)}\,,\\
\mbf{16}_{L(1,0,-2)}-\mbf{16}'_{L(1,0,-2)}\,,
\end{array}\right\}\,,\qquad
\left\{
\begin{array}{l}
\mbf{1}_{L(-4,-1,1)}+\mbf{1}'_{L(-4,-1,1)}\,,\\
\mbf{10}_{L(-2,-1,1)}+\mbf{10}'_{L(-2,-1,1)}\,,\\
\mbf{16}_{L(1,-1,1)}-\mbf{16}'_{L(1,-1,1)}\,,\\
\end{array}\right\}\,,
\eeqnn
or
\beqnn
\left\{
\begin{array}{l}
\mbf{1}_{L(-4,1,1)}+\mbf{1}'_{L(-4,1,1)}\,,\\
\mbf{10}_{L(-2,1,1)}+\mbf{10}'_{L(-2,1,1)}\,,\\
\mbf{16}_{L(1,1,1)}-\mbf{16}'_{L(1,1,1)}
\end{array}\right\}
\eeqnn
of the same gauge group, depending on the $\bb{Z}_{2}$ subgroup of $\mbf{S}_{3}$ that we choose to consider
(see table \ref{SO6VectorSpinorContentNSCosets}).

\paragraph{Case (ii).}
The resulting four-dimensional theory contains gaugini which transform as
\beqnn
(\mbf{3},\mbf{1})_{(0,0)}\,,\qquad (\mbf{1},\mbf{35})_{(0,0)}
\eeqnn
under $K'^{(\mbf{2},\mbf{3})}$, a set of fermion singlets which belong in
\beqnn
(\mbf{1},\mbf{1})_{(0,0)}\,,\qquad(\mbf{1},\mbf{1})_{(0,0)}\,,
\eeqnn
of $K'^{(\mbf{2},\mbf{3})}$ and a set of chiral fermions which belong in one of the linear combinations
\beqnn
\left\{
\begin{array}{l}
(\mbf{1},\mbf{15})_{L(0,-2)}+(\mbf{1},\mbf{15})'_{L(0,-2)}\,,\\
(\mbf{2},\b{\mbf{6}})_{L(0,-2)}-(\mbf{2},\b{\mbf{6}})'_{L(0,-2)}\,,
\end{array}\right\}\,,\qquad
\left\{
\begin{array}{l}
(\mbf{1},\mbf{15})_{L(-1,1)}+(\mbf{1},\mbf{15})'_{L(-1,1)}\,,\\
(\mbf{2},\b{\mbf{6}})_{L(-1,1)}-(\mbf{2},\b{\mbf{6}})'_{L(-1,1)}\,,
\end{array}\right\}\,,
\eeqnn
or
\beqnn
\left\{
\begin{array}{l}
(\mbf{1},\mbf{15})_{L(1,1)}+(\mbf{1},\mbf{15})'_{L(1,1)}\,,\\
(\mbf{2},\b{\mbf{6}})_{L(1,1)}-(\mbf{2},\b{\mbf{6}})'_{L(1,1)}\,,
\end{array}\right\}
\eeqnn
of the same gauge group, depending on the $\bb{Z}_{2}$ subgroup of $\mbf{S}_{3}$ that we choose to consider
(see table \ref{SO6VectorSpinorContentNSCosets}).

Note that in both cases the gauge symmetry of the four-dimensional theory cannot be broken further by a Higgs mechanism due to the
absence of scalars.

Finally, if we have used either the symmetric group of $3$ permutations, $\mbf{S}_{3}$, or its subgroup
$\bb{Z}_{3}\subset\mbf{S}_{3}$, we could not form linear combinations among the two copies of the CSDR-surviving left-handed
fermions and no fermions would survive in four dimensions.

\section{Conclusions}
\label{sec:Conclusions}

The CSDR is a consistent dimensional reduction scheme
\cite{Chatzistavrakidis:2007by,*Chatzistavrakidis:2007pp,*Coquereaux:1984ca,*Chaichian:1986mf,*Dvali:2001qr},
as well as an elegant framework to incorporate in a unified manner the gauge and the ad-hoc Higgs sector of
spontaneously broken four-dimensional gauge theories using the extra dimensions. The kinetic terms of fermions
were easily included in the same unified description. A striking feature of the scheme
concerning fermions was the discovery that chiral ones can be introduced~\cite{Manton:1981es} and
moreover they could result even from vector-like reps of the higher dimensional gauge
theory~\cite{Chapline:1982wy,Kapetanakis:1992hf}. This possibility is due to the presence of
non-trivial background gauge configurations required by the CSDR principle, in accordance with the
index theorem. Another striking feature of the theory is the possibility that the softly broken
sector of the four-dimensional supersymmetric theories can result from a higher-dimensional $\cN=1$
supersymmetric gauge theory with only a vector supermultiplet, when is dimensionally reduced over
non-symmetric coset
spaces~\cite{Manousselis:2000aj,*Manousselis:2001xb,*Manousselis:2001re,*Manousselis:2004xd}.
Another interesting feature useful in realistic model searches is the possibility to deform the
metric in certain non-symmetric coset spaces and introduce more than one
scales~\cite{Kapetanakis:1992hf,Farakos:1986sm,*Farakos:1986cj}. 

Recently there exist a revival of interest in the study of compactifications with internal manifolds
six-dimensional non-symmetric coset spaces possessing an $SU(3)$-structure within the framework of flux 
compactifications. Motivated by this interest we plan to examine the CSDR of the heterotic ten-dimensional
gauge theory in successive steps. In the present work, starting with a supersymmetric $\cN=1$, $E_{8}$ gauge
theory in ten dimensions we made a complete classification of the models obtained in four dimensions after
reducing the theory over all multiply connected six-dimensional coset spaces, resulting by moding out all the
freely acting discrete symmetries on these manifolds, and using the Wilson flux breaking mechanism in an
exhaustive way. The results of our extended investigation have been partially presented in a short
communication~\cite{Douzas:2007zz}. Despite some partial success, our result is that the two mechanisms used
to break the gauge symmetry, i.e. the geometric breaking of the CSDR and the topological of the Hosotani
mechanism are not enough to lead the four-dimensional theory to the SM or some interesting extension
as the MSSM. Limiting ourselves in the old CSDR framework one can think of some new sources of gauge symmetry
breaking, such as new scalars coming from a gauge theory defined even in higher
dimensions~\cite{Koca:1984dr,Jittoh:2008jc}. Much more interesting is to extend our examination in a future
study of the full ten-dimensional $E_{8}\times E_{8}$ gauge theory of the heterotic string. Moreover in that
case one does not have to be restricted in the study of freely acting discrete symmetries of the coset spaces
and can extent the analysis including
orbifolds~\cite{Kim:2006hv,*Kim:2006hw,*Kim:2007mt,*Kim:2007dx,Forste:2005gc, *Nilles:2006np,
Lebedev:2006kn,*Lebedev:2006tr,*Lebedev:2007hv}. More possibilities are offered in
refs~\cite{Blumenhagen:2005pm,*Blumenhagen:2005zg,*Blumenhagen:2006ux,*Blumenhagen:2006wj}.

\section*{Acknowledgements}

We would like to thank A. Kehagias, J. Kim, B. Schellekens, R. Blumenhagen and P. Manousselis for 
interesting discussions. This work was supported by the EPEAEK programme ``Irakleitos" and
co-funded by the European Union (75\%) and the Hellenic State (25\%).

\begin{appendices}

\section{Dimensional reduction over symmetric $6D$ coset spaces}\label{CSDR-SCosets-Results}

\captionsetup{font=normalsize}
\setlength{\tabcolsep}{0.5pt}
\setlength{\arraycolsep}{0.5pt}
\begin{scriptsize}
\begin{longtable}[b]{|c!{\vrule width 1pt}c|c!{\vrule width1pt}c|c|c|}
\caption{{\bf Dimensional reduction over symmetric $6D$ coset spaces}.
\emph{Particle physics models leading to $SO(10)$ GUTs in four dimensions.}
\label{SO10ChannelSCosetsTable-CSDR}}\\
\hline
{\bf Case}
&
{\bf Embedding}
&$\begin{array}{c} 
\mbf{6D}~{\bf Coset~Space}
\end{array}$
& $\mbf{H}$ 
&$\begin{array}{c}
{\bf Surviving}\\
{\bf scalars}\\
{\bf under}~\mbf{H}
\end{array}$
&$\begin{array}{c}
{\bf Surviving}\\
{\bf fermions}\\
{\bf under}~\mbf{H}
\end{array}$
\\[10pt]\hhline{*{6}{=}}                                                               
\endfirsthead
\hline
\multicolumn{6}{|l|}{\tiny continued from previous page}
\\\hline
{\bf Case}
&{\bf Embedding}
&$\begin{array}{c} 
\mbf{6D}~{\bf Coset~Space}
\end{array}$
& $\mbf{H}$ 
&$\begin{array}{c}
{\bf Surviving}\\
{\bf scalars}\\
{\bf under}~\mbf{H}
\end{array}$
&$\begin{array}{c} 
{\bf Surviving}\\
{\bf fermions}\\
{\bf under}~\mbf{H}
\end{array}$
\\[10pt]\hhline{*{6}{=}}
\endhead                                                                                      
\hline\multicolumn{6}{|r|}{\tiny\slshape{continued on next page}}
\\\hline
\endfoot                                                                                        
\hline
\endlastfoot
$\mbf{1a}$
& 
$\begin{array}{l@{}l@{}l@{}}                                                        
E_{8}&\supset& SO(6)\times SO(10)\\                                              
\mbf{248}&=&(\mbf{1}, \mbf{45})+(\mbf{6},\mbf{10})+(\mbf{15},\mbf{1})\\
              &+& (\mbf{4},\mbf{16})+(\b{\mbf {4}},\b{\mbf {16}})
\end{array}$
& 
$\frac{SO(7)}{SO(6)}$
&
 $SO(10)$
&
 $\mbf{10}$
&
$\begin{array}{l}
\mbf{16}_{L}\\
\mbf{16}_{L}'
\end{array}$
\\\hline 
$\mbf{2b}$   
&
 $\begin{array}{l}                                                                 
 E_{8}\supset SO(6)\times SO(10)\\                                        
 SO(6)\backsim SU(4)\supset SU(3)\times U^{I}(1)\\
 \begin{array}{l@{}l@{}l@{}}
 E_{8}&\supset& (SU(3)\times U^{I}(1))\times SO(10)\\
 \mbf{248}&=&(\mbf{1},\mbf{1})_{(0)}+(\mbf{1}, \mbf{45})_{(0)}+(\mbf{8},\mbf{1})_{(0)}\\    
                   &+&(\mbf{3},\mbf{10})_{(-2)}+(\b{\mbf{3}},\mbf{10})_{(2)}\\            
                   &+&(\mbf{3},\mbf{1})_{(4)}+(\b{\mbf{3}},\mbf{1})_{(-4)}\\
                   &+&(\mbf{1},\mbf{16})_{(-3)}+(\mbf{1},\b{\mbf{16}})_{(3)}\\
                   &+&(\mbf{3},\mbf {16})_{(1)}+(\b{\mbf {3}},\b{\mbf {16}})_{(-1)}
 \end{array}
 \end{array}$
&
$\frac{SU(4)}{SU(3)\times U^{I}(1)}$
&
   $ SO(10)\,\Big(\times U^{I}(1)\Big)$
&
$\begin{array}{l}
\mbf{10}_{(-2)}\\
\mbf{10}_{(2)}
 \end{array}$
&
$\begin{array}{l}
 \mbf{16}_{L(3)}\\
 \mbf{16}_{L(-1)}\\
\mbf{16}_{L(3)}'\\
\mbf{16}_{L(-1)}'
 \end{array}$
\\\hline
$\mbf{3d},~\mbf{6d}$
&
$\begin{array}{l}
E_{8}\supset SO(6)\times SO(10)\\                                           
 SO(6)\backsim SU(4)\supset SU'(3)\times U^{II}(1)\\          
 SU'(3)\supset SU^{a}(2)\times U^{I}(1)\\                                      
Y^{I'}=\frac{b}{3}Y^{I}+\frac{b}{3}Y^{II}\\                
Y^{II'}=\frac{2a}{3}Y^{I}-\frac{a}{3}Y^{II}\\[3pt]
\mbox{\textbf{or}}\\[3pt]
E_{8}\supset SO(6)\times SO(10)\\                                 
SO(6)\backsim SU(4)\\                                                        
SU(4) \supset SU^{a}(2)\times SU^{b}(2)\times U^{II}(1)\\          
SU^{b}(2)\supset U^{I}(1)\\
Y^{I'}=-b\,Y^{I}\\                                                                   
Y^{II'}=a\,Y^{II}\\                                                                    
\begin{array}{l@{}l@{}l@{}}
 E_{8} &\supset& (SU^{a}(2)\times U^{I'}(1)\times U^{II'}(1))\\
          &\times& SO(10)\\
\mbf{248}&=&(\mbf{1}, \mbf{1})_{(0, 0)}+(\mbf{1}, \mbf{1})_{(0, 0)}\\             
 	      &+&(\mbf{3}, \mbf{1})_{(0, 0)}+(\mbf{1}, \mbf{45})_{(0, 0)}\\           
  	      &+&(\mbf{1}, \mbf{1})_{(-2 b, 0)}+(\mbf{1}, \mbf{1})_{(2 b, 0)}\\
              &+&(\mbf{2}, \mbf{1})_{(-b, 2 a)}+(\mbf{2}, \mbf{1})_{(b, -2 a)}\\
              &+&(\mbf{2}, \mbf{1})_{(-b, -2 a)}+(\mbf{2}, \mbf{1})_{(b, 2 a)}\\
              &+&(\mbf{1}, \mbf{10})_{(0, -2 a)}+(\mbf{1}, \mbf{10})_{(0, 2 a)}\\
              &+&(\mbf{2}, \mbf{10})_{(b, 0)}+(\mbf{2}, \mbf{10})_{(-b, 0)}\\
              &+&(\mbf{1}, \mbf{16})_{(b, -a)}+(\mbf{1}, \b{\mbf{16}})_{(-b, a)}\\
              &+&(\mbf{1}, \mbf{16})_{(-b, -a)}+(\mbf{1}, \b{\mbf{16}})_{(b, a)}\\
              &+&(\mbf{2}, \mbf{16})_{(0, a)}+(\mbf{2}, \b{\mbf{16}})_{(0, -a)}
\end{array}
\end{array}$
&
$\begin{array}{l}
\left(\frac{SU(3)}{SU^{a}(2)\times U^{I'}(1)}\right)\\
\hfill\times\left(\frac{SU(2)}{U^{II'}(1)}\right)
\end{array}$
&
$\begin{array}{l@{}}
    SO(10)\\
    \Big(\times U^{I'}(1)\times U^{II'}(1)\Big)
\end{array}$
&
$\begin{array}{l}
\mbf{10}_{(0, -2 a)}\\
\mbf{10}_{(0, 2 a)}\\
\mbf{10}_{(b, 0)}\\
\mbf{10}_{(-b, 0)}
\end{array}$
&
$\begin{array}{l}
\mbf{16}_{L(b, -a)}\\
\mbf{16}_{L(-b, -a)}\\
\mbf{16}_{L(0, a)}\\
\mbf{16}_{L(b, -a)}'\\
\mbf{16}_{L(-b, -a)}'\\
\mbf{16}_{L(0, a)}'
\end{array}$
\\\hline
$\mbf{4f},~\mbf{7f}$ 
& 
$\begin{array}{l}                                                                 
 E_{8}\supset SO(6)\times SO(10)\\                                        
 SO(6)\backsim SU(4)\supset SU'(3)\times U^{III}(1)\\                
 SU'(3)\supset SU^{a}(2)\times U^{II}(1)\\
SU^{a}(2)\supset U^{I}(1)\\
Y^{I'}=aY^{I}+\frac{a}{3}Y^{II}+\frac{a}{3}Y^{III}\\
Y^{II'}=-bY^{I}+\frac{b}{3}Y^{II}+\frac{b}{3}Y^{III}\\
Y^{III'}=-\frac{2c}{3}Y^{I}+\frac{c}{3}Y^{III}\\[3pt]
\mbox{\textbf{or}}\\[3pt]
E_{8}\supset SO(6)\times SO(10)\\                                                  
 SO(6)\backsim SU(4)\\                                                                    
SU(4)\supset SU^{a}(2)\times SU^{b}(2)\times U^{III}(1)\\
SU^{a}(2)\supset U^{II}(1)\\
SU^{b}(2)\supset U^{I}(1)\\
Y^{I'}=a\,Y^{I}-a\,Y^{II}\\                                                                 
Y^{II'}=-b\,Y^{I}-b\,Y^{II}\\                                                               
Y^{III'}=-c\,Y^{III}\\                                                                          
 \begin{array}{l@{}l@{}l@{}}
 E_{8}&\supset& SO(10)\\
         &\times& U^{I'}(1)\times U^{II'}(1)\times U^{III'}(1)\\
\mbf{248}&=&\mbf{1}_{(0, 0, 0)}+\mbf{1}_{(0, 0, 0)}+\mbf{1}_{(0, 0, 0)}\\        
              &+&\mbf{45}_{(0, 0, 0)}\\                                                              
              &+&\mbf{1}_{(-2 a, 2 b, 0)}+\mbf{1}_{(2 a, -2 b, 0)}\\                       
              &+&\mbf{1}_{(-2 a, -2 b, 0)}+\mbf{1}_{(2 a, 2 b, 0)}\\
              &+&\mbf{1}_{(-2 a, 0, -2 c)}+\mbf{1}_{(2 a, 0, 2 c)}\\
              &+&\mbf{1}_{(0, -2 b, -2 c)}+\mbf{1}_{(0, 2 b, 2 c)}\\
              &+&\mbf{1}_{(-2 a, 0, 2 c)}+\mbf{1}_{(2 a, 0, -2 c)}\\
              &+&\mbf{1}_{(0, -2 b, 2 c)}+\mbf{1}_{(0, 2 b, -2 c)}\\
              &+&\mbf{10}_{(0, 0, 2 c)}+\mbf{10}_{(0, 0, -2 c)}\\
              &+&\mbf{10}_{(0, 2 b, 0)}+\mbf{10}_{(0, -2 b, 0)}\\
              &+&\mbf{10}_{(2 a, 0, 0)}+\mbf{10}_{(-2 a, 0, 0)}\\
              &+&\mbf{16}_{(a, b, c)}+\b{\mbf{16}}_{(-a, -b, -c)}\\
              &+&\mbf{16}_{(-a, -b, c)}+ \b{\mbf{16}}_{(a, b, -c)}\\
              &+&\mbf{16}_{(-a, b, -c)}+\b{\mbf{16}}_{(a, -b, c)}\\
              &+&\mbf{16}_{(a, -b, -c)}+\b{\mbf{16}}_{(-a, b, c)}
\end{array}
\end{array}$
&
$\begin{array}{l}
\left(\frac{SU(2)}{U^{I'}(1)}\right)\times\left(\frac{SU(2)}{U^{II'}(1)}\right)\\
\hfill\times\left(\frac{SU(2)}{U^{III'}(1)}\right)
\end{array}$
&
$\begin{array}{l@{}}
    SO(10)\,\Big(\times U^{I'}(1)\\
    \times U^{II'}(1)\times U^{III'}(1)\Big)
\end{array}$
&
$\begin{array}{l}
\mbf{10}_{(2 a, 0, 0)}\\
\mbf{10}_{(-2 a, 0, 0)}\\
\mbf{10}_{(0, 2 b, 0)}\\
\mbf{10}_{(0, -2 b, 0)}\\
\mbf{10}_{(0, 0, 2 c)}\\
\mbf{10}_{(0, 0, -2 c)}
 \end{array}$
&
$\begin{array}{l}
\mbf{16}_{L(a, b, c)}\\
\mbf{16}_{L(-a, -b, c)}\\
\mbf{16}_{L(-a, b, -c)}\\
\mbf{16}_{L(a, -b, -c)}\\
\mbf{16}_{L(a, b, c)}'\\
\mbf{16}_{L(-a, -b, c)}'\\
\mbf{16}_{L(-a, b, -c)}'\\
\mbf{16}_{L(a, -b, -c)}'
\end{array}$
\\\hline
$\mbf{5e} $                                                                         
& $\begin{array}{l}                                                                
 E_{8}\supset SO(6)\times SO(10)\\                                      
 SO(6)\backsim SU(4)\\
 SU(4)\supset SU^{a}(2)\times SU^b{}(2)\times U^{I}(1)\\
 \begin{array}{l@{}l@{}l@{}}
 E_{8}&\supset& (SU^{a}(2)\times SU^{b}(2)\times U^{I}(1))\\
         &\times& SO(10)\\
 \mbf{248}&=&(\mbf{1},\mbf{1},\mbf{1})_{(0)}+(\mbf{1},\mbf{1},\mbf{45})_{(0)}\\
                  &+&(\mbf{3},\mbf{1},\mbf{1})_{(0)}+(\mbf{1},\mbf{3},\mbf{1})_{(0)}\\
                  &+&(\mbf{2},\mbf{2},\mbf{1})_{(2)}+(\mbf{2},\mbf{2},\mbf{1})_{(-2)}\\
                  &+&(\mbf{1},\mbf{1},\mbf{10})_{(2)}+(\mbf{1},\mbf{1},\mbf{10})_{(-2)}\\
                  &+&(\mbf{2},\mbf{2},\mbf{10})_{(0)}\\                                                  
                  &+&(\mbf{2},\mbf{1},\mbf{16})_{(1)}+(\mbf{2},\mbf{1},\b{\mbf{16}})_{(-1)}\\
                  &+&(\mbf{1},\mbf{2},\mbf{16})_{(-1)}+(\mbf{1},\mbf{2},\b{\mbf{16}})_{(1)}
 \end{array}
 \end{array}$
& $\frac{Sp(4)\times SU(2)}{SU^{a}(2)\times SU^{b}(2)\times U^{I}(1)}$
&
   $SO(10)\,\Big(\times U^{I}(1)\Big)$
&$\begin{array}{l}
\mbf{10}_{(0)}\\
\mbf{10}_{(2)}\\
\mbf{10}_{(-2)}
\end{array}$
& $\begin{array}{l}
 \mbf{16}_{L(1)}\\
 \mbf{16}_{L(-1)}\\
\mbf{16}_{L(1)}'\\
\mbf{16}_{L(-1)}'
 \end{array}$
\end{longtable}
\end{scriptsize}
\begin{landscape}
\setlength{\tabcolsep}{2pt} 
\setlength{\arraycolsep}{1.5pt} 
\begin{scriptsize}
\begin{longtable}{|c!{\vrule width 1pt}c|c|c|c!{\vrule width 1pt}c|c|}
\caption{{\bf Application of Hosotani breaking mechanism on particle physics models which are listed in
table~\ref{SO10ChannelSCosetsTable-CSDR}}.
\label{SO10ChannelSCosetsTable-CSDR-HOSOTANI}}\\
\hline
{\bf Case}
&
\begin{tabular}{c}
 {\bf Discrete}\\
 {\bf Symme-}\\
 {\bf tries}
\end{tabular}
&
$\mbf{K'}$
&
$\begin{array}{c}
{\bf Surviving}\\
{\bf scalars}\\
{\bf under}~\mbf{K'}
\end{array}$
&
$\begin{array}{c}
{\bf Surviving~fermions}\\
{\bf under}~\mbf{K'}
\end{array}$
&
$\mbf{K}$
&
$\begin{array}{c}
{\bf Surviving~fermions}\\
{\bf under}~\mbf{K}
\end{array}$
\\[10pt]\hhline{*{7}{=}}                                                               
\endfirsthead
\hline
\multicolumn{7}{|l|}{\tiny continued from previous page}
\\\hline
{\bf Case}
&
\begin{tabular}{c}
 {\bf Discrete}\\
 {\bf Symme-}\\
 {\bf tries}
\end{tabular}
&
$\mbf{K'}$
&
$\begin{array}{c}
{\bf Surviving}\\
{\bf scalars}\\
{\bf under}~\mbf{K'}
\end{array}$
&
$\begin{array}{c}
{\bf Surviving~fermions}\\
{\bf under}~\mbf{K'}
\end{array}$
&
$\mbf{K}$
&
$\begin{array}{c}
{\bf Surviving~fermions}\\
{\bf under}~\mbf{K}
\end{array}$
\\[10pt]\hhline{*{7}{=}}
\endhead                                                                                      
\hline\multicolumn{7}{|r|}{\tiny\slshape{continued on next page}}
\\\hline
\endfoot                                                                                        
\hline
\endlastfoot
$\mbf{1a}$
&
$\begin{array}{c}
\mathrm{W}\\
\scriptscriptstyle{(\bb{Z}_{2})}
\end{array}$                                                                                     
&
 $\begin{array}{l@{}}
SU^{(i)}(2) \times SU^{(ii)}(2) \\
\times SU(4)
\end{array}$
&
$(\mbf{1},\mbf{1}, \mbf{6})$
&
 $\begin{array}{l}
 (\mbf{2},\mbf{1},\mbf{4})_{L}-(\mbf{2},\mbf{1}, \mbf{4})'_{L} \\
 (\mbf{1},\mbf{2},\mbf{4})_{L}+(\mbf{1},\mbf{2}, \mbf{4})'_{L}
 \end{array}$
&
$\begin{array}{c}
\mbox{Not}\\
\mbox{Interesting}
\end{array}$
&\\
\ \
&
$\begin{array}{c}
\mathrm{W\times Z}\\
\scriptscriptstyle{(\bb{Z}_{2}\times\bb{Z}_{2})}
\end{array}$
&
 $\mbox{''}$
&
$ (\mbf{2},\mbf{2},\mbf{1})$
&
 $+\lra -$
&
$SU^{diag}(2)\times SU(4)$
&
 $\begin{array}{l}
 (\mbf{2},\mbf{4})_{L}+(\mbf{2}, \mbf{4})'_{L} \\
 (\mbf{2},\mbf{4})_{L}-(\mbf{2}, \mbf{4})'_{L}
 \end{array}$
\\\hline 
$\mbf{2b}$   
 &
$\begin{array}{c}
\mathrm{W}\\
\scriptscriptstyle{(\one)}\\
\end{array}$                                                                                              
 &
 $\mbox{$H$ Unbroken}$
 &
 &
 &
 & \\
\ \
&
 $\begin{array}{c}
\mathrm{W\times Z}\\
\scriptscriptstyle{(\bb{Z}_{4})}
   \end{array}$
&
$\begin{array}{c}
\mbox{Not}\\
\mbox{ Interesting}
\end{array}$
&
&
&
&
\\\hline
$\mbf{3d}$, $\mbf{6d}$
&
$\begin{array}{c}
\mathrm{W}\\
\scriptscriptstyle{(\bb{Z}_{2})}                                                                   
 \end{array}$
&
$\begin{array}{l@{}}
SU^{(i)}(2)\times SU^{(ii)}(2)\\
\times SU(4)\\
 \Big(\times U^{I'}(1)\times U^{II'}(1)\Big)
\end{array}$
 &
$\begin{array}{l}
(\mbf{1},\mbf{1},\mbf{6})_{(b, 0)}\\
(\mbf{1},\mbf{1},\mbf{6})_{(-b, 0)}
\end{array}$
&
$\begin{array}{l}
(\mbf{2},\mbf{1},\mbf{4})_{L(b,-a)}-(\mbf{2},\mbf{1},\mbf{4})'_{L(b,-a)}\\
(\mbf{1},\mbf{2},\mbf{4})_{L(b, -a)}+(\mbf{1},\mbf{2},\mbf{4})'_{L(b, -a)}\\
(\mbf{2},\mbf{1},\mbf{4})_{L(-b,-a)}-(\mbf{2},\mbf{1},\mbf{4})'_{L(-b, -a)}\\
(\mbf{1},\mbf{2},\mbf{4})_{L(-b, -a)}+(\mbf{1},\mbf{2},\mbf{4})'_{L(-b, -a)}\\
(\mbf{2},\mbf{1},\mbf{4})_{L(0, a)}-(\mbf{2},\mbf{1},\mbf{4})'_{L(0, a)}\\
(\mbf{1},\mbf{2},\mbf{4})_{L(0, a)}+(\mbf{1},\mbf{2},\mbf{4})'_{L(0, a)}
\end{array}$
&
$\begin{array}{c}
\mbox{Not}\\
\mbox{Interesting}
\end{array}
$
&\\
\ \
& 
$\begin{array}{c}
\mathrm{W\times Z}\\
\scriptscriptstyle{(\bb{Z}_{2}\times\bb{Z}_{2})} 
\end{array}$
& 
$\mbox{''}$
&
$\begin{array}{l}
(\mbf{2},\mbf{2},\mbf{1})_{(b, 0)}\\
(\mbf{2},\mbf{2},\mbf{1})_{(-b, 0)}
\end{array}$
&
$+\lra-$
&
$\begin{array}{l@{}}
SU^{diag}(2)\times SU(4)\\
\Big(\times U^{I'}(1)\times U^{II'}(1)\Big)
\end{array}$
&
$\begin{array}{l}
(\mbf{2},\mbf{4})_{L(b,-a)}+(\mbf{2},\mbf{4})'_{L(b,-a)}\\
(\mbf{2},\mbf{4})_{L(b, -a)}-(\mbf{2},\mbf{4})'_{L(b, -a)}\\
(\mbf{2},\mbf{4})_{L(-b,-a)}+(\mbf{2},\mbf{4})'_{L(-b, -a)}\\
(\mbf{2},\mbf{4})_{L(-b, -a)}-(\mbf{2},\mbf{4})'_{L(-b, -a)}\\
(\mbf{2},\mbf{4})_{L(0, a)}+(\mbf{2},\mbf{4})'_{L(0, a)}\\
(\mbf{2},\mbf{4})_{L(0, a)}-(\mbf{2},\mbf{4})'_{L(0, a)}
\end{array}$
\\\hline
$\mbf{4f}$, $\mbf{7f}$ 
&
$\begin{array}{c}
\mathrm{W}\\
\scriptscriptstyle{(\bb{Z}_{2})}                                                                              
 \end{array}$
 &
$\begin{array}{l@{}}
 SU^{(i)}(2)\times SU^{(ii)}(2)\\
\times SU(4)\\
\Big(\times U^{I'}(1)\times U^{II'}(1) \\
\times U^{III'}(1)\Big)
\end{array}$
 &
$\begin{array}{l}
(\mbf{1},\mbf{1},\mbf{6})_{(0, 2 b, 0)}\\
(\mbf{1},\mbf{1},\mbf{6})_{(0, -2 b, 0)}\\
(\mbf{1},\mbf{1},\mbf{6})_{(0, 0, 2 c)}\\
(\mbf{1},\mbf{1},\mbf{6})_{(0, 0, -2 c)}
\end{array}$
&
$-$
&
$\begin{array}{c}
\mbox{Not}\\
\mbox{Interesting}
\end{array}$
&\\
\ \
&
&
&
&
&
&\\
\ \
&
$\begin{array}{c}
\mathrm{W}\\
\scriptscriptstyle{(\bb{Z}_{2})^{2}}                                                     
 \end{array}$
&
$\mbox{"}$
&
$\begin{array}{l}
(\mbf{1},\mbf{1},\mbf{6})_{(0, 0, 2 c)}\\
(\mbf{1},\mbf{1},\mbf{6})_{(0, 0, -2 c)}
\end{array}$
&
$\mbox{"}$
&
$\mbox{"}$
&\\
&\ \
&
&
&
&
&\\
\ \
&
$\begin{array}{c}
\mathrm{W\times Z}\\
\scriptscriptstyle{(\bb{Z}_{2}\times \bb{Z}_{2})}                                                 
 \end{array}$
&
$\mbox{"}$
&
$\begin{array}{l}
(\mbf{2},\mbf{2},\mbf{1})_{(0, 2 b, 0)}\\
(\mbf{2},\mbf{2},\mbf{1})_{(0, -2 b, 0)}\\
(\mbf{2},\mbf{2},\mbf{1})_{(0, 0, 2 c)}\\
(\mbf{2},\mbf{2},\mbf{1})_{(0, 0, -2 c)}
\end{array}$
&
$\mbox{"}$
&
$\mbox{"}$
&
\\\hline
$\mbf{5e} $                                                                         
&
$\begin{array}{c}                                                                         
\mathrm{W} \\
\scriptscriptstyle{(\bb{Z}_{2})}
\end{array}$
&
 $\begin{array}{l@{}}
 SU^{(i)}(2)\times SU^{(ii)}(2)\\
 \times SU(4)\\
 \Big(\times U(1)\Big)
\end{array}$
 &
 $\begin{array}{l@{}}
(\mbf{1},\mbf{1},\mbf{6})_{(0)}
\end{array}$
 & 
$\begin{array}{l@{}}
({\mbf 2},{\mbf 1},{\mbf 4})_{L(1)}-({\mbf 2},{\mbf 1},{\mbf 4})'_{L(1)}\\
({\mbf 1},{\mbf 2},{\mbf 4})_{L(1)}+({\mbf 1},{\mbf 2},{\mbf 4})'_{L(1)}\\
({\mbf 2},{\mbf 1},{\mbf 4})_{L(-1)}-({\mbf 2},{\mbf 1},{\mbf 4})'_{L(-1)}\\
({\mbf 1},{\mbf 2},{\mbf 4})_{L(-1)}+({\mbf 1},{\mbf 2},{\mbf 4})'_{L(-1)}
\end{array}$
&
$\begin{array}{c}
\mbox{Not}\\
\mbox{Interesting}
 \end{array}$
&\\
\ \
&
$\begin{array}{c}
\mathrm{W} \\
\scriptscriptstyle{(\bb{Z}_{2})^{2}}
 \end{array}$
&
$\mbox{"}$
&
$\begin{array}{l@{}}
(\mbf{2},\mbf{2},\mbf{1})_{(0)}
\end{array}$
&
$\mbox{"}$
&
$\begin{array}{l@{}}
SU^{diag}(2)\times SU(4)\\
\Big(\times U(1)\Big)
\end{array}$
&
$\begin{array}{l@{}}
({\mbf 2},{\mbf 4})_{L(1)}-({\mbf 2},{\mbf 4})'_{L(1)}\\
({\mbf 2},{\mbf 4})_{L(1)}+({\mbf 2},{\mbf 4})'_{L(1)}\\
({\mbf 2},{\mbf 4})_{L(-1)}-({\mbf 2},{\mbf 4})'_{L(-1)}\\
({\mbf 2},{\mbf 4})_{L(-1)}+({\mbf 2},{\mbf 4})'_{L(-1)}
\end{array}$\\
\ \
&
&
&
&
&
&\\
\ \
&
$\begin{array}{c}
\mathrm{W\times Z} \\
\scriptscriptstyle{(\bb{Z}_{2}\times\bb{Z}_{2})}
 \end{array}$
&
$\mbox{"}$
&
$\mbox{"}$
&
$+\lra-$
&
$\mbox{"}$
&
$\begin{array}{l@{}}
({\mbf 2},{\mbf 4})_{L(1)}+({\mbf 2},{\mbf 4})'_{L(1)}\\
({\mbf 2},{\mbf 4})_{L(1)}-({\mbf 2},{\mbf 4})'_{L(1)}\\
({\mbf 2},{\mbf 4})_{L(-1)}+({\mbf 2},{\mbf 4})'_{L(-1)}\\
({\mbf 2},{\mbf 4})_{L(-1)}-({\mbf 2},{\mbf 4})'_{L(-1)}
\end{array}$
\end{longtable}
\end{scriptsize}
\end{landscape}

\vskip 1.3cm
\section{Dimensional reduction over non-symmetric $6D$ coset spaces}\label{CSDR-NSCosets-Results}

\setlength{\tabcolsep}{1pt}
\setlength{\arraycolsep}{1pt}
\begin{scriptsize}
\begin{longtable}[t]{ |c!{\vrule width 1pt}c|c !{\vrule width 1pt} c|c|c|}
\caption{{\bf Dimensional reduction over symmetric $6D$ coset spaces}.
\emph{Particle physics models leading to  $E_{6}$ GUTs in four dimensions.}
\label{E6ChannelTable-CSDR}}\\
\hline
{\bf Case}
&
{\bf Embedding}
&
$\begin{array}{c} 
\mbf{6D}\\
{\bf Coset~Space}
\end{array}$
& 
$\mbf{H}$ 
&
$\begin{array}{c}
{\bf Surviving}\\
{\bf scalars}\\
{\bf under}~\mbf{H}
\end{array}$
&
$\begin{array}{c} 
{\bf Surviving}\\
{\bf fermions}\\
{\bf under}~\mbf{H}
\end{array}$
\\[10pt]\hhline{*{6}{=}}                                                               
\endfirsthead
\hline
\multicolumn{6}{|l|}{\tiny continued from previous page}
\\\hline
{\bf Case}
&
{\bf Embedding}
&
$\begin{array}{l} 
\mbf{6D}\\
{\bf Coset~Space}
\end{array}$
&
 $\mbf{H}$ 
&
$\begin{array}{c}
{\bf Surviving}\\
{\bf scalars}\\
{\bf under}~\mbf{H}
\end{array}$
&
$\begin{array}{c} 
{\bf Surviving}\\
{\bf fermions}\\
{\bf under}~\mbf{H}
\end{array}$
\\[10pt]\hhline{*{6}{=}}
\endhead                                                                                     
\hline\multicolumn{6}{|r|}{\tiny\slshape{continued on next page}}
\\\hline
\endfoot                                                                                        
\hline
\endlastfoot
$\mbf{2a'}$ 
&
 $\begin{array}{l}                                                         
E_{8}\supset SU'(3)\times E_{6}\\                                      
 \begin{array}{l@{}l@{}l@{}}
 E_{8}&\supset& SU'(3) \times E_{6}\\
\mbf{248}&=&(\mbf{8},\mbf{1})+(\mbf{1},\mbf{78})\\
              &+&(\mbf{3},\mbf{27})+(\b{\mbf{3}},\b{\mbf{27}})
\end{array}
\end{array}$
&
$\frac{G_{2}}{SU'(3)}$
&
$E_{6}$
&
$\begin{array}{l}
\mbf{27}\\
\b{\mbf{27}}
 \end{array}$
&
$\begin{array}{l}
\mbf{78}\\
\mbf{27}_{L}\\
\mbf{27}_{L}'
\end{array}
$
\\\hline
$\mbf{3b'}$
&
$\begin{array}{l}                                                                    
 E_{8}\supset SU'(3)\times E_{6}\\                                              
SU'(3) \supset SU^{a}(2)\times U^{I}(1)\\
Y^{I'}=-Y^{I}\\                                                                             
 \begin{array}{l@{}l@{}l@{}}
 E_{8}&\supset& SU^{a}(2)\times U^{I'}(1) \times E_{6}\\
\mbf{248}&=&(\mbf{1},\mbf{1})_{(0)}+(\mbf{1},\mbf{78})_{(0)}\\        
              &+&(\mbf{3},\mbf{1})_{(0)}\\                                            
              &+&(\mbf{2},\mbf{1})_{(-3)}+(\mbf{2},\mbf{1})_{(3)}\\
              &+&(\mbf{1},\mbf{27})_{(2)}+(\mbf{1},\b{\mbf{27}})_{(-2)}\\
              &+&(\mbf{2},\mbf{27})_{(-1)}+(\mbf{2},\b{\mbf{27}})_{(1)}
\end{array}
\end{array}$
&
$\left(\frac{Sp(4)}{SU^{a}(2)\times U^{I}(1)}\right)_{nonmax}$
&
$E_{6}\,\Big(\times U^{I'}(1)\Big)$
&
$\begin{array}{l}
\mbf{27}_{(2)}\\
\mbf{27}_{(-1)}\\
\b{\mbf{27}}_{(-2)}\\
\b{\mbf{27}}_{(1)}
\end{array}$
&
$\begin{array}{l}
\mbf{1}_{(0)}\\
\mbf{78}_{(0)}\\
\mbf{27}_{L(2)}\\
\mbf{27}_{L(-1)}\\
\mbf{27}_{L(2)}'\\
\mbf{27}_{L(-1)}'\\
\end{array}$
\\\hline       
$\mbf{4c'}$
&
$\begin{array}{l}                                                                 
 E_{8}\supset SU'(3)\times E_{6}\\                                           
SU'(3) \supset SU^{a}(2)\times U^{II}(1)\\
SU^{a}(2)\supset U^{I}(1)\\
 \begin{array}{l@{}l@{}l@{}}
 E_{8}&\supset&  E_{6}\times U^{I}(1) \times U^{II}(1)\\
\mbf{248}&=&\mbf{1}_{(0,0)}+\mbf{1}_{(0,0)}\\
              &+&\mbf{78}_{(0)}\\
              &+&\mbf{1}_{(-2,0)}+\mbf{1}_{(2,0)}\\
              &+&\mbf{1}_{(-1,3)}+\mbf{1}_{(1,-3)}\\
              &+&\mbf{1}_{(1,3)}+\mbf{1}_{(-1,-3)}\\
              &+&\mbf{27}_{(0,-2)}+\b{\mbf{27}}_{(0,2)}\\
              &+&\mbf{27}_{(-1,1)}+\b{\mbf{27}}_{(1,-1)}\\
              &+&\mbf{27}_{(1,1)}+\b{\mbf{27}}_{(-1,-1)}
\end{array}
\end{array}$
&
$\frac{SU(3)}{U^{I}(1)\times U^{II}(1)}$
&
$E_{6}\,\Big(\times U^{I}(1)\times U^{II}(1)\Big)$
&
$\begin{array}{l}                
\mbf{27}_{(0,-2)}\\                
\mbf{27}_{(-1,1)}\\                
\mbf{27}_{(1,1)}\\
\b{\mbf{27}}_{(0,2)}\\
\b{\mbf{27}}_{(1,-1)}\\
\b{\mbf{27}}_{(-1,-1)}\\
\scriptstyle{(a=0,c=-2)}\\
\scriptstyle{(b=-1,d=1)}
\end{array}$
&
$\begin{array}{l}
\mbf{1}_{(0,0)}\\
\mbf{1}_{(0,0)}\\
\mbf{78}_{(0,0)}\\
\mbf{27}_{L(0,-2)}\\
\mbf{27}_{L(-1,1)}\\
\mbf{27}_{L(1,1)}\\
\mbf{27}_{L(0,-2)}'\\
\mbf{27}_{L(-1,1)}'\\
\mbf{27}_{L(1,1)}'
\end{array}$
\\\hline
\end{longtable}
\end{scriptsize}

\setlength{\tabcolsep}{2pt}
\setlength{\arraycolsep}{1.5pt}
\begin{scriptsize}
\begin{longtable}[b]{|c!{\vrule width 1pt}c|c|c|}
\caption{\textbf{Application of Hosotani breaking mechanism on particle physics models which are listed in
table~\ref{E6ChannelTable-CSDR}.}
\emph{The surviving fields are calculated for the embeddings $\bb{Z}_{2}\hookrightarrow E_{6}$ and
$(\bb{Z}_{2}\times\bb{Z}_{2})\hookrightarrow E_{6}$, discussed in
subsections~\ref{TopologicallyInducedGaugeGroupBreaking-E6-Z2} and~\ref{TopologicallyInducedGaugeGroupBreaking-E6-Z2Z2}}
\label{E6ChannelTable-CSDR-HOSOTANI}}\\
\hline
{\bf Case}
&
\begin{tabular}{c}
{\bf Discrete}\\
{\bf Symmetries}
\end{tabular}
&
$\mbf{K'}$
&
$\begin{array}{c}
{\bf Surviving~fermions}\\
{\bf under}~\mbf{K'}
\end{array}$
\\[10pt]\hhline{*{4}{=}}                                                               
\endfirsthead
\hline
\multicolumn{4}{|l|}{\tiny continued from previous page}
\\\hline
{\bf Case}
&
\begin{tabular}{c}
{\bf Discrete}\\
{\bf Symmetries}
\end{tabular}
&
$\mbf{K'}$
&
$\begin{array}{c}
{\bf Surviving~fermions}\\
{\bf under}~\mbf{K'}
\end{array}$
\\[10pt]\hhline{*{4}{=}}
\endhead                                                                                     
\hline\multicolumn{4}{|r|}{\tiny\slshape{continued on next page}}
\\\hline
\endfoot                                                                                        
\hline
\endlastfoot
$\mbf{2a'}$ 
&
$\begin{array}{c}
\mathrm{W}\\
\scriptstyle{(\bb{Z}_{2})}\\
\scriptstyle{[embedding~(\mbf{1})]}
\end{array}$                                                                                              
&
$SO(10)\times U(1)$
&
$\begin{array}{c}
\mbf{1}_{(0)}\\
\mbf{45}_{(0)}\\
\begin{array}{l}
\mbf{1}_{L(-4)}+\mbf{1}'_{L(-4)}\\
\mbf{10}_{L(-2)}+\mbf{10}'_{L(-2)}\\
\mbf{16}_{L(1)}-\mbf{16}'_{L(1)}
\end{array}
\end{array}$\\
\ \
&
&
&\\
\ \
&
$\begin{array}{c}
\mathrm{W}\\
\scriptstyle{(\bb{Z}_{2})}\\
\scriptstyle{[embeddings}\\
\scriptstyle{(\mbf{2}),~(\mbf{3})]}
\end{array}$                                                                                              
&
$SU(2)\times SU(6)$
&
$\begin{array}{c}
(\mbf{3},\mbf{1})\\
(\mbf{1},\mbf{35})\\
\begin{array}{l}
(\mbf{1},\mbf{15})_{L}+(\mbf{1},\mbf{15})'_{L}\\
(\mbf{2},\b{\mbf{6}})_{L}-(\mbf{2},\b{\mbf{6}})'_{L}
 \end{array}
 \end{array}$
\\\hline
$\mbf{3b'}$
&
$\begin{array}{c}
\mathrm{W}\\
\scriptstyle{(\bb{Z}_{2})}\\
\scriptstyle{[embedding~(\mbf{1})]}
\end{array}$                                                                                              
&
$SO(10)\times U(1)\,\Big(\times U^{I}(1)\Big)$
&
$\begin{array}{c}
\mbf{1}_{(0,0)}\\
\mbf{1}_{(0,0)}\\
\mbf{45}_{(0,0)}\\
\begin{array}{l}
\mbf{1}_{L(-4,2)}+\mbf{1}'_{L(-4,2)}\\
\mbf{10}_{L(-2,2)}+\mbf{10}'_{L(-2,2)}\\
\mbf{16}_{L(1,2)}-\mbf{16}'_{L(1,2)}\\
\mbf{1}_{L(-4,-1)}+\mbf{1}'_{L(-4,-1)}\\
\mbf{10}_{L(-2,-1)}+\mbf{10}'_{L(-2,-1)}\\
\mbf{16}_{L(1,-1)}-\mbf{16}'_{L(1,-1)}
\end{array}
\end{array}$\\
\ \
&
&
&\\
\ \
&
$\begin{array}{c}
\mathrm{W}\\
\scriptstyle{(\bb{Z}_{2})}\\
\scriptstyle{[embeddings}\\
\scriptstyle{(\mbf{2}),~(\mbf{3})]}
\end{array}$                                                                                              
&
$SU(2) \times SU(6)\,\Big(\times U^{I}(1)\Big)$
&
$\begin{array}{c}
(\mbf{1},\mbf{1})_{(0)}\\
(\mbf{3},\mbf{1})_{(0)}\\
(\mbf{1},\mbf{35})_{(0)}\\
\begin{array}{l}
(\mbf{1},\mbf{15})_{L(2)}+(\mbf{1},\mbf{15})'_{L(2)}\\
(\mbf{2},\b{\mbf{6}})_{L(2)}-(\mbf{2},\b{\mbf{6}})'_{L(2)}\\
(\mbf{1},\mbf{15})_{L(-1)}+(\mbf{1},\mbf{15})'_{L(-1)}\\
(\mbf{2},\b{\mbf{6}})_{L(-1)}-(\mbf{2},\b{\mbf{6}})'_{L(-1)}
 \end{array}
\end{array}$\\
\ \
&
&
&\\
\ \
& 
$\begin{array}{c}                                                    
\mathrm{W\times Z}\\                                                  
\scriptstyle{(\bb{Z}_{2}\times \bb{Z}_{2})}\\
\scriptstyle{[embedding~(\mbf{2'})]}\\
\end{array}$
&
$\begin{array}{c}
SU(2)\times SU(6)\,\Big(\times U^{I}(1)\Big)
\end{array}$
&
$\begin{array}{c}
(\mbf{1},\mbf{1})_{(0)}\\
(\mbf{3},\mbf{1})_{(0)}\\
(\mbf{1},\mbf{35})_{(0)}\\
\begin{array}{l}
(\mbf{1},\mbf{15})_{L(2)}-(\mbf{1},\mbf{15})'_{L(2)}\\
(\mbf{2},\b{\mbf{6}})_{L(2)}+(\mbf{2},\b{\mbf{6}})'_{L(2)}\\
(\mbf{1},\mbf{15})_{L(-1)}-(\mbf{1},\mbf{15})'_{L(-1)}\\
(\mbf{2},\b{\mbf{6}})_{L(-1)}+(\mbf{2},\b{\mbf{6}})'_{L(-1)}
 \end{array}
\end{array}$\\
\ \
&
&
&\\
\ \
&
$\begin{array}{c}                                                    
\mathrm{W\times Z}\\                                                  
\scriptstyle{(\bb{Z}_{2}\times \bb{Z}_{2})}\\
\scriptstyle{[embedding~(\mbf{3'})]}
\end{array}$
&
$\begin{array}{c}
SU^{(i)}(2)\times SU^{(ii)}(2)\times SU(4)\times U(1)
\,\Big(\times U^{I}(1)\Big)
\end{array}$
&
$\begin{array}{c}
(\mbf{1},\mbf{1},\mbf{1})_{(0,0)}\\
(\mbf{1},\mbf{1},\mbf{1})_{(0,0)}\\
(\mbf{3},\mbf{1},\mbf{1})_{(0,0)}\\
(\mbf{1},\mbf{3},\mbf{1})_{(0,0)}\\
(\mbf{1},\mbf{1},\mbf{15})_{(0,0)}\\
\begin{array}{l}
(\mbf{1},\mbf{1},\mbf{1})_{L(4,2)}+(\mbf{1},\mbf{1},\mbf{1})'_{L(4,2)}\\    %
(\mbf{2},\mbf{2},\mbf{1})_{L(2,2)}+(\mbf{2},\mbf{2},\mbf{1})'_{L(2,2)}\\
(\mbf{1},\mbf{1},\mbf{6})_{L(-2,2)}+(\mbf{1},\mbf{1},\mbf{6})'_{L(-2,2)}\\      %
(\mbf{2},\mbf{1},\mbf{4})_{L(-1,2)}-(\mbf{2},\mbf{1},\mbf{4})'_{L(-1,2)}\\
(\mbf{1},\mbf{2},\b{\mbf{4}})_{L(1,2)}-(\mbf{1},\mbf{2},\b{\mbf{4}})_{L(1,2)}\\      %
(\mbf{1},\mbf{1},\mbf{1})_{L(4,-1)}+(\mbf{1},\mbf{1},\mbf{1})'_{L(4,-1)}\\       %
(\mbf{2},\mbf{2},\mbf{1})_{L(2,-1)}+(\mbf{2},\mbf{2},\mbf{1})'_{L(2,-1)}\\
(\mbf{1},\mbf{1},\mbf{6})_{L(-2,-1)}+(\mbf{1},\mbf{1},\mbf{6})'_{L(-2,-1)}\\         %
(\mbf{2},\mbf{1},\mbf{4})_{L(-1,-1)}-(\mbf{2},\mbf{1},\mbf{4})'_{L(-1,-1)}\\
(\mbf{1},\mbf{2},\b{\mbf{4}})_{L(1,-1)}-(\mbf{1},\mbf{2},\b{\mbf{4}})_{L(1,-1)}\\         
\end{array}
\end{array}$
\\\hline       
$\mbf{4c'}$
& 
$\begin{array}{c}
\mathrm{W}\\
\scriptstyle{(\bb{Z}_{2})}\\
\scriptstyle{[embedding~(\mbf{1})]}
\end{array}$                                                                                              
&
$SO(10)\times U(1)\,\Big(\times U^{I}(1)\times U^{II}(1)\Big)$
&
$\begin{array}{c}
\mbf{1}_{(0,0,0)}\\
\mbf{1}_{(0,0,0)}\\
\mbf{1}_{(0,0,0)}\\
\mbf{45}_{(0,0,0)}\\
\mbox{and}\\
\begin{array}{l}
\mbf{1}_{L(-4,0,-2)}+\mbf{1}'_{L(-4,0,-2)}\\
\mbf{10}_{L(-2,0,-2)}+\mbf{10}'_{L(-2,0,-2)}\\
\mbf{16}_{L(1,0,-2)}-\mbf{16}'_{L(1,0,-2)}\\
\end{array}\\
\mbox{or}\\
\begin{array}{l}
\mbf{1}_{L(-4,-1,1)}+\mbf{1}'_{L(-4,-1,1)}\\
\mbf{10}_{L(-2,-1,1)}+\mbf{10}'_{L(-2,-1,1)}\\
\mbf{16}_{L(1,-1,1)}-\mbf{16}'_{L(1,-1,1)}\\
\end{array}\\
\mbox{or}\\
\begin{array}{l}
\mbf{1}_{L(-4,1,1)}+\mbf{1}'_{L(-4,1,1)}\\
\mbf{10}_{L(-2,1,1)}+\mbf{10}'_{L(-2,1,1)}\\
\mbf{16}_{L(1,1,1)}-\mbf{16}'_{L(1,1,1)}
\end{array}1
\end{array}$\\
\ \
&
&
&\\
\ \
&$\begin{array}{c}
\mathrm{W}\\
\scriptstyle{(\bb{Z}_{2})}\\
\scriptstyle{[embeddings}\\
\scriptstyle{(\mbf{2}),~(\mbf{3})]}
\end{array}$                                                                                              
&
$SU(2)\times SU(6)\,\Big(\times U^{I}(1)\times U^{II}(1)\Big)$
&
$\begin{array}{c}
\mbf{1}_{(0,0)}\\
\mbf{1}_{(0,0)}\\
(\mbf{3},\mbf{1})_{(0,0)}\\
(\mbf{1},\mbf{35})_{(0,0)}\\
\mbox{and}\\
\begin{array}{l}
(\mbf{1},\mbf{15})_{L(0,-2)}+(\mbf{1},\mbf{15})'_{L(0,-2)}\\
(\mbf{2},\b{\mbf{6}})_{L(0,-2)}-(\mbf{2},\b{\mbf{6}})'_{L(0,-2)}
\end{array}\\
\mbox{or}\\
\begin{array}{l}
(\mbf{1},\mbf{15})_{L(-1,1)}+(\mbf{1},\mbf{15})'_{L(-1,1)}\\
(\mbf{2},\b{\mbf{6}})_{L(-1,1)}-(\mbf{2},\b{\mbf{6}})'_{L(-1,1)}
\end{array}\\
\mbox{or}\\
\begin{array}{l}
(\mbf{1},\mbf{15})_{L(1,1)}+(\mbf{1},\mbf{15})'_{L(1,1)}\\
(\mbf{2},\b{\mbf{6}})_{L(1,1)}-(\mbf{2},\b{\mbf{6}})'_{L(1,1)}
\end{array}
\end{array}$
\\\hline
\end{longtable}
\end{scriptsize}

\end{appendices}


\end{document}